\documentclass[11pt,a4paper]{article}
\usepackage{jheppub}
\usepackage{setspace} 
\usepackage{multirow}
\usepackage{colonequals}

\usepackage[page,toc]{appendix}
\usepackage{tensor}
\usepackage{etoolbox}
\usepackage{subcaption}
\AtBeginEnvironment{align}{\setcounter{subeqn}{0}}
\newcounter{subeqn} %

\usepackage{pdfpages}
\numberwithin{equation}{section}
\usepackage{cancel}
\usepackage{subcaption}
\usepackage{xparse} 
\usepackage{empheq} 

\usepackage{tabu} 
\usepackage{adjustbox} 
\usepackage{simplewick} 
\usepackage{tocloft} 
\usepackage[utf8]{inputenc} 
\usepackage{feynmp-auto} 

\newcommand{\beq}{\begin{equation}}
\newcommand{\eeq}{\end{equation}}

\newcommand{\cO}{\mathcal{O}}
\newcommand{\cJ}{\mathcal{J}}
\newcommand{\pder}[2]{\frac{\partial#1}{\partial#2}} 
\DeclareMathOperator{\tr}{tr}

\DeclareMathOperator{\sign}{sign}
\renewcommand\Re{\operatorname{Re}}

\DeclareDocumentCommand \at { o m }
{
           \IfNoValueTF {#1}
             {\big |_{#2}}
             {\left.#1\right|_{#2}}
}

\newcount\colveccount
\newcommand*\colvec[1]{
        \global\colveccount#1
        \begin{pmatrix}
        \colvecnext
}
\def\colvecnext#1{
        #1
        \global\advance\colveccount-1
        \ifnum\colveccount>0
                \\
                \expandafter\colvecnext
        \else
                \end{pmatrix}
        \fi
}

\newcommand*\pPqskip{8mu}
\catcode`,\active
\newcommand*\pPq{\begingroup
        \catcode`\,\active
        \def ,{\mskip\pPqskip\relax}%
        \dopPq
}
\catcode`\,12
\def\dopPq#1#2#3#4#5{%
        {}_{#1}\Phi_{#2}\biggl[\genfrac..{0pt}{}{#3}{#4};#5\biggr]%
        \endgroup
}

\title{Towards a full solution of the large N double-scaled SYK model}
\author{Micha Berkooz\texorpdfstring{${}^a$}{},}
\author{Mikhail Isachenkov\texorpdfstring{${}^{a, b}$}{},}
\author{Vladimir Narovlansky\texorpdfstring{${}^a$}{},}
\author{Genis Torrents\texorpdfstring{${}^a$}{}}
\emailAdd{micha.berkooz@weizmann.ac.il}
\emailAdd{isachenkov@ihes.fr}
\emailAdd{n.vladi@gmail.com}
\emailAdd{genis.torrents@weizmann.ac.il}
\affiliation{${}^a$Department of Particle Physics and Astrophysics, \\ Weizmann Institute of Science, Rehovot 7610001, Israel\\
${}^b$Institut des Hautes \'Etudes Scientifiques,\\
35 Route de Chartres, 91440 Bures-sur-Yvette, France
}

\abstract{
We compute the exact, all energy scale, 4-point function of the large $N$ double-scaled SYK model, by using only combinatorial tools and relating the correlation functions to sums over chord diagrams.
We apply the result to obtain corrections to the maximal Lyapunov exponent at low temperatures.
We present the rules for the non-perturbative diagrammatic description of correlation functions of the entire model. The latter indicate that the model can be solved by a reduction of a quantum deformation of SL$(2)$, that generalizes the Schwarzian to the complete range of energies. }

\begin{document}

%
%

\maketitle

\section{Introduction and summary of results}

The Sachdev-Ye-Kitaev (SYK) model \cite{Sachdev:1992fk,Sachdev:2010um,Kitaev:2015lct,Polchinski:2016xgd,Maldacena:2016hyu} is a quantum-mechanical model in $0+1$ dimensions, constructed out of $N$ Majorana fermions with all-to-all interactions, and random (disordered) couplings. Denote the $N$ Majorana fermions by $\psi_i$, $i=1,\cdots ,N$, satisfying the algebra $\{ \psi_i,\psi_j\}=2\delta _{ij} $. The Hamiltonian of the system is given by
\begin{equation} \label{eq:double_scaled_SYK}
H = i^{p/2} \sum _{1 \le i_1<\cdots <i_{p}\le N } J_{i_1 i_2 \cdots i_p} \psi_{i_1} \cdots \psi_{i_p} ,
\end{equation}
with the $J_{i_1\cdots i_p} $ being random couplings (whose statistics is specified below). 

Over the last few years this model has attracted considerable attention as a toy testing ground of quantum gravity, since it is a rare case which is both tractable, in the IR and other limits, and has a maximal chaos exponent. In particular, it possesses a conformal regime in the IR, and a pattern of symmetry encoded by the Schwarzian low energy effective action, which is the effective dynamics of gravity on $AdS_2$ coupled to a scalar field \cite{Almheiri:2014cka,Jensen:2016pah,Maldacena:2016upp}.

Many additional interesting models related to SYK were also studied.
Higher dimensional SYK-like models were considered in
\cite{parcollet1999non,Berkooz:2016cvq,Gu:2016oyy,Berkooz:2017efq,Murugan:2017eto} (see also \cite{Narovlansky:2018muj,Aharony:2018mjm} for disorder effects relevant to these theories).
There are also theories with similar behavior to SYK, but without disorder
\cite{Gurau:2010ba,Witten:2016iux,Klebanov:2016xxf}.
The relation of SYK to random matrix theory was discussed in
\cite{Cotler:2016fpe,Gharibyan:2018jrp}.
Higher correlation functions were studied as well
\cite{Gross:2017hcz,Gross:2017aos}, and results beyond the leading order in $N$ were obtained
\cite{Garcia-Garcia:2016mno,Garcia-Garcia:2017pzl,a:2018kvh}.

One usually studies the SYK model in the limit $N\rightarrow\infty$ and $p$ fixed. This large $N$ limit is solved by observing that the Feynman diagrams are dominated by melonic diagrams, leading to Schwinger-Dyson consistency equations for the 2-point function $\langle\psi_i(t) \psi_i(0)\rangle$; these equations can then be solved in the IR using a conformal Ansatz \cite{Sachdev:1992fk,Sachdev:2010um,Kitaev:2015lct,Polchinski:2016xgd,Maldacena:2016hyu}. Consistency equations were obtained also for the 4-point function \cite{Kitaev:2015lct,Polchinski:2016xgd,Maldacena:2016hyu}. As mentioned above, at low energies, an emergent reparametrization symmetry (which is broken both spontaneously and explicitly) selects the Schwarzian theory as the low energy limit of  the SYK model.  This fact was used to get results on the low energy limit of SYK, including the 4-point function \cite{Bagrets:2016cdf, Bagrets:2017pwq,Mertens:2017mtv,Lam:2018pvp}.
We will find it particularly useful to compare our results to \cite{Mertens:2017mtv}.

In this work, our goal is to provide a full all-energy solution to the model in a slightly different scaling, which is the double-scaled SYK \cite{erdHos2014phase,Cotler:2016fpe}. In this limit, $p$ is not held fixed (as $N\rightarrow\infty$) but rather is proportional to $\sqrt{N} $ (and is always assumed to be even), i.e.,
\begin{equation} \label{eq:double_scaling}
N \to \infty ,\qquad \lambda  = \frac{2p^2}{N} = \text{fixed} .
\end{equation}
We will also use interchangeably $\lambda $ and $q$, which is defined by
\begin{equation} \label{eq:q_def}
q \equiv e^{-\lambda } .
\end{equation}
Despite the different scaling, the model can still be used to study gravitational physics. In fact, in \cite{Cotler:2016fpe} this model was triple-scaled along with $E-E_{\text{min}} \rightarrow 0$ to obtain a model with the same IR physics as the usual SYK scaling. We will be interested in computing the exact correlation functions in the full double-scaled model rather than just in the IR (using combinatorial tools, and bypassing the Schwinger-Dyson equations).

The random couplings $J_{i_1\cdots i_p} $ are usually assumed to be independent and Gaussian. However, it is enough for us to take them to be independent random variables, with zero mean and uniformly bounded moments independent of $N$ \cite{erdHos2014phase}.\footnote{More precisely, the moments of $\binom{N}{p} ^{1/2} J_{i_1\cdots i_p} $ are uniformly bounded by a number independent of $N$.} They even need not be identically distributed, as in this limit only their standard deviation matters. They are normalized according to 
\begin{equation} \label{eq:J_variance}
\langle J_{i_1 \cdots i_p} ^2\rangle_J = \binom{N}{p} ^{-1} \cJ^2 ,
\end{equation}
where $\langle * \rangle_J$ stands for the ensemble average over the random couplings $J$.
The variance \eqref{eq:J_variance} is compatible with the one used in \cite{erdHos2014phase}, but in the scaling \eqref{eq:double_scaling} it differs from the usual SYK conventions \cite{Maldacena:2016hyu} by a constant factor of $\lambda $ only.
Without loss of generality, we will set $\cJ=1$, corresponding to normalizing the couplings such that $ \langle \tr H^2\rangle_J = 1$;\footnote{The trace here is normalized by $\tr 1 = 1$.} $\cJ$ can be restored easily by dimensional analysis.

Actually, there is a large class of models with the same reduction to chord diagrams, as the SYK model in the limit \eqref{eq:double_scaling}.
I.e., they all boil down to a very similar counting problem after averaging over the random couplings in this limit.
One of such different microscopic realizations of this universality class is analyzed in \cite{erdHos2014phase,Berkooz:2018qkz}. It consists of $N$ sites, with a spin $1/2$ variable (a qubit) at each site, on which the Pauli matrices $\sigma ^{(a)} $ ($a=1,2,3$) act. The Hamiltonian is given by an all-to-all, length $p$, random Hamiltonian
\begin{equation}\label{ErdosModel}
\tilde H = \sum _{1 \le i_1 < \cdots < i_p \le N} \sum_{a_1,\cdots ,a_p=1}^3 \alpha _{a_1,\cdots ,a_p,(i_1,\cdots ,i_p)} \sigma _{i_1} ^{(a_1)} \cdots \sigma _{i_p} ^{(a_p)} .
\end{equation}
The random couplings $\alpha $ are again independent, have zero mean, uniformly bounded moments, and variance
\begin{equation}
\langle \alpha _{a_1,\cdots ,a_p,(i_1,\cdots ,i_p)} ^2 \rangle = 3^{-p} \binom{N}{p} ^{-1}.
\end{equation}
The analysis of this model is completely analogous to that of \eqref{eq:double_scaled_SYK} in the double-scaling limit \eqref{eq:double_scaling}; the results are the same when expressed in terms of $\lambda $ or $q$, with the only difference being that $\lambda = \frac{4p^2}{3N} $ in this case. 

\subsection{Summary of results, and outline}

The main results in the paper are the following
\begin{itemize}
\item We set up the more general formalism of bi-local operators in the entire double-scaled SYK model (and not just in its low energy Schwarzian limit),
\item We compute the exact 4-point function in the model,
\item We compute the corrections to the chaos exponent as a function of the temperature,
\item We recast the formulas in a way that suggests  that all correlation functions can be computed using a quantum group symmetry.
\end{itemize}

In \autoref{sec:review} we review the class of operators under consideration and the reduction of their correlation functions to counting chord diagrams.  

Then, in \autoref{sec:bi_local_operators} we notice that in the averaged theory, the operators appear as bi-local operators (this being manifest in the chord diagrams). We explain the basic ingredients needed to evaluate higher-point functions of such operators, demonstrating them for the 2-point function, and mentioning a general feature of those bi-local operators, namely time-translation invariance.

In \autoref{sec:4_point_function} we use this formalism to write the uncrossed and the crossed 4-point functions in double-scaled SYK.
The 2-point, 4-point, and higher-point functions can be described compactly by non-perturbative diagrammatic rules given in \autoref{sec:diag_rules}. Using these diagrammatic rules, we notice that as $q \to 1^-$ and at low energies, the correlation functions calculated in this paper become those obtained for the Schwarzian.
One of the applications of having the 4-point function is the ability to compute the chaos (Lyapunov) exponent in double-scaled SYK. This is done in \autoref{sec:chaos} for small $\lambda $, after obtaining an (exact) integral expression for the R-matrix defined by the crossed 4-point function.

In \autoref{sec:quantum-def} we suggest a relation of double-scaled SYK to the Hamiltonian reduction of the motion of a quantum particle that lives on a quantum group deforming $SU(1,1)$. The finite energy spectrum that we find for finite values of $\lambda $ is consistent with the finite spectrum of the Casimir of its dual quantum group. We comment on the steps required to perform the reduction in this quantum group setting. In addition, analogously to the Schwarzian case relating the R-matrix to the $6j$-symbol of $SU(1,1)$, the R-matrix in double-scaled SYK is closely related to the $6j$-symbol of $\mathcal{U}_{q^{1/2}}(su(1,1))$. It suggests that the entire model can be solved by considering symmetry with respect to this quantum group.

In \autoref{sec:Largeq} we explain the relation of the double-scaled model to the large $p$ limit of the SYK model studied previously, and elaborate on the general structure of corrections that arise when both $p$ and $N$ are finite. Finally, \autoref{sec:LastSec} contains a summary and some future directions of research.

\section{Computing using chord diagrams} \label{sec:review}

In this section we review how to solve the model using chord diagrams (the reader familiar with these techniques can skip to the next section). Our approach in solving any of these models consists of two steps:
The first involves writing a combinatorial expression for the desired quantity, such as the partition function or correlation function, in terms of chord diagrams. This was done in \cite{erdHos2014phase} for the partition function and in \cite{Berkooz:2018qkz} for 2-point functions. The second step is to get an analytic expression from a combinatorial counting, following \cite{Berkooz:2018qkz}. We review each of these in turn.

Physics wise, we can think about chord diagrams as ``information flow'' or ``correlation flow'' diagrams. If we insert a small random perturbation in the system, we can ``fish it out'' later only by precisely the same object, as any other observable will not be correlated with the initial perturbation. The chords in the chord diagram keep track of such correlations.

\subsection{Chord diagrams} \label{subsection:chord_diagrams}

\begin{figure}[h]
\centering
\includegraphics[width=0.3\textwidth]{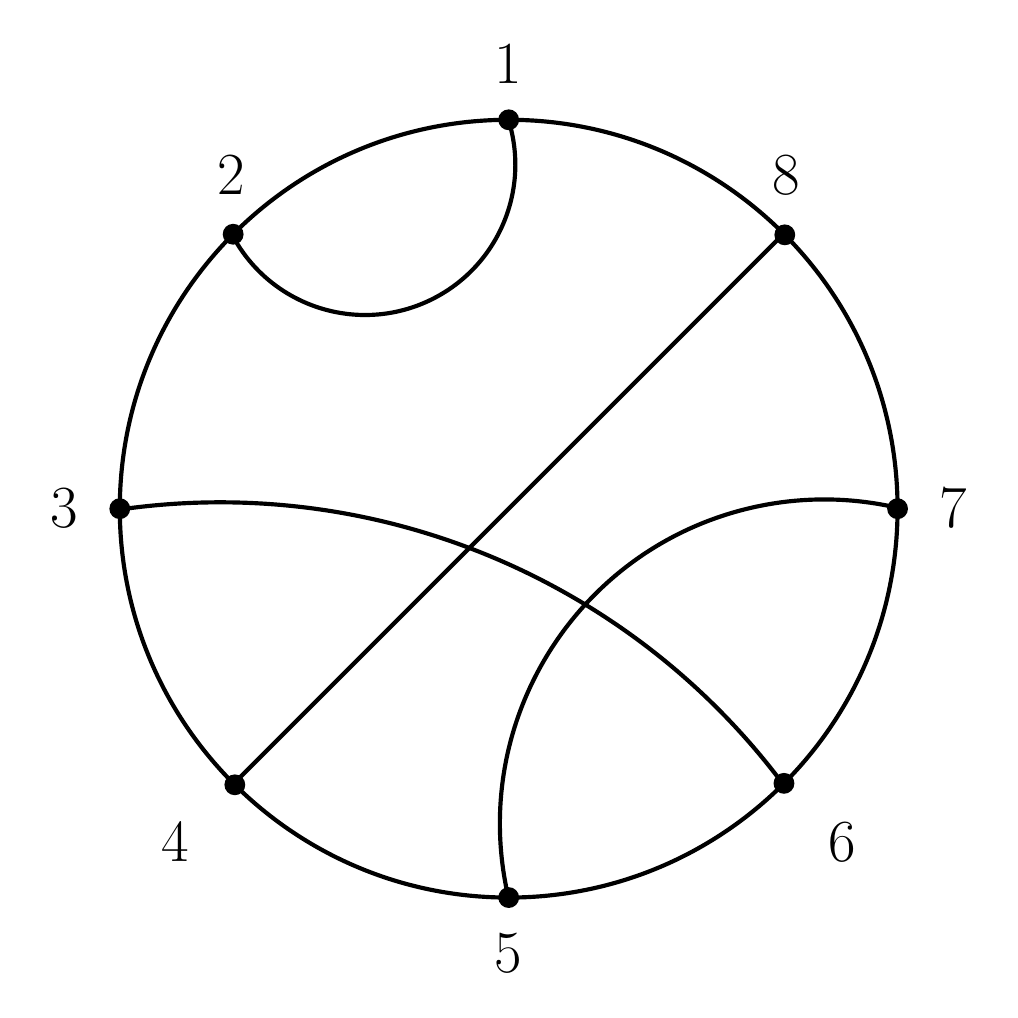}
\caption{An example of a chord diagram for $k=8$. With the numbering in the figure, this diagram means that $I_1=I_2$, $I_3=I_6$, $I_4=I_8$ and $I_5=I_7$ (and otherwise all of these are distinct). }
\label{fig:chord_diagram}
\end{figure}

Let us start by explaining what are chord diagrams, and how they arise in the double-scaling limit.
A chord diagram is a segment or a circle, with nodes marked on it, connected in pairs by chords (see \autoref{fig:chord_diagram}). These chords may intersect each other, and much of the work has to do with counting these intersections. 

 For the partition function, full details of the reduction to chord diagrams of the model \eqref{ErdosModel} appear in \cite{erdHos2014phase}, and its adaptation to Majorana fermions  \eqref{eq:double_scaled_SYK} appears in \cite{Cotler:2016fpe}. Let us start by considering the moments
\begin{equation} \label{eq:moments_def}
m_k =  \langle \tr H^k \rangle_J.
\end{equation}
From \eqref{eq:moments_def} we can then get the ensemble averaged partition function. Here and below the trace is normalized such that $\tr(1)=1$. Let us denote sets of $p$ distinct sites $i_1,\cdots ,i_p$, without ordering, by capital $I$ (with subscripts whenever there are several of those). We would like to evaluate ($k$ even)
\begin{equation} \label{eq:m_k_expr}
m_k = i^{kp/2} \sum _{I_1,\cdots ,I_k} \langle J_{I_1} \cdots J_{I_k} \rangle_J \cdot \tr\bigl( \psi_{I_1} \cdots \psi_{I_k}\bigr),
\end{equation}
where $\psi_I$ stands for the appropriate string $\psi_{i_1} \cdots \psi_{i_k} $.
We assume that the moments of the random variables $\binom{N}{p} ^{1/2} J_{i_1\cdots i_p} $ are independent of $N$. As a result, the ensemble average of the $J$'s gives an $N$ dependence of $\binom{N}{p} ^{-k/2} $. Because of the independence and zero mean of the $J$'s, among $I_1,\cdots ,I_k$ no distinct $I$ can appear once. 
Now, clearly, we have the most number of possibilities for choosing the various $\{ I_j \}$ if each distinct $I$ appears there exactly twice. Indeed, if this is not the case, then we have at most $m=\lfloor \frac{k-1}{2} \rfloor$ distinct elements in $\{ I_j \}$ and accordingly the number of terms we have in the sum is bounded above by $\binom{N}{p} ^m$ (times the number of possibilities of assigning those $m$ distinct values to $I_1,\cdots ,I_k$, which is $k$-dependent but $N$-independent). Therefore, we get an upper bound on the contribution of terms with $m$ distinct elements in $\{I_j\}$ to $m_k$, having the $N$ dependence $\binom{N}{p} ^{m-k/2} $; this goes to zero as $N \to \infty $. This is the gist of the argument made in \cite{erdHos2014phase}.

We see then, that we can restrict ourselves to having exactly $k/2$ distinct elements among the $\{ I_j \}$ in the sum.
This simple observation already accounts for the description of the problem in terms of chord diagrams: Start by representing an $H^k$ by $k$ nodes on a circle -- each node describing an insertion of $H$. The nodes are placed on a circle since the trace is cyclic. Each node is labelled by an index $j=1,\cdots, k$ (and corresponds to a string of $p$ individual $\psi_{i} $ factors). We then connect the nodes in pairs using chords, to designate which pairs have identical sets of sites $I_j$. See \autoref{fig:chord_diagram} for an example of a chord diagram. The averaging of the $J_{I_j} $ pairs simply gives $\binom{N}{p} ^{-k/2}$. 

Having \eqref{eq:m_k_expr} split into a sum over chord diagrams, we need to evaluate the expression for each of the latter, i.e., we need to compute
\begin{equation}
i^{kp/2} \binom{N}{p} ^{-k/2} \sum _{I_1,\cdots ,I_{k/2} \text{ distinct}} \tr \bigl(\psi_{I_1} \psi_{I_2}...\bigr) 
\label{eq:ChordDiagramGeneral}
\end{equation}
where the distribution of $I_1,\cdots ,I_{k/2} $ among the $k$ strings of $\psi$'s is according to the chord diagram.
For this purpose, we need to disentangle the chord diagram by exchanging nodes, so that eventually the chords connect neighboring nodes. This corresponds to commuting the strings of $\psi$'s. Upon commuting $\psi_{I_j} $ and $\psi_{I_{j'}} $, we get simply a factor of $(-1)^{|I_j \cap I_{j'}|} $ (where $|I_j \cap I_{j'}|$ is the number of sites in common to $I_j$ and $I_{j'}$). It is clear that if $p$ is small enough (for example, fixed), then with probability approaching 1, as $N \to \infty $, there will not be a chance that two randomly chosen sets of size $p$ will intersect. In our case, with $p$ scaling according to \eqref{eq:double_scaling}, as shown in \autoref{sec:Lemmas} (following \cite{erdHos2014phase}) the number of sites in the intersection $I_j \cap I_{j'}$ is Poisson distributed with mean $p^2/N$. Furthermore, different intersections are distinct with probability approaching 1 (for example, 3 different $I_j$'s do not have a common intersection).
The $\binom{N}{p} ^{-k/2} $ prefactor precisely turns counting of appearances of a certain type in the sum into probabilities of such events.
Therefore, each intersection in the chord diagram gives a factor (summing over the possibilities for the number $m$ of sites in the intersection)
\begin{equation}
\sum _{m=0} ^{\infty } \frac{(p^2/N)^m}{m!} e^{-p^2/N} (-1)^m =  e^{-\lambda } = q.
\end{equation}
After commuting the terms, the pairs are neighboring, each giving just a sign, such that all of these signs together cancel the $i^{kp/2} $.

We find that $m_k$ is given by a sum over chord diagrams, with each intersection\footnote{Connecting nodes with straight lines within the chord diagram is convenient, but of course the number of intersections can be defined without resorting to a specific way of drawing. Stated differently, the division of $1,\cdots ,k$ into pairs can be represented by a map $\pi :\{1,\cdots ,k\} \to \{1,\cdots ,k/2\}$ such that $|\pi ^{-1} (j)|=2$ (for every $1 \le j \le k/2$), and then each intersection is a pair $1 \le r,s \le k/2$ such that there exist $1 \le a<b<c<d \le k$ with $\pi (a)=\pi (c)=r$ and $\pi (b)=\pi (d)=s$.} of 2 chords simply assigned a factor of $q$. This is the {\it chord partition function}:
\begin{equation} \label{eq:moments_through_chords}
m_k = \sum _{\text{chord diagrams}} q^{\text{\# intersections}} .
\end{equation}

\subsection{Observables}

The natural observables in the SYK model are polynomials of the $\psi$ fields (time derivatives of $\psi$ are obtained by commutators of $\psi$ with $H$), i.e. operators of the form ${\cal O}_{\tilde I}=\Pi_{i\in {\tilde I}} \psi_i$, where ${\tilde I}$ is a set of indices of some fixed length $p_{\tilde I}$. Their physical behaviour depends on their length: here it is important to distinguish cases where $p_{\tilde I}$ is held fixed in the limit $N\rightarrow\infty$ from those in which it scales like $\sqrt{N}$ (or even linearly in $N$).

The standard observables, $\psi_i$ themselves or any finite strings of $\psi$'s (whose length is held fixed as $N\rightarrow\infty$), become trivial in a sense that will be made more precise as part of the discussion in \autoref{sec:Largeq}. 
Instead, we will consider operators with length $p_A$ proportional to $\sqrt{N} $ (but not necessarily of the same size as $p$), with $A$ labelling the observable when we have several of them. Furthermore, we will also take each observable operator to be random\footnote{The computations can be adapted for non-random operators as well.}, i.e.,
\begin{equation}
M_A = i^{p_A/2} \sum _{1 \le i_1 < \cdots < i_{p_A}  \le N} J^{(A)} _{i_1 \cdots i_{p_A} } \psi_{i_1} \cdots \psi_{i_{p_A} } .
\label{eq:M_Observables}
\end{equation}
 The $J^{(A)}_{i_1 \cdots i_{p_A} }  $ are again independent, have zero mean, uniformly bounded moments (which for $\binom{N}{p_A} ^{1/2} J^{(A)} _{i_1 \cdots i_{p_A} } $ are independent of $N$), and
 \begin{equation} \label{eq:operators_couplings_variance}
 \langle J^{(A)} _{i_1 \cdots i_{p_A} } J^{(B)} _{j_1 \cdots j_{p_B} } \rangle_J = \binom{N}{p_A} ^{-1}  \delta ^{AB} \delta _{i_1,j_1} \delta _{i_2,j_2} \cdots .
 \end{equation}
We will take them to be independent of the random couplings in the Hamiltonian.\footnote{It does not matter if we introduce additional constants $\cJ^{(A)} $ for these operators, as the dependence on them in a given correlation function is trivial.}

 The argument for choosing random operators in this way is that we think about these operators as the analogue of single-trace operators familiar in the context of the AdS/CFT correspondence. The latter are not very different from the local energy-momentum tensor. So we would like to discuss probes which have similar statistical properties as the Hamiltonian, but with a different dimension. This is implemented by \eqref{eq:M_Observables} (for a more detailed discussion see \cite{Berkooz:2018qkz}).

In order to evaluate the 2-point function, instead of calculating the moments $m_k$ as above, we need to consider the objects
\begin{equation} \label{eq:2pf_ingredients}
\langle \tr H^{k_3} M H^{k_2} M H^{k_1} \rangle_J .
\end{equation}
By summing such terms over the $k_j$'s, with appropriate weights, we can get in an obvious way the 2-point function of the operator $M$ at any temperature and time separation.

\begin{figure}[h]
\centering
\includegraphics[width=0.3\textwidth]{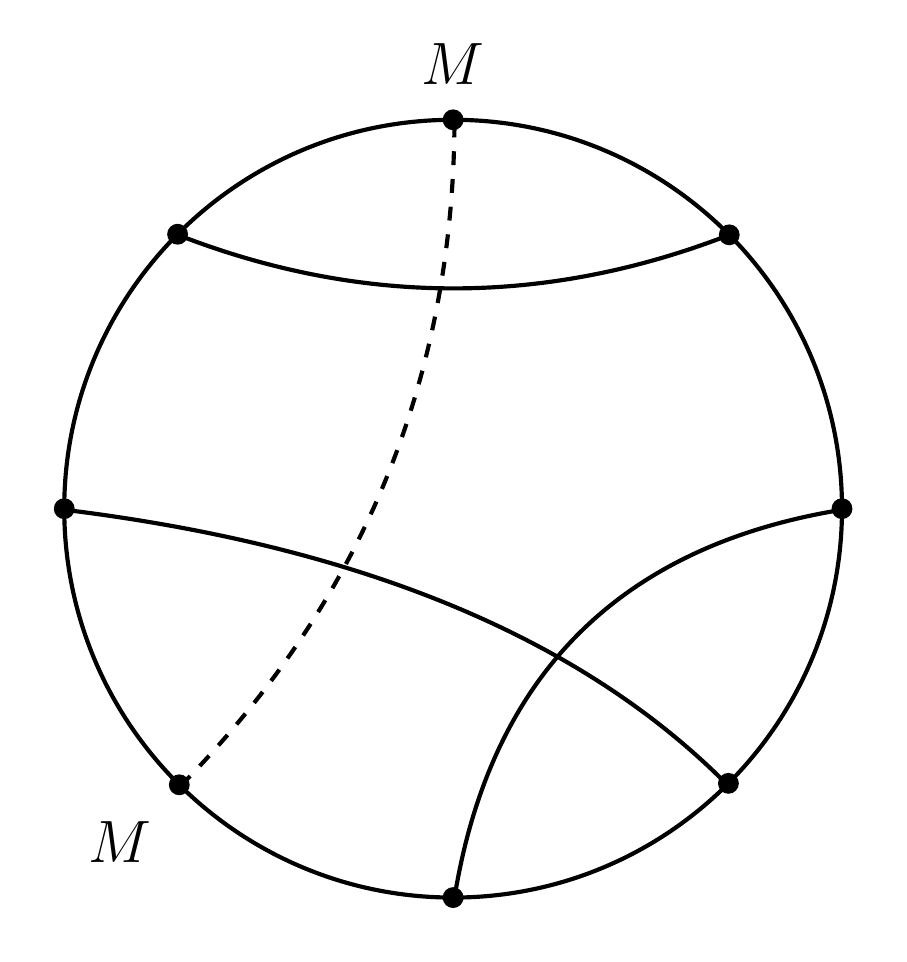}
\caption{A chord diagram contributing to $ \langle \tr M H^2 M H^4 \rangle_J$. A chord connecting 2 $M$-operators is distinguished from the Hamiltonian chords, and is represented here by a dashed line.}
\label{fig:chord_diagram_with_operators}
\end{figure}

The first step in the calculation of \eqref{eq:2pf_ingredients}, or the correlator for any number of such operators $M_A$, is to get the chord diagram description, in complete analogy to what was done in \autoref{subsection:chord_diagrams}. We write \eqref{eq:2pf_ingredients} in an expression analogous to \eqref{eq:m_k_expr}, and carry out two ensemble averages --  both on the $J$'s, the random couplings in the Hamiltonian, and the $J^{(A)}$'s, the random coefficients in $M_A$, and these two ensemble averages are separate and independent. For the same reason as before, even if we have several insertions of $M$'s, the only contributions come from those terms in which the insertions of the operators come in pairs, and the insertions from the Hamiltonian terms come in pairs as well. For an arbitrary number of operator insertions $M_A$, only operators of the same flavor $A$ can be paired because of \eqref{eq:operators_couplings_variance}. This means that for each pairing (of all the Hamiltonian and operator couplings) we get an appropriate chord diagram with nodes corresponding to the Hamiltonian insertions, and separate nodes corresponding to the operator insertions. For example, in \eqref{eq:2pf_ingredients}, we will have $k_3$ Hamiltonian nodes, then a single $M$ operator node, then $k_2$ Hamiltonian nodes, another $M$ operator node, and lastly $k_1$ Hamiltonian nodes. The two operator $M$ nodes must be paired with each other.  An example of a chord diagram contributing to \eqref{eq:2pf_ingredients} is shown in \autoref{fig:chord_diagram_with_operators}.
We will represent all the chords that connect operator ($M_A$) insertions in the same way, as dashed lines, but specify explicitly the type of operators at the nodes.

In commuting the different strings of $\psi$'s, we still get a $(-1)^{|I_j \cap I_{j'}|} $ for each commutation. The only difference is the probability distribution of the number of individual $\psi_i$'s in the intersection of two chords. We mentioned before that the number of $\psi_i$'s in the intersection of two sets of size $p$ is Poisson distributed with parameter $p^2/N$. Similarly, the intersection of the sets of size $p$ with the sets of size $p_A$ is Poisson distributed with parameter $pp_A / N$ (\autoref{sec:Lemmas}). Summing over the possibilities for the number of $\psi_i$'s in the intersection of sets of size $p$ and sets of size $p_A$, weighted by the Poisson probability distribution, and by the $(-1)^{|I_j \cap I_{j'}|} $ from the commutation, gives $e^{-2pp_A / N} $.

To summarize, in order to calculate an expression $\langle \tr \left( \cdots H^{k_3} M_{B} H^{k_2} M_{A} H^{k_1} \right) \rangle_J$, one should sum over all the chord diagrams with several types of chords -- $M_{A,B ,\cdots}$-chords connecting the nodes that correspond to operator insertions on the circle (a different kind of chord for every flavor of the insertion), and usual Hamiltonian chords connecting Hamiltonian nodes. Each chord diagram is assigned a factor of $q$ for every intersection of two Hamiltonian chords, a factor of
\begin{equation}
\tilde q_A = e^{- 2p \cdot p_A/N } 
\end{equation}
for an intersection of an Hamiltonian chord, or `$H$-chord' in short, with an `$M_A$-chord', and a factor of
\begin{equation}
\tilde q_{AB} =e^{-2p_A \cdot p_B / N}
\end{equation}
for an intersection of an $M_A$-chord with an $M_B$-chord. In short
\begin{equation}
\begin{split}
&  \langle \tr \left( \cdots H^{k_3} M_{B} H^{k_2} M_{A} H^{k_1} \right) \rangle_J = \\
& \qquad = \sum _{\text{chord diagrams}} q^{\text{\# $H$-$H$ intersections}} \prod _A \tilde q_A^{\text{\# $H$-$M_A$ intersections}} \prod _{A,B} \tilde q_{AB} ^{\text{\# $M_A$-$M_B$ intersections}} .
\end{split}
\end{equation}

\subsection{Analytic evaluation}

\begin{figure}[h]
\centering
\includegraphics[width=0.9\textwidth]{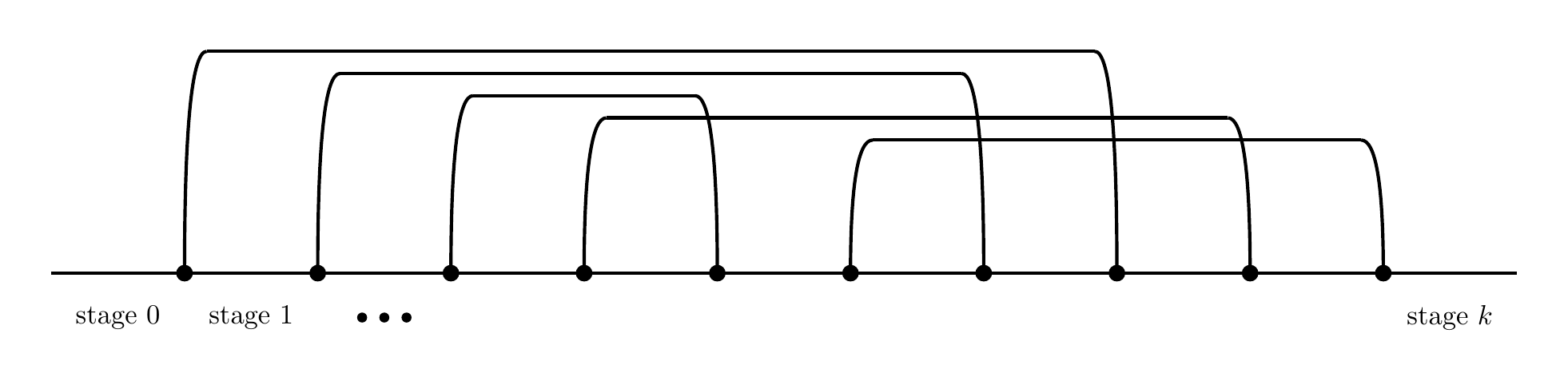}
\caption{Cutting open a chord diagram.}
\label{fig:cutting_open_chord_diagram}
\end{figure}

Now we will review how to evaluate the chord partition function \eqref{eq:moments_through_chords}, following \cite{Berkooz:2018qkz} (in the next section we will evaluate the 2-point function using a similar method, generalizing \cite{Berkooz:2018qkz}). Here it is more convenient to cut open the chord diagrams at any chosen (fixed) point (it does not matter which one we choose), so that the $k$ nodes lie on a line rather than a circle (\autoref{fig:cutting_open_chord_diagram}).

Introduce an auxiliary Hilbert space $ \mathcal{H}_{aux} $, with basis vectors $|l\rangle$ enumerated by $l=0,1,2,\cdots $, where $l$ represents number of open chords. We will consider the evolution of the system from left to right. At each stage between two adjacent nodes, $i=1,\cdots, k-1$, there is a particular number of open chords, see \autoref{fig:cutting_open_chord_diagram}. Let us label by $i=0$ and $i=k$ the stage before the first node and the stage after the last node, respectively, in which there are always no open chords.
To sum the chord diagrams, it is natural to introduce partial sums over chord diagrams as follows.
Denote by $v_l^{(i)} $ the sum over all possible (open) chord diagrams from stage $0$ up to stage $i$, having $l$ open chords at stage $i$, each chord diagram weighted by $q$ to the power of the number of intersections (until this point).

\begin{figure}[h]
\centering
\includegraphics[width=0.5\textwidth]{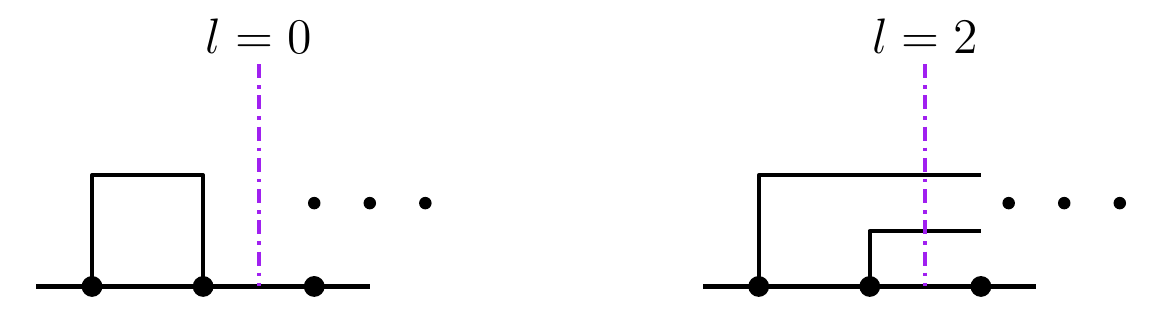}
\caption{An example for partial open chords, until stage 2.}
\label{fig:example_v}
\end{figure}

For example, for $i=2$ there are two possible open chord diagrams (see \autoref{fig:example_v}), giving at stage 2 either $l=0$ or $l=2$ chords. In both of them there are no intersections, and therefore, writing $v_l^{(i)} $ in the auxiliary Hilbert space, we have $v^{(2)} =(1,0,1,0,0,\cdots )$.

\begin{figure}[t!]
    \centering
    \begin{subfigure}[t]{0.45\textwidth}
          \centering
          \includegraphics[height=0.25\textwidth]{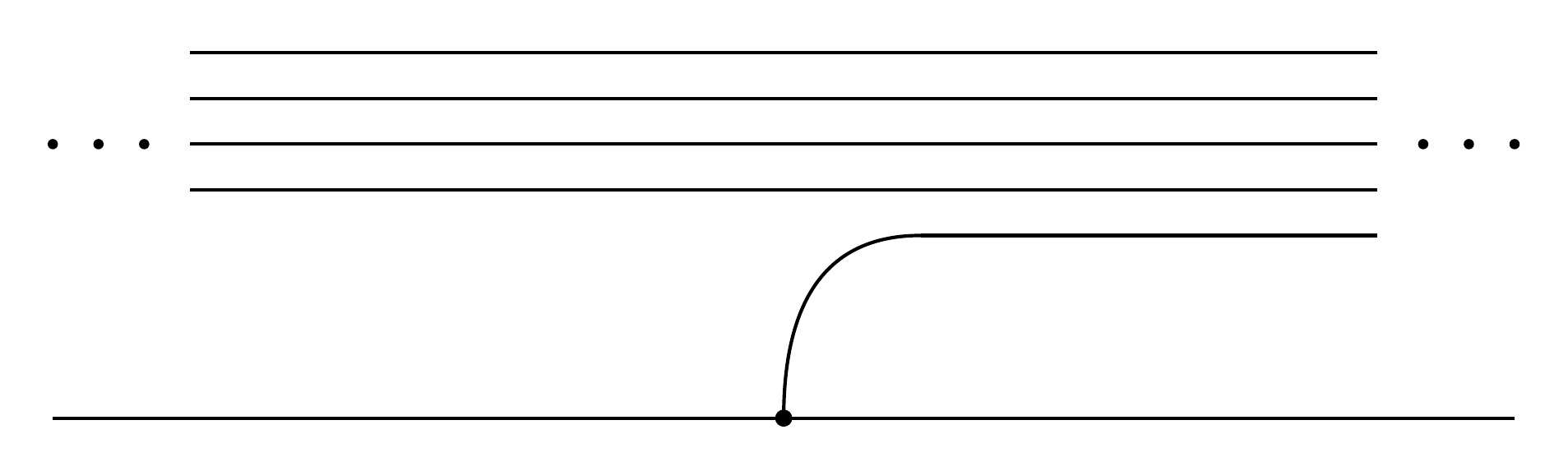}
          \caption{}
          \label{fig:constructing_chords_case_1}
    \end{subfigure}
    \begin{subfigure}[t]{0.45\textwidth}
          \centering
          \includegraphics[height=0.25\textwidth]{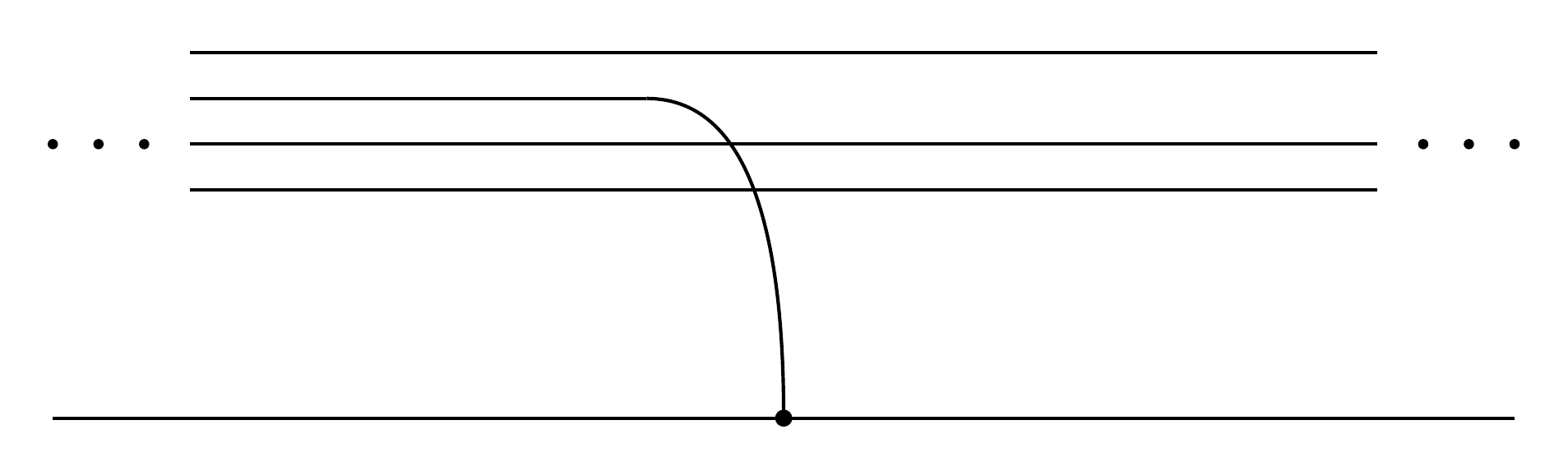}
          \caption{}
          \label{fig:constructing_chords_case_2}
    \end{subfigure}
    \caption{At any given node, either a chord opens (as in the left), or a chord is closed (as in the right), crossing all the chords below it.}
\end{figure}

$v_l^{(i)} $ satisfy a simple recursion relation. Given that there are $l$ open chords at stage $i$, there are two possibilities for the number of chords in the previous stage. The first possibility is that there are $l-1$ chords at the previous stage, and then in the following node a new chord is opened (see \autoref{fig:constructing_chords_case_1}). We keep such new opened chords below all the other chords that are accumulated from previous stages, in order to avoid unnecessary intersections (and introduce intersections only when necessary).  The second possibility is that in the previous stage there are $l+1$ chords, and one is closed in the following node. Any open chord out of the $l+1$ chords can close at this point, and by doing so it will intersect all the chords below it (see \autoref{fig:constructing_chords_case_2}). This gives us $l+1$ possibilities for chord diagrams, each giving an extra factor of $q^0,q^1,\cdots ,q^l$ according to the choice of which chord is closed. Therefore
\begin{equation}
v_l^{(i+1)} = v_{l-1} ^{(i)} + (1+q+q^2+\cdots q^l) v_{l+1} ^{(i)} = v^{(i)} _{l-1} + \frac{1-q^{l+1} }{1-q} v^{(i)} _{l+1} .
\end{equation}

In the auxiliary Hilbert space with basis $|l\rangle$ ($l=0,1,2,\cdots $), this can be written as $v^{(i+1)} =T v^{(i)} $ where the transfer matrix $T$ is
\begin{equation} \label{eq:T_matrix}
T = \begin{pmatrix}
0 & \frac{1-q}{1-q} & 0 & 0 & \cdots \\
1 & 0 & \frac{1-q^2}{1-q} & 0 & \cdots \\
0 & 1 & 0 & \frac{1-q^3}{1-q} & \cdots \\
0 & 0 & 1 & 0 & \cdots \\
\vdots & \vdots & \vdots & \vdots & \ddots
\end{pmatrix} .
\end{equation}
At stage $i=0$ there are no open chords and therefore $v^{(0)} =|0\rangle$. Moreover, since all the chords have to close by node $k$, in order to get those chord diagrams we should project $v^{(k)} $ on $|0\rangle$. Thus,
\begin{equation}
m_k = \sum _{\text{chord diagrams}} q^{\text{\# intersections}} = \langle 0| T^k |0 \rangle .
\end{equation}
Comparing this to \eqref{eq:moments_def}, it is natural to identify $T$ with the Hamiltonian of the auxiliary system. Any function of $H$ inside a trace in the original Hilbert space, can be replaced by the same function of $T$ taken between the states $\langle 0|$ and $| 0\rangle$.  As suggested in \cite{Berkooz:2018qkz}, we can think about ${\cal H}_{aux}$ and $T$ as encoding the dual gravitational background.

This (infinite) transfer matrix $T$ can be diagonalized (in fact, there is a similarity transformation bringing it to a symmetric form \cite{Berkooz:2018qkz}), with the eigenvalues playing the role of energies $E$. It was shown in \cite{Berkooz:2018qkz} that these energies are parametrized by $\theta \in [0,\pi ]$  and are given by
\begin{equation}
E(\theta ) = \frac{2 \cos \theta }{\sqrt{1-q} } .
\end{equation}
We saw that the auxiliary Hilbert space is essentially the same as the Hilbert space of a harmonic oscillator. Actually, we will see later that it is related to a $q$-deformation of the harmonic oscillator algebra (as already noted in \cite{Berkooz:2018qkz}). The eigenstates of the auxiliary system $T$ (parameterized by $\theta $) are proportional to the continuous $q$-Hermite polynomials $H_l$:
\begin{equation} \label{eq:T_eigenvectors}
\langle l | \psi^{(\theta )} _T \rangle = \sqrt{(q;q)_{\infty } } | (e^{2i\theta } ;q)_{\infty } | (1-q)^{l/2} \frac{H_l(\cos \theta |q)}{\sqrt{2\pi } (q;q)_l} .
\end{equation}
The $q$-Pochhammer symbol $(a_1,a_2,\cdots ;q)_n$ is defined by
\begin{equation}
\begin{split}
& (a_1,a_2,\cdots ;q)_n = (a_1;q)_n (a_2;q)_n \cdots ,\\
& (a;q)_n = \prod _{k=1} ^n (1-aq^{k-1} ),
\end{split}
\end{equation}
and the $q$-Hermite polynomials are defined by the recursion relation \eqref{eq:q_Hermite_def}, with more details on them given in \autoref{app:special_functions}.

Any matrix element $\langle l_1|T^k|l_2\rangle$ can be evaluated by using the diagonal basis\footnote{This is most conveniently derived using a similarity transformation to get to the symmetric form of $T$ for which the eigenvectors are orthonormal.}
\begin{equation} \label{eq:Tab_expression}
\langle l_1 | T^k | l_2 \rangle = (1-q)^{(l_1-l_2)/2} \int _0^{\pi } \frac{d\theta}{2\pi} (q, e^{\pm 2i\theta };q)_{\infty } \, \left( \frac{2\cos(\theta )}{\sqrt{1-q} } \right)^{k} \frac{H_{l_1}(\cos \theta |q) H_{l_2}(\cos \theta |q)}{(q;q)_{l_1}}  .
\end{equation}
In this expression and below, we adopt the convention that the appearance of each $\pm $ stands for a product of the corresponding term with a plus sign by the same term with a minus sign; for instance $(e^{\pm 2i\theta } ;q)_{\infty } \equiv (e^{2i\theta } ;q)_{\infty } (e^{-2i\theta } ;q)_{\infty } $.

In particular, this leads to an analytic expression for the moments
\begin{equation} \label{eq:partition_function_moment_integral_form}
m_k =  \int _0^{\pi } \frac{d\theta}{2\pi} (q, e^{\pm 2i\theta };q)_{\infty } \, \left( \frac{2\cos(\theta )}{\sqrt{1-q} } \right)^{k} ,
\end{equation}
which were written earlier, using a different method, in \cite{erdHos2014phase}. These values of $m_k$ are the moments of what is known as the $q$-normal distribution defined in \cite{szablowski2010} (note that there are several different distributions with this name). The pointwise (in energy) $q \to 1^-$ limit of the $q$-normal distribution is given by the usual normal distribution, while the $q \to 0^+$ limit of it is the Wigner semicircle distribution. 

The partition function is obtained by replacing the power of the energy by an exponential of it. For the partition function, as well as some correlation functions, we carry out the integral over $\theta $ in \autoref{sec:non_integral_forms}, and obtain them as sums over Bessel functions.

Note that it may be more convenient to think about the Hamiltonian of the auxiliary system ${\cal H}_{aux} $ as a symmetric version of $T$ (so that it is Hermitian in this basis), obtained by a similarity transformation $\hat T = PTP^{-1} $ with $P_{l,l'} =\sqrt{(q;q)_l} (1-q)^{-l/2} \delta _{l,l'} $ \cite{Berkooz:2018qkz}.

\subsection{Relation to quantum groups}

The $T$ operator \eqref{eq:T_matrix} acting on the bosonic Fock space $\ell^2(\mathbb{Z}_{\geq 0})$, spanned by $|l\rangle$, can be naturally seen as living in a representation of so called (Arik-Coon) $q$-oscillator algebra $\mathcal{D}$ \cite{Arik-Coon}. This algebra is formed by three generators $A, A^+, N$ with relations: 
\begin{align} \label{eq:q_oscillator_algebra}
[N,A]=-A, \quad [N, A^+]=A^+, \quad AA^+-qA^+A=1.
\end{align}
In fact, there is more structure: it can be built up to a proper (real form of) quantum group, i.e. a Hopf $*$-algebra. In particular, the $*$-structure is given by $A^*=A^+$, $N^*=N$.

The representation of $\mathcal{D}$ in question is via
\begin{align}
N|n\rangle=n|n\rangle, \quad A|n\rangle = [n]_q^{1/2}|n-1\rangle, \quad A^+|n\rangle = [n+1]^{1/2}_q|n+1\rangle,
\end{align}
where $[n]_q=(1-q^{n} )/(1-q)$ is the $q$-number.
The ${\hat T}$ operator is a position operator ${\hat T}=A+A^+$.

As we will discuss in some details in \autoref{sec:quantum-def}, double-scaled SYK is related to the quantum group $U_q(sl(2,\mathbb{C}))$, generated by $K^{\pm 1}, E, F$ obeying 
\begin{equation*}
K K^{-1}=K^{-1} K=1, \quad K E K^{-1}= q E, \quad K F K^{-1}= q^{-1} F, \quad EF - FE=\frac{K^2-K^{-2}}{q-q^{-1}},
\end{equation*}
see appendix \ref{app:u_qsu(1,1)} for a complete definition.
It is known that our $q$-oscillator algebra \eqref{eq:q_oscillator_algebra} can be obtained as a simple contraction of this algebra \cite{Kulish1993}:
\begin{align}
A=\lim_{f \to 0} f (q-q^{-1})^{1/2} E, \quad A^+=\lim_{f \to 0} f (q-q^{-1})^{1/2} F, \quad q^{-N}=\lim_{f \to 0} f K.
\end{align}

\section{Bi-local observables in double-scaled SYK} \label{sec:bi_local_operators}

In this section we will develop some basic tools which are needed in order to compute exact general $n$-point functions in the double-scaling limit, i.e.
\begin{equation}\label{n-pt sec 3}
\langle \tr  e^{-\beta H} M_1(t_1) M_2(t_2) \cdots M_n(t_n)\rangle_J .
\end{equation}
The main focus will be the bi-local operators in the theory, which come about when two insertions of given observables are connected in a chord (as will be explained below). This object can then be inserted in various ways into a correlation function to compute a more general $n$-point function. 

As discussed in \autoref{sec:review}, evaluating \eqref{n-pt sec 3} is done by expanding each time evolution operator, $e^{-\beta H}$ or $e^{it H}$,  in a power series in $H$, computing each term of the form
\begin{equation}\label{eq:Obs_Expand}
\langle \tr  H^{k_1} M_1 H^{k_2} M_2 H^{k_3} \cdots M_n H^{k_{n+1}}\rangle_J
\end{equation} 
using chord diagrams and then resumming these expressions\footnote{With required precautions, regarding e.g. absolute convergence when changing order of summations.}. Of course, with the transfer matrix \eqref{eq:T_matrix} we implement this by directly exponentiating $H$, and this can be conveniently done in the eigenbasis of $T$.

Both the $H$ and the $M_A$ operators, of length $p$ and $p_A$ respectively, involve random couplings. We carry out the ensemble average over these random coefficients, which pairs up the $H$ and the $M_A$'s. The basic appearance of an operator is therefore in a pair, i.e. as part of a bi-local operator. In this section we will give a general construction of bi-local operators using combinatorics (in \autoref{sec:propagation_M_region}), which can be inserted into a large class of $n$-point functions. In \autoref{sec:2_point_function} we will check the formalism on the 2-point function, and in the following sections we study the 4-point function.
Bi-local operators appear also in the discussion of the Schwarzian \cite{Mertens:2017mtv} but their origin here is different and comes from the statistics of the random coefficients in the operator.
In \autoref{sec:trans_inv} we point out a general feature of the bi-local operator. 

\subsection{Chord construction for bi-local operators} \label{sec:propagation_M_region}

A convenient starting point is the 2-point function of an operator $M$, made up of a string of $\psi$'s of length $p'$, with an associated parameter $\tilde q=e^{-2pp'/N} $. To get the thermal 2-point function, as mentioned above, one needs to evaluate \eqref{eq:2pf_ingredients} using chord diagrams; one such chord diagram contribution was shown in \autoref{fig:chord_diagram_with_operators}. To do this, we have to cut open the chord diagram. Any cut is allowed and should give the same result. 

\begin{figure}[h]
\centering
\includegraphics[width=0.8\textwidth]{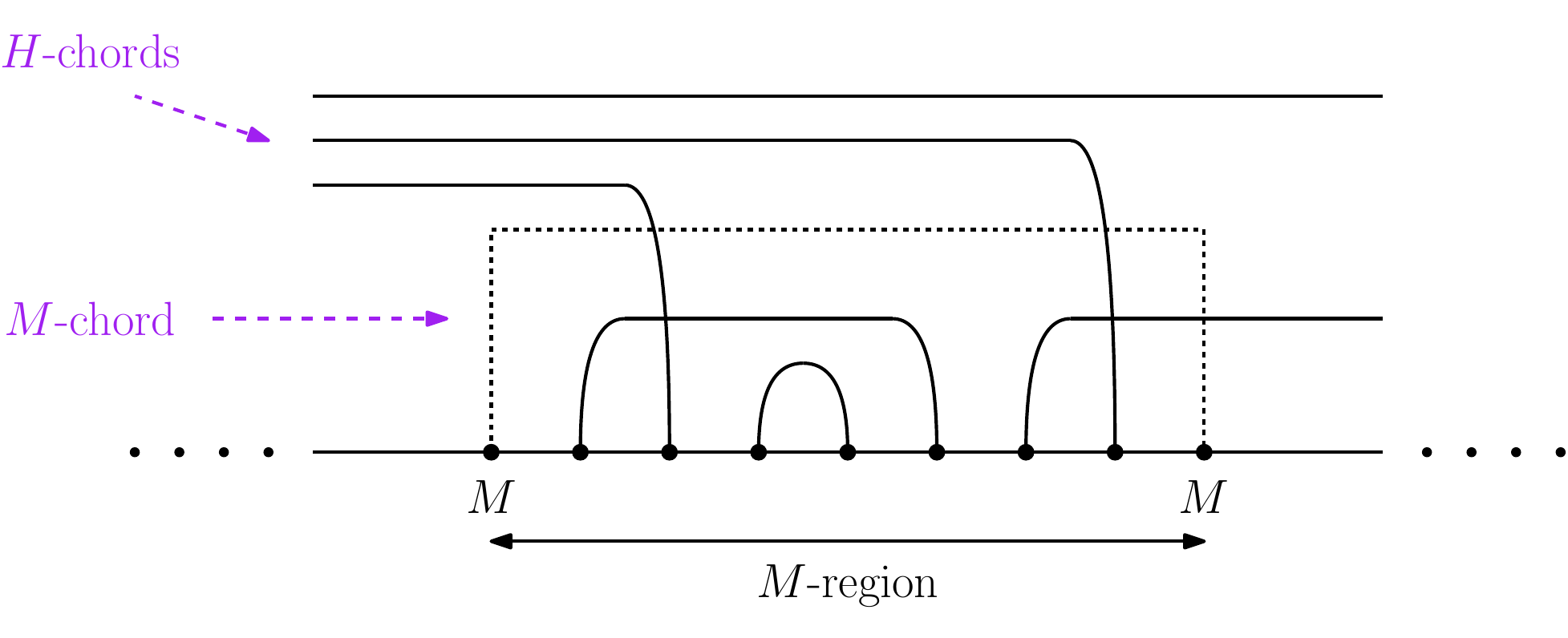}
\caption{A subregion consisting of two $M$ nodes which are contracted.}
\label{fig:propagation_M_region}
\end{figure}

In \cite{Berkooz:2018qkz}, the cut was chosen to be placed right after one $M$ insertion; this choice makes the calculation very similar to what was done in \autoref{sec:review}. We will not review this here. Instead, we will explain how \eqref{eq:2pf_ingredients} can be evaluated with a cut at an arbitrary point. 
Actually, we will describe the evolution, in the auxiliary Hilbert space, over a region (in the chord diagram) between two $M$ nodes that are paired (we shall also use below `contracted' for `paired'). 
There can be anything to the left and to the right of these two insertions. This means that we can then take this bi-local operator and insert it into more general $n$-point functions. We will refer to the region made out of the stages between the two $M$ nodes as the `$M$-region' for brevity -- this is shown, for a particular example, in \autoref{fig:propagation_M_region}. 

In the approach of \autoref{sec:review}, we were keeping track at each stage of the state of the system (number of chords in the case of the partition function) and the correct intersection factors up to that stage. One could, in principle, follow that approach in the case at hand, which would require us to keep track of the number of $H$-chords below and above the $M$-chord at each stage. We will not do that here and instead use a somewhat ``less local'' approach, 
that exonerates us from
keeping track of the correct intersection factors at any given stage, but 
still accounts for them in the end.

Let us use the same ${\cal H}_{aux}$ as before, counting the total number of $H$-chords at a given stage (ignoring the $M$-chords). Moreover, let us introduce a factor of $\tilde q$ for each $H$-chord incoming into the $M$-region, and another factor of $\tilde q$ for each $H$-chord leaving the $M$-region (say as we go from left to right).
This is achieved by an insertion of the matrix
\begin{equation} \label{eq:def_S_matrix}
S = \begin{pmatrix}
1 & 0 & 0 & \cdots \\
0 & \tilde q & 0 & \cdots \\
0 & 0 & \tilde q^2 & \cdots \\
\vdots & \vdots & \vdots & \ddots
\end{pmatrix} \qquad 
\end{equation}
right before and right after the $M$-region.

\begin{figure}[h]
\centering
\includegraphics[width=0.8\textwidth]{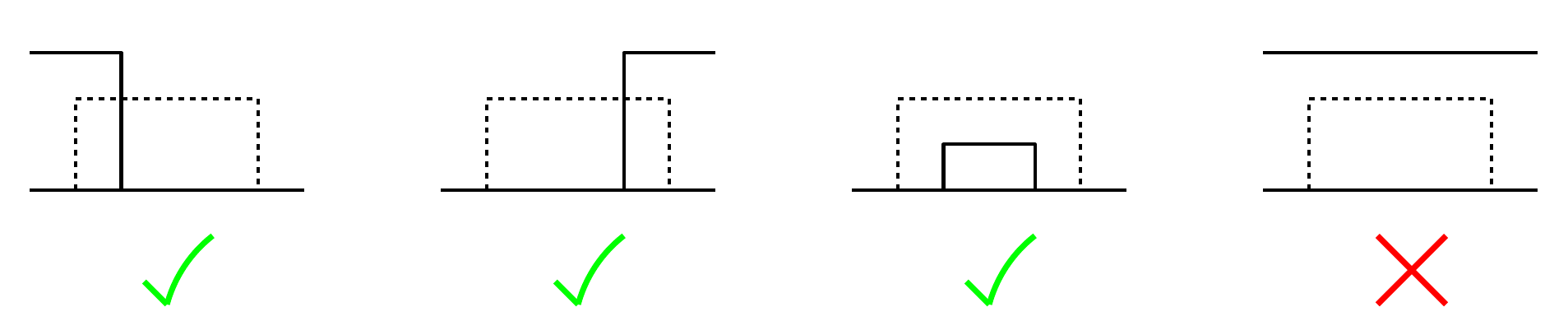}
\caption{Counting scheme for correlators. The last case is assigned a value of $\tilde q^2$ in our counting, while it should get no $\tilde q$ factors.}
\label{fig:counting_correlators}
\end{figure}

This almost accounts correctly for the $\tilde q$ factors. Namely, chords incoming to the $M$-region and closing inside it, chords created in the $M$-region and leaving it, and chords created and closed inside the $M$-region, all get the correct power of $\tilde q$. The only problem is that a chord incoming to the $M$-region but not closing inside it (i.e., 
going through the whole $M$-region) is assigned a factor of $\tilde q^2$, while it was supposed to give no $\tilde q$ factors (since it does not cross the $M$-chord), see \autoref{fig:counting_correlators}.

Let us fix the number of chords which are incoming to the $M$-region (not necessarily crossing the $M$-chord) to be $l$.
The chord diagrams with one chord out of $l$ not closing in the $M$-region receive an unnecessary factor of $\tilde q^2$ as mentioned. It is useful to separate this chord from the others, so that it appears above all other chords.
We can correct this unnecessary factor if, by hand, we  consider the evolution, inside the $M$-region, of an incoming set of $l-1$ chords, multiplied by $1-\tilde q^2$, adding one more chord to them (this additional chord does not give any contribution inside the $M$-region since it does not intersect any other chord there). But then, in the sum of these two classes, the chord diagrams with $2,3,\cdots $ chords not closing inside the $M$-region get an incorrect $\tilde q$ dependence. It should now be clear how to deal with all of those. We should add all the chord diagrams with $l-i$ incoming chords, to which $i$ chords are added eventually, multiplied by an appropriately chosen coefficient.

\begin{figure}[h]
\centering
\includegraphics[width=0.8\textwidth]{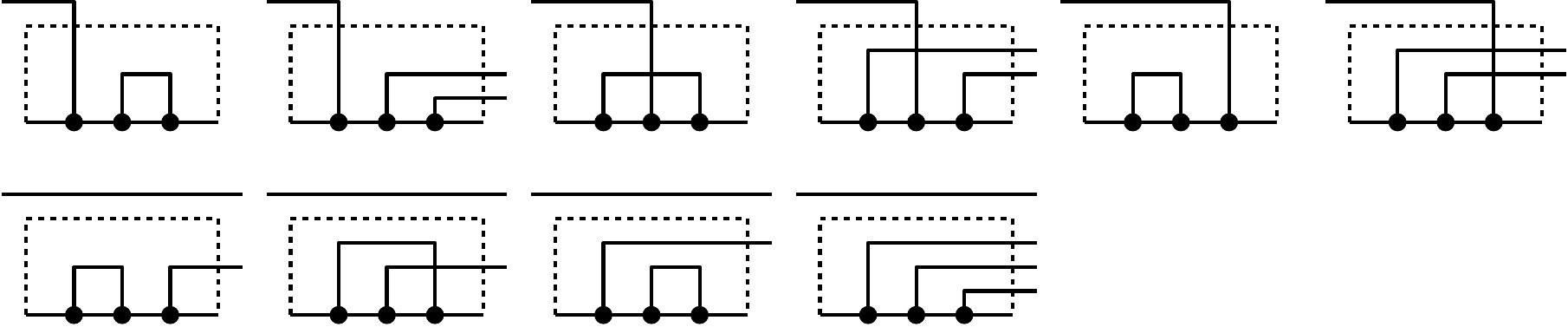}
\caption{Chord diagrams contributing in a region between two contracted $M$ nodes, with three $H$ nodes in between them. }
\label{fig:new_counting_example}
\end{figure}

As a simple example, consider the case in which there are three $H$ nodes in between the paired $M$ nodes. The chord diagrams in this region are shown in \autoref{fig:new_counting_example}.
The chord diagrams in the first row get in our counting the correct value of $\tilde q(2+q+\tilde q^2+\tilde q^2q+\tilde q^2q^2)$. The diagrams on the second row, however, are assigned a value of $\tilde q^2 (2\tilde q+\tilde q q +\tilde q^3)$ where the $\tilde q^2$ comes from the $\tilde q$ factor assigned to the incoming and outgoing chord. If we add the chord diagrams with no incoming chords to this region, multiplied by $1-\tilde q^2$, we would get precisely the correct value (the additional chord that passes by the region does not give any intersections).

Going from $l$ chords to $l-1$ chords (and so on), as we need before the $M$-region, is conveniently implemented using the matrix
\begin{equation}
U = \begin{pmatrix}
0 & 1 & 0 & \cdots \\
0 & 0 & 1 & \cdots \\
0 & 0 & 0 & \ddots \\
\vdots & \vdots & \ddots & \ddots
\end{pmatrix} \qquad \left( U_{l,l'} = \delta _{l',l+1} \right).
\end{equation}
Similarly, to restore the missing chords (going from $l$ to $l+1$ chords) one uses
\begin{equation}
D = \begin{pmatrix}
0 & 0 & 0 & \cdots \\
1 & 0 & 0 & \cdots \\
0 & 1 & 0 & \ddots \\
\vdots & \vdots & \ddots & \ddots
\end{pmatrix} \qquad \left( D_{l,l'} = \delta _{l,l'+1} \right).
\end{equation}

\begin{figure}[h]
\centering
\includegraphics[width=0.7\textwidth]{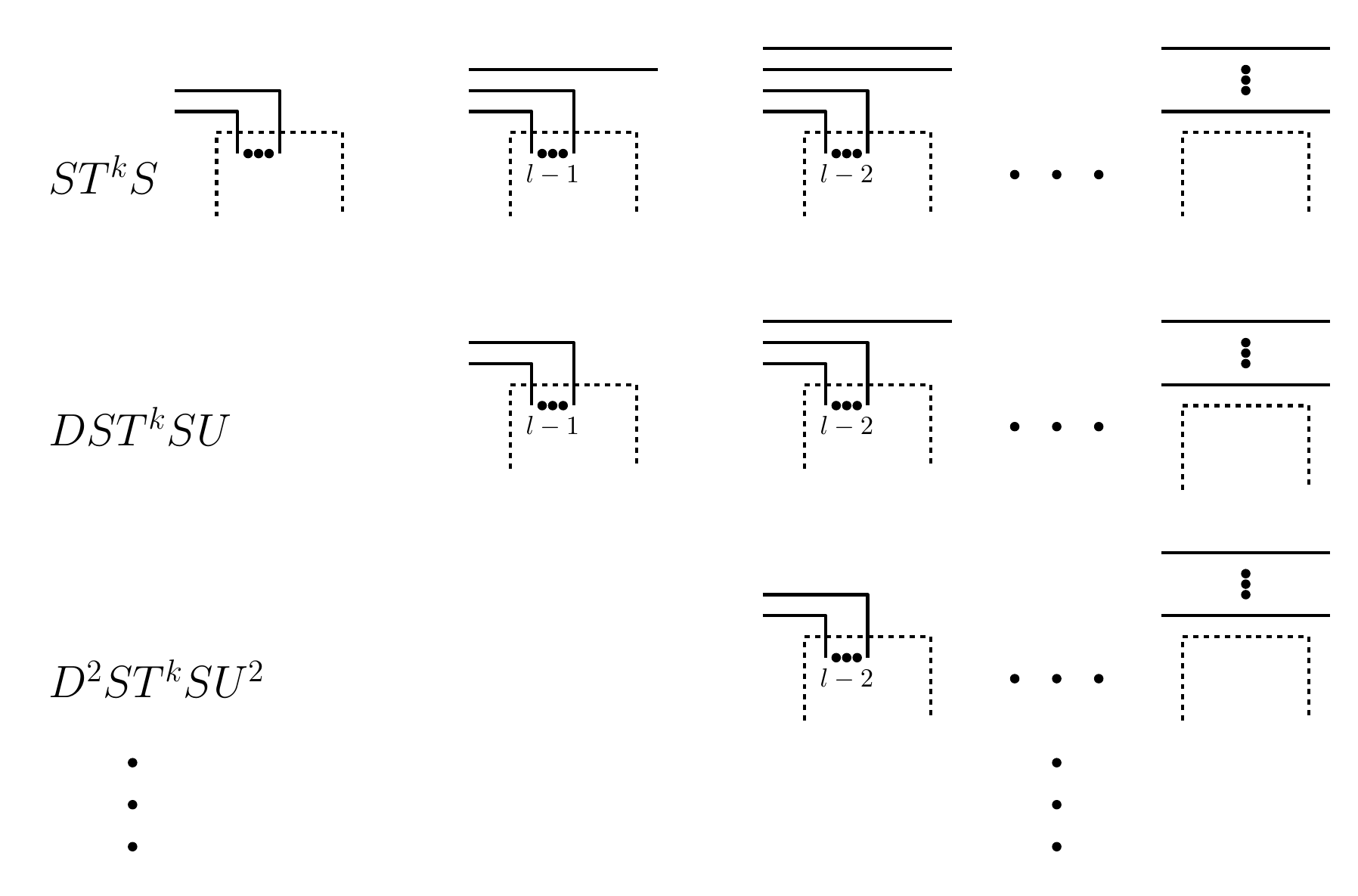}
\caption{A schematic drawing of the different contributions to the transition matrix.
Each diagram in the figure stands for the sum of the diagrams with the indicated number of chords closing in the $M$-region. The $i$'th row stands for the correcting diagrams with $l-i+1$ chords entering the $M$-region (and then $i$ additional chords are added). In the $j$'th column, we see all the diagrams where $l-j+1$ chords close in the $M$-region.
}
\label{fig:1M_line_without_n1n2}
\end{figure}

According to the arguments above, for $l$ incoming chords to the $M$-region, consisting of $k$ Hamiltonian nodes, the transition matrix through the entire $M$-region can be written as
\begin{equation} \label{eq:propagation_M_region_preliminary}
ST^k S + P^{(l)} _1 D S T^k S U+ P^{(l)} _2 D^2 S T^k S U^2 + \cdots + P^{(l)} _l D^l S T^k S U^l ,
\end{equation}
where $P^{(l)} _i$ are coefficients, depending on $q,\tilde q$, that we should determine.
The different contributions, according to the number of chords closing in the $M$-region, of the terms in \eqref{eq:propagation_M_region_preliminary} are shown in \autoref{fig:1M_line_without_n1n2}.

\begin{figure}[h]
\centering
\includegraphics[width=0.7\textwidth]{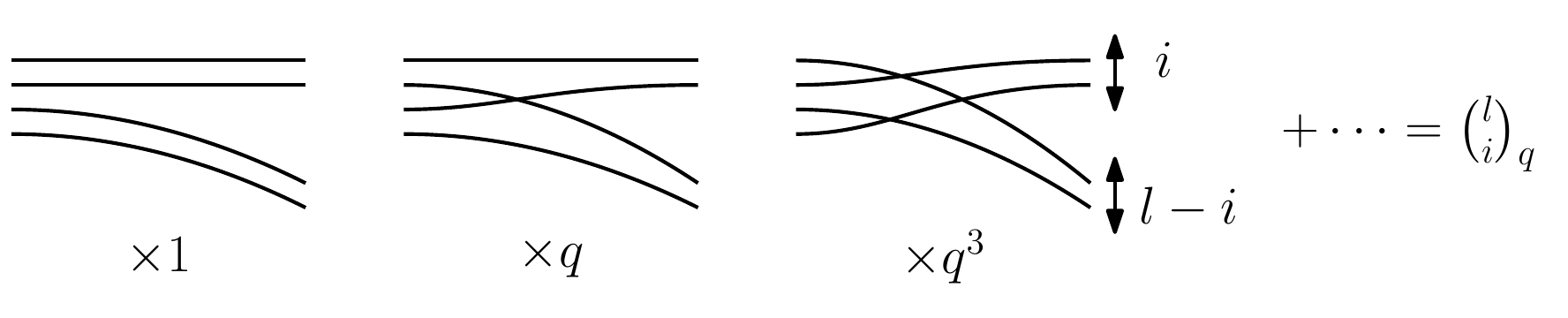}
\caption{Summing over the possibilities for separations of $i$ out of $l$ chords weighted by $q^{\text{\# intersections}} $ is given by the $q$-binomial coefficient.}
\label{fig:q_binomial}
\end{figure}

Note that above, incoming to the $M$-region, we have separated the $i$ chords (out of $l$) that do not close in the $M$-region. We then took them to be above all the other chords. However, they could be any $i$ out of the $l$ chords that arrived from the left.
As we separate these $i$ chords, and take them above the remaining $l-i$ chords, they will intersect the latter, see \autoref{fig:q_binomial}. The number of possibilities for choosing $i$ out of $l$ objects is the usual binomial coefficient $\binom{l}{i} $. If we sum over the possibilities for choosing $i$ out of $l$ chords, with each one multiplied by $q$ to the power of the number of intersections, we get the $q$-binomial coefficient, defined by
\begin{equation}
\binom{l}{i}_q \equiv \frac{(1-q^l)(1-q^{l-1}) \cdots (1-q^{l-i+1} )}{(1-q)(1-q^2)\cdots (1-q^i)} = \frac{(q;q)_l}{(q;q)_i(q;q)_{l-i} } 
\end{equation}
for $i \le l$ and $\binom{l}{i} _q=0$ otherwise. This is shown in \autoref{sec:Lemmas}.

In order to determine the coefficients $P^{(l)} _i$ we proceed as follows. Recall that we have an overall $l$ chords coming in, and consider further the subset of diagrams in which $l-j$ chords close inside the $M$-region.  Each such diagram is part of the sets of diagrams represented in the $(j+1)$'th column of \autoref{fig:1M_line_without_n1n2}. The weight associated with this diagram in \eqref{eq:propagation_M_region_preliminary} can be written as the correct weight, plus an excessive value of
\begin{equation} \label{eq:eqs_for_Pil}
(\tilde q^{2j} -1) \binom{l}{l-j}_q + \tilde q^{2(j-1)} \binom{l-1}{l-j}_q P_1^{(l)} + \tilde q^{2(j-2)} \binom{l-2}{l-j}_q P_2^{(l)} + \cdots + P_j^{(l)} = 0
\end{equation}
(times the weight of the diagram).
We get a (triangular) system of equations with $j=1,\cdots ,l$. The solution turns out to be quite simple
\begin{equation} \label{eq:Pil_expression}
P_i^{(l)} = \frac{(q;q)_l (\tilde q^2;q)_i}{(q;q)_{l-i} (q;q)_i}
\end{equation}
(see, for instance, Ex. 1.6, (iii) of \cite{gasper2004basic}, with $a \mapsto\tilde  q^2$).
We define $P_0=1$, being compatible with \eqref{eq:Pil_expression}.

To summarize, we found that given an incoming state of $l$ chords, the propagation through the entire $M$-region is realized by
\begin{equation}\label{LineL}
\sum _{i=0} ^l P_i^{(l)} D^i S T^k S U^i .
\end{equation}

\subsection{The 2-point function} \label{sec:2_point_function}

As a test of the previous subsection, we can see how to reproduce the known expressions for the 2-point function, given in \cite{Berkooz:2018qkz}, for which we should first evaluate expressions of the form $\langle \tr H^{k_3} M H^{k_2} M H^{k_1}\rangle_J$.

The simplest comparison is to recall that in \cite{Berkooz:2018qkz} the convenient choice of $k_3=0$ was made, relying on the cyclicity of the trace. By this we mean that we cut  the chord diagram right before the leftmost $M$ node. In this case, there are no chords incoming into the $M$-region, and in particular no $H$-chords.
The evolution over this region acts on $|0\rangle$ (annihilated by all $U^i$, $i>0$) by:
\begin{equation} \label{LineL_first_node}
P_0^{(0)} S T^k S |0\rangle = ST^k |0\rangle ,
\end{equation}
reproducing the approach used in \cite{Berkooz:2018qkz}.

This, however, does not show the use of the full formula \eqref{LineL}. As a check, we now apply \eqref{LineL} to evaluate \eqref{eq:2pf_ingredients} for an arbitrary cut of the chord diagram. Also, the averaging of the operators' ($M_A$) couplings will be indicated explicitly by contraction symbols. Inserting a complete basis of states $|l\rangle$ before the first $M$-node, and using \eqref{LineL}, we get that the moments \eqref{eq:2pf_ingredients} of the 2-point function of $M$ are given by
\begin{equation} \label{eq:2_point_function_expr1}
\begin{split}
& \langle \tr H^{k_3}
\contraction{}{M}{H^{k_2}}{M}
 M H^{k_2} M H^{k_1}
\rangle_J = \langle0| T^{k_3} \left( \text{evolution in $M$-region} \right) T^{k_1} |0\rangle = \\
& = \sum _{l,m=0} ^{\infty } \sum _{i=0} ^l P_i^{(l)} \langle 0| T^{k_3} |m\rangle \langle m | D^i S T^{k_2} S U^i |l\rangle \langle l | T^{k_1} |0\rangle = \\
& = \sum _{l,m=0} ^{\infty } \sum _{i=0} ^{\min(m,l)} P_i^{(l)} \tilde q^{l+m-2i} \langle0 | T^{k_3} |m\rangle \langle m-i | T^{k_2} | l-i \rangle \langle l | T^{k_1} | 0\rangle = \\
& = \sum _{i=0} ^{\infty } \sum _{m,l=i} ^{\infty } P_i^{(l)} \tilde q^{l+m-2i} \langle0 | T^{k_3} |m\rangle \langle m-i | T^{k_2}  | l-i\rangle \langle l |T^{k_1} | 0 \rangle = \\
&= \sum _{m,l=0} ^{\infty } \tilde q^{m+l} \langle m| T^{k_2} |l \rangle \cdot \sum _{i=0} ^{\infty } P_i^{(l+i)} \langle 0 |T^{k_3} | m+i \rangle \langle l+i | T^{k_1}  | 0 \rangle .
\end{split}
\end{equation}

Substituting \eqref{eq:Pil_expression} and \eqref{eq:Tab_expression} into \eqref{eq:2_point_function_expr1} we find
\begin{equation}
\begin{split}
& \langle \tr H^{k_3}
\contraction{}{M}{H^{k_2}}{M}
 M H^{k_2} M H^{k_1} \rangle_J= \\
 & = \sum _{i,m,l=0} ^{\infty } \tilde q^{m+l} \frac{(\tilde q^2;q)_i }{(q;q)_i (q;q)_m (q;q)_l  }  \\
 & \qquad \cdot  \int _0^{\pi } \prod_{j=1}^3 \left\{\frac{d\theta_j}{2\pi}\left(q, e^{\pm 2i\theta_j};q\right)_{\infty} \left( \frac{2 x_j}{\sqrt{1-q} } \right)^{k_j}\right\} \cdot \\
 & \qquad \qquad \qquad \cdot H_{l+i} (x_1|q)  H_l(x_2|q) H_m(x_2|q) H_{m+i} (x_3|q) ,
\end{split}
\end{equation}
where $x_j = \cos(\theta _j)$;
after interchanging sums and integrals and using \eqref{qHermite-bilinear-shifted} twice, we get an expression in terms of the Al Salam-Chihara polynomials $Q_n$ defined in \eqref{ASC}:
\begin{align} \label{eq:two_point_function_QQ_form}
 \int _0^{\pi } \prod_{j=1}^3 & \left\{\frac{d\theta_j}{2\pi}\left(q, e^{\pm 2i\theta_j};q\right)_{\infty} \left( \frac{2 x_j}{\sqrt{1-q} } \right)^{k_j}\right\} \prod_{k=1}^2\frac{\left(\tilde q^2 ;q\right)_{\infty}}{\left(\tilde q\, e^{i\left(\pm \theta_k \pm \theta_{k+1}\right)};q\right)_{\infty}}\\
& \times \sum_{i=0}^{\infty}\frac{Q_{i}\left(x_1|\tilde q\, e^{\pm i\theta_2}, q\right)Q_{i}\left(x_3|\tilde q\, e^{\pm i\theta_2}, q\right)}{\left(q,\tilde q^2;q\right)_{i}}. \nonumber
\end{align}
With \eqref{eq:sum_Q_a_eqs_b}, equation \eqref{eq:two_point_function_QQ_form} reduces to the form of the 2-point function obtained in \cite{Berkooz:2018qkz} 
\begin{equation} \label{eq:2pf_moments}
\begin{split}
& \langle \tr H^{k_3}
\contraction{}{M}{H^{k_2}}{M}
 M H^{k_2} M H^{k_1} \rangle_J= \\
& = \int _0^{\pi } \prod _{j=1} ^2 \left\{ \frac{d\theta _j}{2\pi } (q,e^{\pm 2i\theta _j} ;q)_{\infty } \left( \frac{2x_j}{\sqrt{1-q} } \right)^{k'_j} \right\} \frac{(\tilde q^2;q)_{\infty } }{(\tilde q e^{i(\pm \theta _1 \pm \theta _2)} ;q)_{\infty } } ,
\end{split}
\end{equation}
where $k'_1=k_1+k_3$ and $k'_2=k_2$. To go from moments to the correlation function, we exponentiate the energies
\begin{equation} \label{eq:2_point_function}
\begin{split}
& \langle \tr
\contraction{}{M}{e^{-\beta _2 H}}{M}
 M e^{-\beta _2 H} M e^{-\beta _1 H} \rangle_J= \\
 & = \int _0^{\pi } \prod _{j=1} ^2 \left\{ \frac{d\theta _j}{2\pi } (q,e^{\pm 2i\theta _j} ;q)_{\infty } \exp \left( - \frac{2x_j \beta_j}{\sqrt{1-q} } \right) \right\} \frac{(\tilde q^2;q)_{\infty } }{(\tilde q e^{i(\pm \theta _1 \pm \theta _2)} ;q)_{\infty } } .
\end{split}
\end{equation}

\subsection{Time translation invariance of the bi-local operator} \label{sec:trans_inv}

A general property of two $M$ insertions contracted with each other (with no other $M_A$ operator insertions in between), is that this bi-local operator, which we denote here by $\cO $, is invariant under time translations (of its center of mass, keeping the separation between the two $M$ insertions fixed). That is, we can move the contracted $M$ pair in \autoref{fig:propagation_M_region} as a unit to the left or to the right independently of what other operator insertions we have on both sides. A similar statement holds for the bi-local operators in the Schwarzian theory in \cite{Mertens:2017mtv}. We can show that $[{\cal O},T]=0$, or equivalently ${\cal O} | \psi^{(\theta )} _T \rangle \propto | \psi^{(\theta )} _T \rangle$, where the latter state is an eigenvector of $T$ defined in \eqref{eq:T_eigenvectors}.

The operator $\cO $ is given by \eqref{LineL} when acting on a state of $l$ incoming lines. The matrix elements of $\cO $ are then
\begin{equation}
\langle m| \cO  | l \rangle = \sum _{i=0} ^{\min(m,l)} P_i^{(l)} \tilde q ^{l+m-2i} \langle m-i | T^k | l-i\rangle .
\end{equation}
Acting with $\cO $ on a state \eqref{eq:T_eigenvectors} of energy $E(\theta )$ gives (we change $l \to l-i$)
\begin{equation}
\begin{split}
& \langle m | \cO  | \psi^{(\theta )} _T \rangle = \sum _{i=0} ^m \sum _{l=0} ^{\infty } \frac{\sqrt{(q;q)_{\infty } } }{\sqrt{2\pi } } | (e^{2i\theta } ;q)_{\infty } | (1-q)^{m/2} H_{l+i} (\cos \theta |q) \frac{(\tilde q^2;q)_i}{(q;q)_l (q;q)_i} \tilde q^{m-i+l} \cdot \\
& \cdot \int _0^{\pi } \frac{d\theta '}{2\pi } (q,e^{\pm 2i\theta '} ;q)_{\infty } \left( \frac{2\cos \theta '}{\sqrt{1-q} } \right)^k \frac{H_l(\cos \theta '|q) H_{m-i} (\cos \theta '|q)}{(q;q)_{m-i} }.
\end{split}
\end{equation}
Now we use \eqref{qHermite-bilinear-shifted} to get
\begin{equation}
\begin{split}
& \langle m | \cO  | \psi^{(\theta )} _T \rangle = \sum _{i=0} ^m \sqrt{\frac{(q;q)_{\infty } }{2\pi } } |(e^{2i\theta } ;q)_{\infty } | (1-q)^{m/2} \tilde q^{m-i} \frac{1}{(q;q)_i} \cdot \\
& \cdot  \int _0^{\pi } \frac{d\theta '}{2\pi } (q,e^{\pm 2i\theta '} ;q)_{\infty } \left( \frac{2\cos \theta '}{\sqrt{1-q} } \right)^k \frac{H_{m-i} (\cos \theta '|q)}{(q;q)_{m-i} } \frac{(\tilde q^2;q)_{\infty } }{(\tilde q e^{i(\pm \theta  \pm \theta ')} ;q
)_{\infty } } Q_i(\cos \theta |\tilde qe^{\mp i \theta '} ,q),
\end{split}
\end{equation}
which, using \eqref{Hermite-ASC}, becomes
\begin{equation}
\begin{split}
&\langle m| \cO | \psi_T^{(\theta )} \rangle= \left( \int _0^{\pi } \frac{d\theta '}{2\pi } (q,e^{\pm 2i\theta '} ;q)_{\infty } \left( \frac{2\cos \theta '}{\sqrt{1-q} } \right)^k \frac{(\tilde q^2;q)_{\infty } }{(\tilde q e^{i(\pm \theta  \pm \theta ')} ;q)_{\infty } } \right) \cdot \langle m | \psi_T^{(\theta )} \rangle . 
\end{split}
\end{equation}
This is proportional to the eigenvector of $T$ with the same energy,
showing that the bi-local operator $\cO$ preserves the latter.

\section{4-point function} \label{sec:4_point_function}

The next 3 sections will be devoted to evaluating
the thermal 4-point function
\begin{equation}
\langle \tr e^{-\beta H} M(t_1) M(t_2)M(t_3) M(t_4) \rangle_J .
\end{equation}
In this section we take the formulas for the combinatorial construction of bi-local operators, from the previous section, and compute the 4-point function explicitly in terms of hypergeometric functions. In \autoref{sec:diag_rules} we summarize the construction using a set of ``Feynman rules'', whose propagators and vertices are these hypergeometric functions, and verify that in the appropriate $q\rightarrow 1^-$ limit we obtain the Schwarzian theory. Then in  \autoref{sec:chaos}, we evaluate the chaos exponent at low energy, and corrections to it.

\begin{figure}[h]
\centering
\includegraphics[width=1\textwidth]{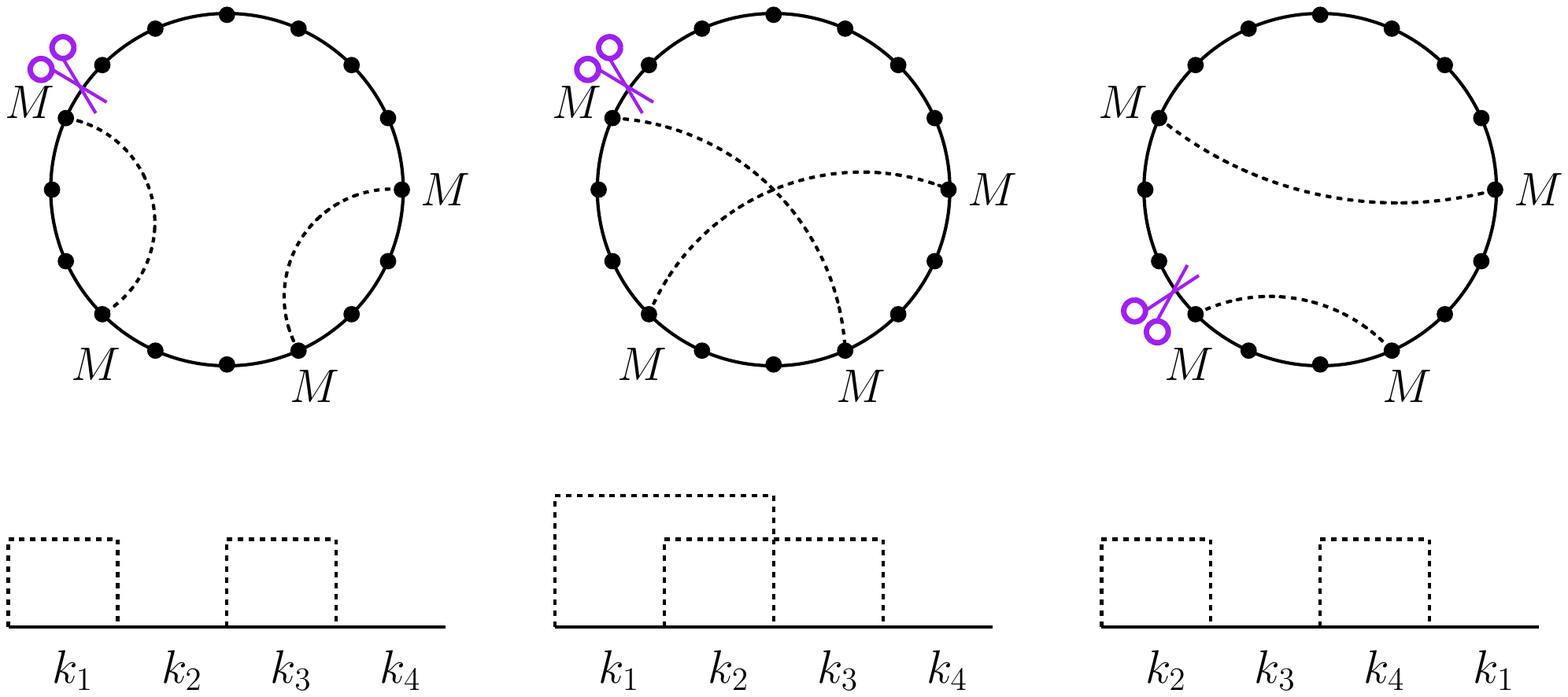}
\caption{The three possible contractions in a 4-point function. Only the $M$-nodes are shown. By choosing appropriate cuts, we get that the first and third contractions are the same.}
\label{fig:4_point_function_3_contractions}
\end{figure}

As before, we start by calculating the moments
\begin{equation}
\langle  \tr H^{k_4} M H^{k_3} M H^{k_2} M H^{k_1} M \rangle_J .
\end{equation}
After averaging over the random coefficients, there are three different pairwise contractions contributing to these moments. These contractions, along with a convenient set of cutting points are shown in \autoref{fig:4_point_function_3_contractions}. Cutting the first contraction as shown in the figure, we can evaluate this moment by employing the analysis of \autoref{sec:propagation_M_region}. For the third contraction, cutting it at the same point as in the first diagram would give an $M$-chord inside another $M$-chord, which is not of the same form as the case considered in \autoref{sec:propagation_M_region}. However, if we cut it as shown in the figure, we get the same sort of diagram as for the first contraction.

\begin{figure}[h]
\centering
\includegraphics[width=0.8\textwidth]{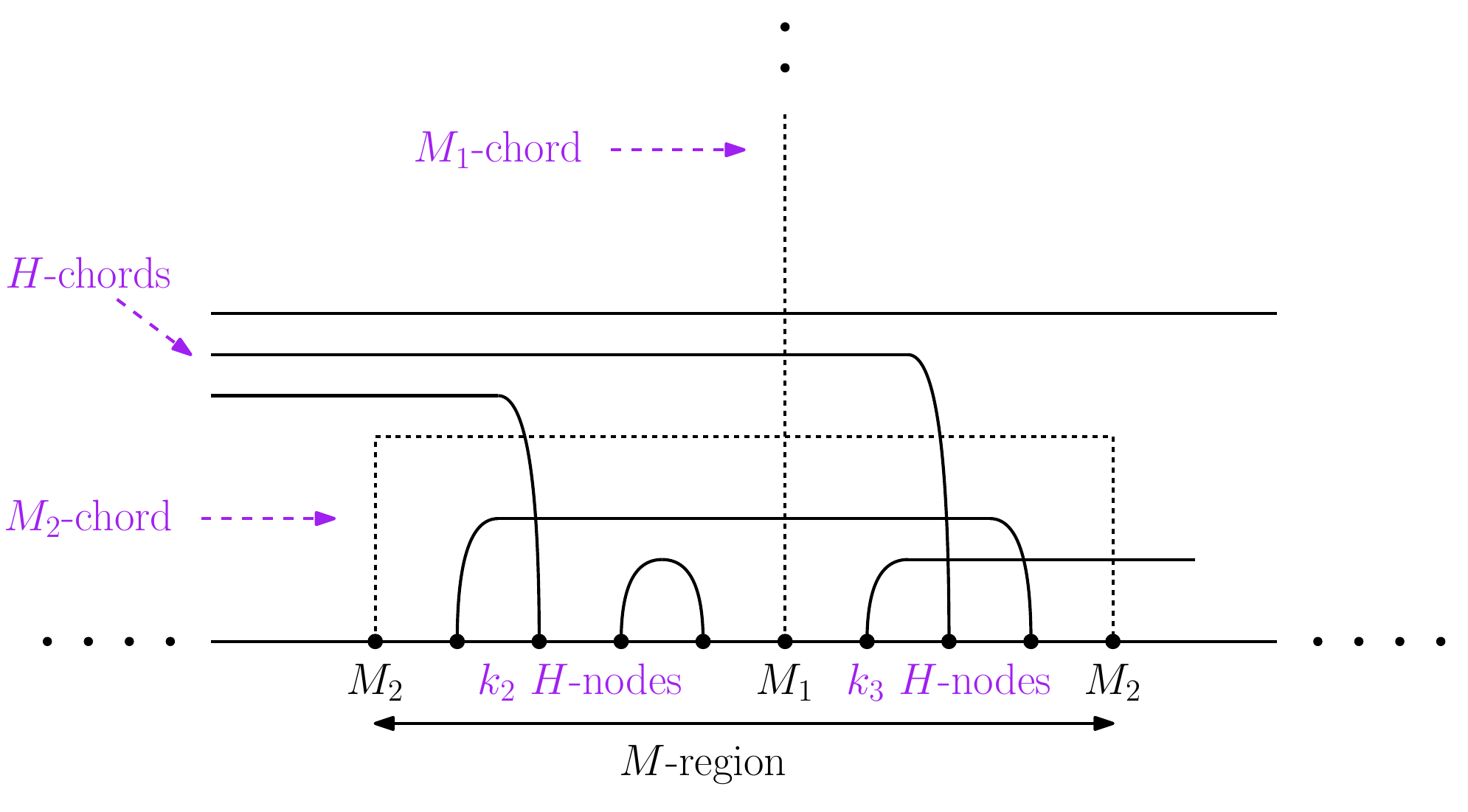}
\caption{A subregion consisting of two $M_2$ nodes which are contracted, with another $M_1$ node between them. We will think of the $M_1$-chord as being high enough, so that there are no $H$-chords above it.}
\label{fig:propagation_M_region_modified}
\end{figure}

We are left with the second contraction.  In fact, let us consider a slight generalization where the two $M$-chords can be of a different flavor, $M_1$ and $M_2$. 
A major simplification is obtained by making the cut right before one of the $M$ nodes. 
The reason for this is seen by considering the structure of the resulting chord diagrams at the bottom of \autoref{fig:4_point_function_3_contractions} (second diagram). Since the first node opens the first $M$-chord, all the successive $H$-chords can be taken to be below this $M$-chord; they will intersect it only if they reach the closing $M$-node of this $M$-chord (between $k_2$ and $k_3$ in the figure). Therefore, we can think about this configuration as the configuration shown in \autoref{fig:propagation_M_region_modified}, where we imagine this $M$-chord 
as a chord that goes arbitrarily high and can only intersect other chords at the closing node. 

The ingredients used in \autoref{sec:propagation_M_region} are enough to understand the generalization to the case at hand. In order to propagate through the $M$-region (by which we now mean the region between the contracted $M_2$-pair, see \autoref{fig:propagation_M_region_modified}) we can still use the same auxiliary Hilbert space. Entering and getting out of the $M$-region, each chord is assigned a value of $\tilde q_2$ (this is implemented by the matrix $S_2$, where $S_i$ is the same matrix as $S$ defined in \eqref{eq:def_S_matrix} with $\tilde q \to \tilde q_i$). The difference, compared to the previous case, is that now we should simply multiply each chord open at the $M_1$ node by $\tilde q_1$ (again using an $S_1$ matrix). This propagation is given by $S_2 T^{k_3} S_1 T^{k_2} S_2$.

Exactly as before, this does not count correctly diagrams with chords that do not close inside the $M$-region. To fix that, given $l$ incoming $H$-chords, we should add similar objects with less than $l$ incoming chords and appropriate compensating coefficients. Whenever we add those diagrams with $l-i$ incoming chords, we eventually mean that there are $i$ additional chords not closing in the $M$-region (but rather just passing by it) and so we should remember to include explicitly a factor of $\tilde q_1^i$ from the intersection with the $M_1$-chord. Therefore, for the total propagation through the $M$-region having $l$ incoming chords, we should use instead of \eqref{eq:propagation_M_region_preliminary} the following expression
\begin{equation}
\begin{split}
& S_2 T^{k_3} S_1 T^{k_2} S_2+ P_1^{(l)} \tilde q_1 D S_2 T^{k_3} S_1 T^{k_2} S_2 U + P_2^{(l)} \tilde q_1^2 D^2 S_2 T^{k_3} S_1 T^{k_2} S_2 U^2+ \\
& \qquad +  \cdots + P_l^{(l)} \tilde q_1^l D^l  S_2 T^{k_3} S_1 T^{k_2} S_2 U^l.
\end{split}
\end{equation}
The equations for $P_i^{(l)} $ are given by the same equations \eqref{eq:eqs_for_Pil} with $\tilde q \to \tilde q_2$ and so $P_i^{(l)}$ are the same as before, given by \eqref{eq:Pil_expression} (again with $\tilde q_2$). To indicate the use of $\tilde q_2$ in the expressions for $P_i^{(l)} $ we will denote them by $P^{(l)} _{i,2} $. Recalling that the intersection between the $M_1$ and $M_2$ chords gives a factor $\tilde q_{12} $, a propagation through such an $M$-region for $l$ incoming chords is given by
\begin{equation} \label{eq:Propagation_M_line_with_intermediate_M_line}
\tilde q_{12} \sum _{i=0} ^l P_{i,2}^{(l)} \tilde q_1^i D^i S_2 T^{k_3} S_1 T^{k_2} S_2 U^i.
\end{equation}

To conclude, there are two distinct contractions contributing to the 4-point function of an operator $M$, and we have just seen the needed ingredients to evaluate each. 
In the following two subsections we instead consider two different orderings of a 4-point function of two distinct random operators $M_1$ and $M_2$, each appearing twice. These will isolate a particular contraction from the two distinct ones. The two orderings are evaluated in the following, and clearly the full 4-point function of a single operator $M$ can be written by assembling the different expressions according to \autoref{fig:4_point_function_3_contractions} (i.e., having the first contraction appearing twice, with the appropriate parameters).

\subsection{Uncrossed 4-point function $\langle M_1 M_1 M_2 M_2\rangle$}

The first ordering we use is $M_1 M_1 M_2 M_2$. Only the first contraction on the bottom of \autoref{fig:4_point_function_3_contractions} contributes to this correlation function, written using \eqref{LineL}.
Since we cut the diagrams right before the first $M_1$ node, the contracted $M_1$ pair is actually given by the simplified vector \eqref{LineL_first_node}. Denote the $l_1,l_2$ component of the matrix $T^k$ by
\begin{equation}
\langle l_1 | T^k |l_2 \rangle = T^k _{l_1 l_2} .
\end{equation}
Then we get for this correlation function
\begin{equation}
\begin{split}
& \langle \tr H^{k_4}
\contraction{}{M_2}{H^{k_3}}{M_2}
M_2 H^{k_3} M_2
H^{k_2}
\contraction{}{M_1}{H^{k_1}}{M_1}
M_1 H^{k_1} M_1 \rangle_J = \\
& = \sum _{l=0} ^{\infty } \sum _{i=0} ^l P_{i,2}^{(l)} \langle 0 |T^{k_4} D^i S_2 T^{k_3} S_2 U^i |l\rangle \langle l | T^{k_2} S_1 T^{k_1} | 0 \rangle = \\
& = \sum _{l,m,n=0} ^{\infty } \sum _{i=0} ^l P_{i,2}^{(l)} \langle 0 |T^{k_4} | m \rangle \langle m | D^i S_2 T^{k_3} S_2 U^i |l \rangle \langle l | T^{k_2} S_1 |n \rangle \langle n | T^{k_1} |0\rangle = \\
&= \sum _{i,l,m,n=0} ^{\infty } P_{i,2}^{(l+i)} \tilde q_2^{l+m} \tilde q_1^n T^{k_4} _{0,m+i} T^{k_3} _{m,l} T^{k_2} _{l+i,n} T^{k_1} _{n,0} = \\
& = \sum _{i,l,m,n=0} ^{\infty } \tilde q_2^{l+m} \tilde q_1^n \frac{(\tilde q_2^2;q)_i}{(q;q)_i(q;q)_l(q;q)_n(q;q)_m} \int _0^{\pi } \prod _{j=1} ^4 \left\{ \frac{d\theta _j}{2\pi } (q,e^{\pm 2i\theta _j} ;q)_{\infty } \left( \frac{2x_j}{\sqrt{1-q} } \right)^{k_j} \right\} \cdot \\
& \cdot H_n(x_1|q) H_n(x_2|q) H_{l+i} (x_2|q) H_l(x_3|q) H_m(x_3|q) H_{m+i} (x_4|q)
\end{split}
\end{equation}
(in the fourth equality we changed the summation order and shifted $l \to l+i$ and $m \to m+i$).

Now we start simplifying this expression. First perform the sum over indices using \eqref{qHermite-bilinear-shifted} to get
\begin{equation}\label{4pt-q1q2-ASC}
\begin{split}
& \int _0^{\pi } \prod _{j=1} ^4 \left\{ \frac{d\theta _j}{2\pi } (q,e^{\pm 2i\theta _j} ;q)_{\infty } \left( \frac{2x_j}{\sqrt{1-q} } \right)^{k_j} \right\} \, \prod_{k=1}^3\frac{\left(\tilde q_k^2 ;q\right)_{\infty}}{\left(\tilde q_k \, e^{i\left(\pm \theta_k \pm \theta_{k+1}\right)};q\right)_{\infty}} \\
& \cdot \sum_{i=0}^{\infty}
\frac{Q_i(x_2|\tilde q_2e^{\mp i\theta _3} ,q) Q_i(x_4|\tilde q_2 e^{\mp i\theta _3} ,q) }{(q;q)_i (\tilde q_2^2;q)_i }  
\end{split}
\end{equation}
where we introduced $\tilde q_3 \equiv \tilde q_2$ in order to shorten the formula.
We may also perform the remaining sum over $i$
using \eqref{eq:sum_Q_a_eqs_b} and get
\begin{equation}
\begin{split}
& \langle \tr H^{k_4}
\contraction{}{M_2}{H^{k_3}}{M_2}
M_2 H^{k_3} M_2
H^{k_2}
\contraction{}{M_1}{H^{k_1}}{M_1}
M_1 H^{k_1} M_1 \rangle_J = \\
&= \int _0^{\pi } \prod _{j=1} ^3 \left\{ \frac{d\theta _j}{2\pi } (q,e^{\pm 2 i \theta _j} ;q)_{\infty } \left( \frac{2x_j}{\sqrt{1-q} } \right)^{k'_j}  \right\} \prod _{k=1} ^2 \frac{\left(\tilde q_k^2;q\right)_{\infty } }{\left( \tilde q_k e^{i(\pm \theta _k \pm \theta _{k+1} )} ;q \right)_{\infty } } ,
\end{split}
\end{equation}
where $k'_1=k_1$, $k'_2 = k_2+k_4$ and $k'_3=k_3$. This 4-point function depends only on $k_2+k_4$ and not on their separate values, i.e., it is independent of the separation of the two pairs of contracted $M$ nodes. This is indeed a consequence of what was shown in \autoref{sec:trans_inv}. The 4-point function is then (real times for the operators are obtained by assigning an imaginary part to the $\beta _j$ appearing below)
\begin{equation}\label{4-pt-q1q2TO-int}
\begin{split}
& \langle \tr e^{-\beta _4 H}
\contraction{}{M_2}{e^{-\beta _3 H}}{M_2}
M_2 e^{-\beta _3 H} M_2
e^{-\beta _2 H}
\contraction{}{M_1}{e^{-\beta _1 H}}{M_1}
M_1 e^{-\beta _1 H} M_1 \rangle_J = \\
&= \int _0^{\pi } \prod _{j=1} ^3 \left\{ \frac{d\theta _j}{2\pi } (q,e^{\pm 2 i \theta _j} ;q)_{\infty } \exp \left(-\frac{2x_j \beta '_j}{\sqrt{1-q} } \right)  \right\} \prod _{k=1} ^2 \frac{\left(\tilde q_k^2;q\right)_{\infty } }{\left( \tilde q_k e^{i(\pm \theta _k \pm \theta _{k+1} )} ;q \right)_{\infty } } ,
\end{split}
\end{equation}
where again $\beta '_2=\beta _2+\beta _4$ and $\beta '_i=\beta _i$ for $i=1,3$.

In fact, \eqref{LineL} can be used to evaluate any $2n$-point function given by a similar sequence of contracted $M$ pairs. This is discussed in \autoref{app:2n_point_function}.

\subsection{Crossed 4-point function $\langle M_1 M_2 M_1 M_2\rangle$}

Next we consider the other ordering $M_1 M_2 M_1 M_2$. From now on we will not write explicitly the constant $\tilde q_{12} $ (since it appears as an overall constant in this correlation function)\footnote{When assembling the full 4-point function of an operator $M$, however, the coupling corresponding to $\tilde q_{12} $ is not an overall constant.}. Using \eqref{eq:Propagation_M_line_with_intermediate_M_line} this 4-point function is given by
\begin{equation} \label{eq:crossed_4_pt_initial}
\begin{split}
& \langle \tr H^{k_4}
\contraction[2ex]{}{M_2}{H^{k_3} M_1 H^{k_2} }{M_2}
\contraction{M_2 H^{k_3}}{M_1}{H^{k_2} M_2 H^{k_1}}{M_1}
M_2 H^{k_3} M_1 H^{k_2} M_2 H^{k_1} M_1 \rangle_J = \\
&= \sum _{l,m,n=0} ^{\infty } \sum _{i=0} ^l P^{(l)} _{i,2} \tilde q_1^i \langle 0 | T^{k_4} | m\rangle \langle m| D^i S_2 T^{k_3} S_1 |n \rangle \langle n | T^{k_2} S_2 U^i | l \rangle \langle l | T^{k_1}  | 0 \rangle = \\
&= \sum _{l,m,n,i=0} ^{\infty } P^{(l+i)} _{i,2} \tilde q_1^{i+n} \tilde q_2^{l+m} T^{k_4} _{0,m+i} T^{k_3} _{m,n} T^{k_2} _{n,l} T^{k_1} _{l+i,0} = \\
&= \sum _{l,m,n,i=0} ^{\infty } \frac{(\tilde q_2^2;q)_i \tilde q_1^{i+n} \tilde q_2^{l+m} }{(q;q)_l (q;q)_i (q;q)_m (q;q)_n} \int _0^{\pi }  \prod _{j=1} ^4 \left\{ \frac{d\theta _j}{2\pi } (q,e^{\pm 2i\theta _j} ;q)_{\infty } \left( \frac{2x_j}{\sqrt{1-q} } \right)^{k_j} \right\} \cdot \\
& \qquad \cdot H_{l+i} (x_1|q) H_l(x_2|q) H_n(x_2|q) H_n(x_3|q) H_m(x_3|q) H_{m+i} (x_4|q)
\end{split}
\end{equation}
(once again we have changed the summation indices $l \to l+i$ and $m \to m+i$ in the third line).

Using \eqref{qHermite-bilinear-shifted} as before, and denoting $\tilde q_0 \equiv \tilde q_2$ to get a short form for the equation, we find that \eqref{eq:crossed_4_pt_initial} equals
\begin{equation}
\begin{split}
& \int _0^{\pi }  \prod _{j=1} ^4 \left\{ \frac{d\theta _j}{2\pi } (q,e^{\pm 2i\theta _j} ;q)_{\infty } \left( \frac{2x_j}{\sqrt{1-q} } \right)^{k_j} \right\} \prod _{j=0} ^2 \frac{(\tilde q_j^2;q)_{\infty } }{(\tilde q_j e^{i(\pm  \theta _{j+1} \pm \theta _{j+2} )} ;q)_{\infty } } \cdot \\
& \qquad \cdot \sum _{i=0} ^{\infty } \frac{\tilde q_1^i}{(q,\tilde q_2^2;q)_i} Q_i(x_1|\tilde q_2 e^{\mp i \theta _2} ,q) Q_i(x_4|\tilde q_2 e^{\mp i \theta _3} ,q) .
\end{split}
\end{equation}

By \eqref{ASC-bilinear}, we get the result
\begin{equation}\label{eq:4-pt-q1q2OTO-int}
\begin{split}
& \langle \tr H^{k_4}
\contraction[2ex]{}{M_2}{H^{k_3} M_1 H^{k_2} }{M_2}
\contraction{M_2 H^{k_3}}{M_1}{H^{k_2} M_2 H^{k_1}}{M_1}
M_2 H^{k_3} M_1 H^{k_2} M_2 H^{k_1} M_1 \rangle_J = \\
& \int _0^{\pi }  \prod _{j=1} ^4 \left\{ \frac{d\theta _j}{2\pi } (q,e^{\pm 2i\theta _j} ;q)_{\infty } \left( \frac{2x_j}{\sqrt{1-q} } \right)^{k_j} \right\}  \frac{\left(\tilde q_1 e^{-i(\theta_2+\theta_3)}, \tilde q_1 \tilde q_2 e^{i(\theta_3\pm \theta_1)}, \tilde q_1 \tilde q_2 e^{i(\theta_2\pm \theta_4)} ;q \right)_{\infty } }{\left(\tilde q_1 \tilde q_2^2 e^{i( \theta_2+\theta_3)} ;q \right)_{\infty } } \\
& \qquad \times  \frac{\left(\tilde q_1^2, \tilde q_2^2, \tilde q_2^2;q \right)_{\infty } }{\left(\tilde q_1 e^{i(\pm  \theta _{2} \pm \theta _{3} )}, \tilde q_1 e^{i(\pm  \theta _{1} \pm \theta _{4} )}, \tilde q_2 e^{i(\pm  \theta _{1} \pm \theta _{2} )}, \tilde q_2 e^{i(\pm  \theta _{3} \pm \theta _{4} )} ;q \right)_{\infty } } \\
& \qquad \times  {}_8W_7\left( \frac{\tilde q_1 \tilde q_2^2 e^{i(\theta_2+\theta_3)} }{q}; \tilde q_1 e^{i(\theta_2+\theta _3)} ,\tilde q_2 e^{i(\theta _2 \pm \theta _1)}, \tilde q_2 e^{i(\theta _3 \pm \theta _4)} ;q, \tilde q_1 e^{-i(\theta_2+\theta _3)} \right),
\end{split}
\end{equation}
which represents the moments. The basic hypergeometric series ${}_8W_7$ is defined in \eqref{8W7}.
Then, the 4-point function is just
\begin{equation} \label{eq:OTO_4_point_function}
\begin{split}
& \langle \tr e^{-\beta _4 H}
\contraction[2ex]{}{M_2}{e^{-\beta _3 H} M_1 e^{-\beta _2 H} }{M_2}
\contraction{M_2 e^{-\beta _3 H}}{M_1}{e^{-\beta _2 H} M_2 e^{-\beta _1 H} }{M_1}
M_2 e^{-\beta _3 H} M_1 e^{-\beta _2 H} M_2 e^{-\beta _1 H} M_1 \rangle_J = \\
& \int _0^{\pi }  \prod _{j=1} ^4 \left\{ \frac{d\theta _j}{2\pi } (q,e^{\pm 2i\theta _j} ;q)_{\infty } \exp \left(- \frac{2x_j \beta _j}{\sqrt{1-q} } \right) \right\}  \frac{\left(\tilde q_1 e^{-i(\theta_2+\theta_3)}, \tilde q_1 \tilde q_2 e^{i(\theta_3\pm \theta_1)}, \tilde q_1 \tilde q_2 e^{i(\theta_2\pm \theta_4)} ;q \right)_{\infty } }{\left(\tilde q_1 \tilde q_2^2 e^{i( \theta_2+\theta_3)} ;q \right)_{\infty } } \\
& \qquad \times  \frac{\left(\tilde q_1^2, \tilde q_2^2, \tilde q_2^2;q \right)_{\infty } }{\left(\tilde q_1 e^{i(\pm  \theta _{2} \pm \theta _{3} )}, \tilde q_1 e^{i(\pm  \theta _{1} \pm \theta _{4} )}, \tilde q_2 e^{i(\pm  \theta _{1} \pm \theta _{2} )}, \tilde q_2 e^{i(\pm  \theta _{3} \pm \theta _{4} )} ;q \right)_{\infty } } \\
& \qquad \times  {}_8W_7\left( \frac{\tilde q_1 \tilde q_2^2 e^{i(\theta_2+\theta_3)} }{q}; \tilde q_1 e^{i(\theta_2+\theta _3)} ,\tilde q_2 e^{i(\theta _2 \pm \theta _1)}, \tilde q_2 e^{i(\theta _3 \pm \theta _4)} ;q, \tilde q_1 e^{-i(\theta_2+\theta _3)} \right).
\end{split}
\end{equation}

\section{The diagrammatic rules of the full model} \label{sec:diag_rules}

\begin{fmffile}{Fdiagrams}
\fmfcmd{%
vardef cross_bar (expr p, len, ang) =
((-len/2,0)--(len/2,0))
rotated (ang + angle direction length(p)/2 of p)
shifted point length(p)/2 of p
enddef;
style_def crossed expr p =
cdraw p;
ccutdraw cross_bar (p, 3mm, 45);
ccutdraw cross_bar (p, 3mm, -45)
enddef;}

The chord diagrams in our context were introduced as a combinatorial tool to organize the calculation of moments in double-scaled SYK. However, it turns out that they are much more significant than that. First, chords connecting probe operators (the $M$'s) can be interpreted as trajectories in a bulk made out of the $H$ chords, and hence they may have an intrinsic holographic meaning. Second, they naturally organize the diagrammatic expansion of the theory, allowing for a more general calculation of correlation functions. In this section we work out these diagrammatic rules, in analogy to what was done in  \cite{Mertens:2017mtv, Mertens:2018fds, Lam:2018pvp, Blommaert:2018oro} for the Schwarzian theory. 

\begin{figure}[h]
\centering
\includegraphics[width=0.3\textwidth]{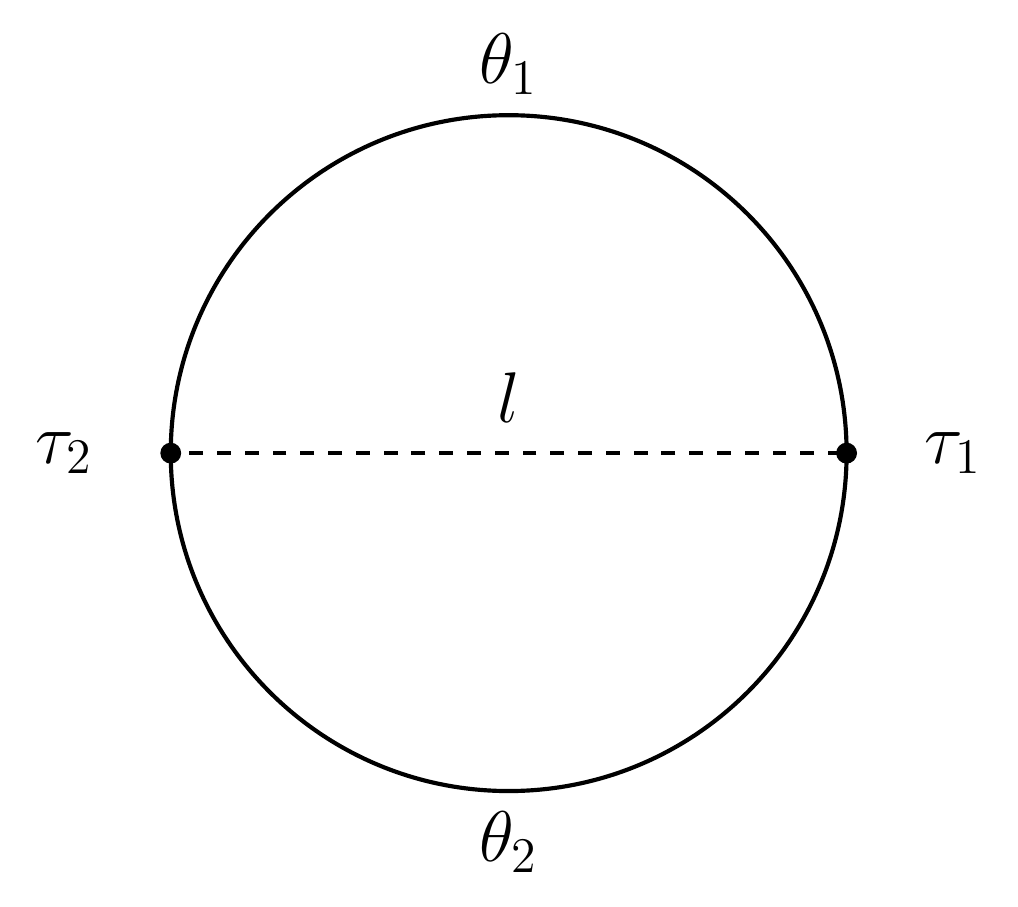}
\caption{A skeleton chord diagram for the 2-point function.}
\label{fig:diagrams_2_point_function}
\end{figure}

For this, we will need just the skeleton of the chord diagrams, which is a similar diagram but made only out of the $M$-chords, with the explicit $H$ insertions dropped. We insert the $M$'s at specific Euclidean times $\tau $, and exponentiate the intermediate Hamiltonians to reflect these times. That is, on each arc we insert  $e^{-\Delta \tau \cdot T} $ (see \autoref{fig:diagrams_2_point_function}). Correspondingly to marking the Euclidean times of the operators on the circle, the circumference of the circle is taken to be $1/T$ ($T$ being the temperature).

Then, the following diagrammatic rules can be used to construct correlation functions:
\begin{itemize}
\item We saw that each segment along the circle represents an evolution using the $T$ matrix during a Euclidean time $\Delta \tau $. Therefore it is natural to associate to such a propagator, for an eigenstate of $T$ with energy $E(\theta ) = \frac{2 \cos \theta }{\sqrt{1-q} } $, the factor

\begin{equation} \label{eq:diagrammatic_rules_propagator}
\begin{gathered}
  \begin{fmfgraph*}(50,50)
    \fmfleft{i1}
    \fmfright{o1}
    \fmfdot{i1,o1}
    \fmflabel{$\tau _2$}{i1}
    \fmflabel{$\tau _1$}{o1}
    \fmf{plain,left,label=$\theta $}{i1,o1}
  \end{fmfgraph*}
\end{gathered}  \quad \quad = e^{-\Delta \tau  \cdot E(\theta )} .
\end{equation}
\item Next, we sum over the energy eigenstates that can propagate in such a segment. Equivalently, we can sum over the corresponding $\theta $. The measure for integration over $\theta $ is
\begin{equation} \label{eq:diagrammatic_rules_measure}
d\mu (\theta ) = \frac{d\theta }{2\pi } (q,e^{\pm 2i\theta } ;q)_{\infty }.
\end{equation}
This measure is the one used in all of the correlation functions we have evaluated (and originates in the measure for $q$-Hermite polynomials).
\item Each operator insertion between two segments $\theta _1,\theta _2$ is naturally thought of as a matrix element $\langle \psi_T^{(\theta _1)} | O | \psi_T^{(\theta_2 )} \rangle$. Let us parametrize the parameter $\tilde q_A$ associated with the operator by $l_A$, defined as
\begin{equation}
\tilde q_A=q^{l_A} .
\end{equation}
The corresponding vertex in the skeleton chord diagram is assigned the value

\begin{equation} \label{eq:diagrammatic_rules_vertex}
\begin{gathered}
  \begin{fmfgraph*}(50,50)
    \fmfleft{i1}
    \fmfright{o1,o2,o3}
    \fmfdot{o2}
    \fmf{plain,label=$l_A$}{o2,i1}
    \fmf{plain,right=0.1,label=$\theta _2$}{o1,o2}
    \fmf{plain,right=0.1,label=$\theta _1$}{o2,o3}
  \end{fmfgraph*}
\end{gathered}  \quad \quad =\sqrt{\frac{(\tilde q_A^2;q)_{\infty } }{\left( \tilde q_A e^{i(\pm \theta _1 \pm \theta _2)} ;q\right)_{\infty } } } ,
\end{equation}
as can be seen by looking at the expression \eqref{eq:2_point_function} for the 2-point function.
\item As we saw in \autoref{sec:trans_inv}, a contracted pair of operators conserves the energy. Thus, the same $\theta $ variable is used before and after such a contracted pair, if there is no additional operator insertions in between. (An example is given below.)

\end{itemize}

\begin{figure}[t!]
    \centering
    \begin{subfigure}[t]{0.4\textwidth}
          \centering
          \includegraphics[height=0.8\textwidth]{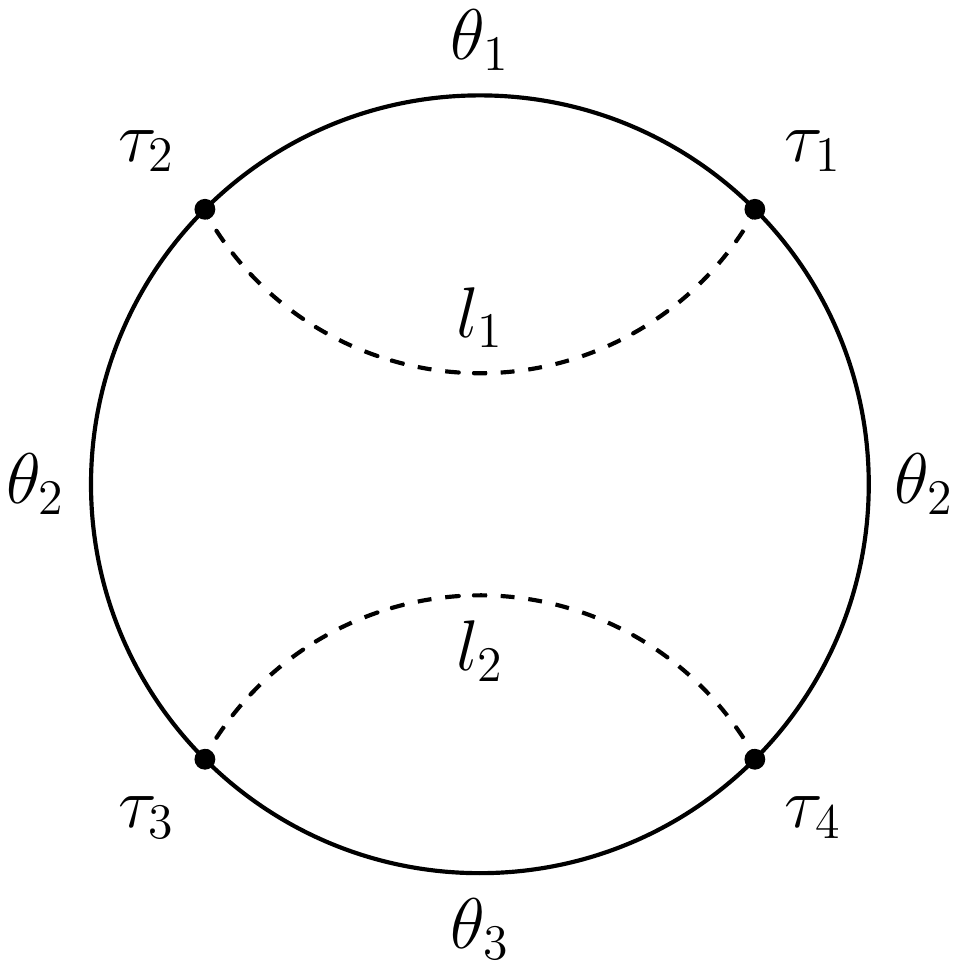}
          \caption{}
          \label{fig:diagrams_4_point_function_ordered}
    \end{subfigure}
    \begin{subfigure}[t]{0.4\textwidth}
          \centering
          \includegraphics[height=0.8\textwidth]{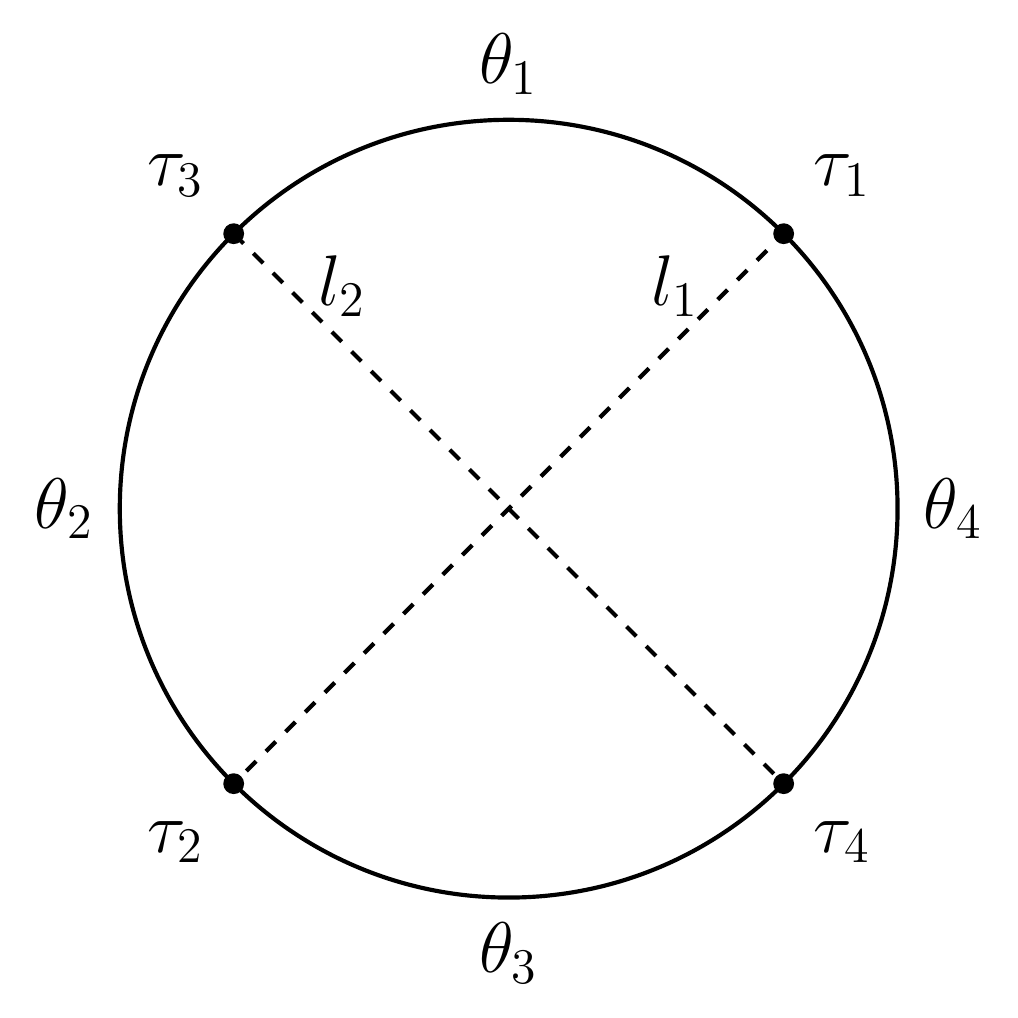}
          \caption{}
          \label{fig:diagrams_4_point_function_OTO}
    \end{subfigure}
    \caption{Skeleton chord diagrams of the 4-point functions $\langle M_1M_1M_2M_2\rangle$ and $\langle M_1 M_2 M_1 M_2\rangle$.}
\end{figure}

These rules indeed reproduce (by construction) the 2-point function. Let us next consider the 4-point functions, starting with the uncrossed one. The corresponding skeleton chord diagram is shown in \autoref{fig:diagrams_4_point_function_ordered}. Note that, because of energy conservation, on both sides of the two contracted operators, the same variable $\theta _2$ appears. The diagrammatic rules above precisely reproduce the expression \eqref{4-pt-q1q2TO-int}.

For the OTO, or more generally crossed, 4-point function expression  we need to add one more rule. Equation \eqref{eq:OTO_4_point_function} can be written as
\begin{equation} \label{eq:OTO_in_terms_of_R}
\begin{split}
& \int _0^{\pi } \prod _{j=1} ^4 \left\{ \frac{d\theta _j}{2\pi } (q,e^{\pm 2i\theta _j} ;q)_{\infty } \right\} e^{-\beta _1 E(\theta _1) - \beta _2E(\theta _2) -\beta _3 E(\theta _3) -\beta _4 E(\theta _4)} \cdot \\
& \cdot \left( \frac{(\tilde q_1^2;q)_{\infty } }{(\tilde q_1 e^{i(\pm \theta _1 \pm \theta _4)} ;q)_{\infty } } 
\frac{(\tilde q_1^2;q)_{\infty } }{(\tilde q_1 e^{i(\pm \theta _2 \pm \theta _3)} ;q)_{\infty } } 
\frac{(\tilde q_2^2;q)_{\infty } }{(\tilde q_2 e^{i(\pm \theta _1 \pm \theta _2)} ;q)_{\infty } } 
\frac{(\tilde q_2^2;q)_{\infty } }{(\tilde q_2 e^{i(\pm \theta _3 \pm \theta _4)} ;q)_{\infty } } \right)^{1/2}  \cdot R^{(q)} _{\theta _4\theta _2} \begin{bmatrix}\theta _3 & l_2 \\ \theta _1 & l_1 \end{bmatrix} .
\end{split}
\end{equation}
with
\begin{equation} \label{eq:R_matrix_full_expr}
\begin{split}
& R^{(q)} _{\theta _4\theta _2} \begin{bmatrix}\theta _3 & l_2 \\ \theta _1 & l_1 \end{bmatrix} 
= \frac{\left( \tilde q_1 e^{-i(\theta _2+\theta _3)} ,\tilde q_1 \tilde q_2 e^{i(\theta _3 \pm \theta _1)} ,\tilde q_1\tilde q_2e^{i(\theta _2 \pm \theta _4)} ;q\right)_{\infty } }{\left( \tilde q_1 \tilde q_2^2 e^{i(\theta _2+\theta _3)} ;q\right)_{\infty } } \cdot \\
& \cdot \frac{(\tilde q_2^2;q)_{\infty } }{\left[ \left( \tilde q_1 e^{i(\pm \theta _2 \pm \theta _3)} ,\tilde q_1 e^{i(\pm \theta _1 \pm \theta _4)} ,\tilde q_2 e^{i(\pm \theta _1 \pm \theta _2)} ,\tilde q_2 e^{i(\pm \theta _3 \pm \theta _4)} ;q\right)_{\infty } \right]^{1/2} }  \cdot \\
& \cdot {}_8W_7\left( \frac{\tilde q_1\tilde q_2^2 e^{i(\theta _2+\theta _3)} }{q} ;\tilde q_1 e^{i(\theta _2+\theta _3)} ,\tilde q_2e^{i(\theta _2\pm \theta _1)} ,\tilde q_2 e^{i(\theta _3 \pm \theta _4)} ;q,\tilde q_1e^{-i(\theta _2+\theta _3)} \right) .
\end{split}
\end{equation}
The set of rules introduced before, applied to \autoref{fig:diagrams_4_point_function_OTO}, give this expression apart from the quantity $R$. It can be thought of as coming from the commutator of operators bringing us from the uncrossed correlation function to the crossed one. The additional rule is therefore 
\begin{itemize}
\item Every crossing of two internal lines comes with a factor of $R$.
\end{itemize}

These diagrammatic rules also give the correct expressions for some of the higher-point functions. For example, any $2n$-point function of the form $\langle e^{-\beta H} M_1(t_1) M_1(t'_1)  \cdot \ldots \cdot M_n(t_n) M_n(t'_n) \rangle_J$ is given simply by using these diagrammatic rules, similarly to the uncrossed 4-point function, and the integration over the $\theta $ variables in the resulting expression can be performed, see \autoref{app:2n_point_function}.

Note that these skeleton chord diagrams are very reminiscent of the hyperbolic disc, with the boundary being our auxiliary quantum mechanical system with Hamiltonian given by the $T$ matrix. The crossing of the chords is like an interaction in the bulk, giving the R matrix. We see that a holographic set of diagrams arises very naturally in the language of chord diagrams.

\end{fmffile}

\subsection{$q \to 1^-$ and low energy limit of the correlation functions}

As a check, we would like to show that the diagrammatic rules reduce, in the limit of $q \to 1^-$ and low energies, to the ones obtained for the ordinary SYK model using the Schwarzian theory \cite{Mertens:2017mtv}.

To show the relation to the Schwarzian, let us introduce a variable $y$ defined by
\begin{equation} \label{eq:y_variables}
\theta =\pi - \lambda y .
\end{equation}
We will use these $y$ variables extensively below. Here we would like to show that they are the closest analog of $k$ of \cite{Mertens:2017mtv}. There, the energies in the Schwarzian case take the form $E(k)=k^2/(2C)$ for $C\sim N$  (or $k^2/(2M)$ in the notation of \cite{Bagrets:2016cdf})\footnote{A simple way to understand this spectrum is that as shown in \cite{Bagrets:2016cdf}, the Schwarzian theory is a Liouville quantum mechanics, and asymptotically the potential vanishes so that the energies are those of a free particle in quantum mechanics.}. For us, in the limit of low energy and $q\rightarrow 1^-$
\begin{equation}
E(y)-E_{min}=\lambda^{3/2} y^2\ ,
\end{equation}
and the identification which leads to the same energies in this limit, is
\begin{equation}
{ \frac{1}{2} } \lambda^{{- }3/2}\rightarrow\ C\ =M,\ \ y\rightarrow k.
\end{equation}
This identification is important in order to understand the time scales in the model. For example, in \cite{Bagrets:2016cdf}, $M$ was identified as setting the time scale in which large fluctuations of the Goldstone modes in the Schwarzian action overtake the purely conformal behavior. For us this is set by $\lambda^{ - 3/2}$, which was verified in \cite{Berkooz:2018qkz} for the 2-point function.

We will also need the $q$-deformation of the Gamma function, which is the $\Gamma _q$ function defined by
\begin{align} \label{eq:Q_Pochhammer_Gamma}
\left(q^x;q\right)_{\infty} = \frac{(q;q)_{\infty}(1-q)^{1-x}}{\Gamma_q(x)}, \quad \quad \lim_{q\to 1^-} \Gamma_q(x) = \Gamma(x).
\end{align} 
Let us see that the diagrammatic rules above indeed become those of the Schwarzian:
\begin{itemize}
\item 
With the identification of the energies above, the propagator \eqref{eq:diagrammatic_rules_propagator} becomes $e^{-\Delta \tau \cdot  E(k)} $ as $\lambda  \to 0$ which is the one in the Schwarzian.
\item 
The measure of integration \eqref{eq:diagrammatic_rules_measure} over intermediate states is given in terms of $y$ by
\begin{equation}
\int _{0} ^{\pi } \frac{d\theta }{2\pi } (q,e^{\pm 2i\theta } ;q)_{\infty } \cdots = \frac{\lambda }{2\pi } (q;q)_{\infty } ^3 (1-q)^2 \int _0^{\pi /\lambda } dy \frac{1}{\Gamma _q(\pm 2iy)} \cdots .
\end{equation}
In the limit $\lambda  \to 0$ this reduces to
\begin{equation}
\frac{2\pi ^2}{\lambda  (q;q)_{\infty } ^3 (1-q)^2} \int _0^{\pi } \frac{d\theta }{2\pi } (q,e^{\pm 2i\theta } ;q)_{\infty } \cdots \overset{\lambda  \to 0}{\rightarrow } \int _0^{\infty  } dy^2\, \sinh (2\pi y) \cdots.
\end{equation}
This is precisely the measure in the case of the Schwarzian with $y \to k$ (we have inserted a $q$-dependent factor on the left-hand side in order to get a finite limit, instead of tracking overall normalizations throughout the computation).
\item 
The value assigned to the vertex \eqref{eq:diagrammatic_rules_vertex} can also be written in terms of the $y$ variables
\begin{equation}
\frac{(q;q)_{\infty } ^3}{(1-q)^{-3+2l} }  \frac{(\tilde q^2;q)_{\infty } }{(\tilde q e^{i(\pm \theta _1 \pm \theta _2)} ;q)_{\infty } } =  \frac{\Gamma_q (l \pm iy_1 \pm i y_2)}{\Gamma_q (2l)} \overset{\lambda  \to 0}{\rightarrow } \frac{\Gamma (l \pm iy_1 \pm i y_2)}{\Gamma (2l)}  , \\
\end{equation}
again giving the Schwarzian result for $y \to k$.

\end{itemize}

These show that the 2-point function and the uncrossed 4-point function reduce to the results from the Schwarzian as $\lambda  \to 0$ and low energies. To show this also for the crossed 4-point function, it remains to check the R-matrix. We have verified that $R$, multiplied by $(q;q)_{\infty } ^3 (1-q)^3$, reduces to the expression in \cite{Mertens:2017mtv} as $q \to 1^-$. This will also be shown for the integral representation of $R$ in the next section.

\section{Chaos and the R-matrix} \label{sec:chaos}

In this section we use the result for the 4-point function to calculate the chaos exponent in double-scaled SYK in the limit of low energy and $q\rightarrow 1^-$. We will verify that we obtain the maximal chaos exponent, and show how to compute corrections (in temperature) to it.

\subsection{Integral form of the R-matrix}

For the calculation of the Lyapunov exponent, it will be useful to obtain an integral representation of the R-matrix. This is done in the present subsection. The Lyapunov exponent can then be argued to come from a specific pole in that integral representation \cite{Mertens:2017mtv,Lam:2018pvp}.

The ${}_8W_7$ function that we have in \eqref{eq:R_matrix_full_expr} is very well-poised and we can use the Mellin-Barnes-Agarwal integral representation \eqref{8W7-MB} for it. However, it will be useful first to take advantage of the symmetries of ${}_8W_7$ (this will also lead to an expression that is manifestly a generalization of the integral form of the R-matrix in \cite{Mertens:2017mtv}). This function has an obvious $S_5$ permutation symmetry in its parameters, which is known to enhance to a $W\left(D_5\right)$ symmetry (the Weyl group of the $D_5$ Lie algebra).
Two nontrivial elements of the double coset $S_5\backslash W\left(D_5\right) \slash S_5$ (i.e.\ excluding equivalence class of the identity) correspond to two symmetry identities \eqref{8W7-Bailey}, \eqref{eq:8W7_second_D5}.
We will find it useful to first perform the transformation $ \begin{pmatrix}1 & 2 &3 &4 &5 \\ 1 & 3 & 2 & 4 & 5\end{pmatrix} \in S_5$, then employ the second identity \eqref{eq:8W7_second_D5}, and then perform $ \begin{pmatrix}1 &2 &3 &4 &5 \\ 1&3&4&2&5\end{pmatrix} \in S_5$. Using these transformations, the ${}_8W_7$ in \eqref{eq:R_matrix_full_expr} becomes
\begin{equation}
\begin{split}
& {}_8W_7\left( \frac{\tilde q_1\tilde q_2^2 e^{i(\theta _2+\theta _3)} }{q} ;\tilde q_1 e^{i(\theta _2+\theta _3)} ,\tilde q_2e^{i(\theta _2\pm \theta _1)} ,\tilde q_2 e^{i(\theta _3 \pm \theta _4)} ;q,\tilde q_1e^{-i(\theta _2+\theta _3)} \right) = \\
& = \frac{\left( \tilde q_1\tilde q_2^2 e^{i(\theta _2+\theta _3)} ,\tilde q_1e^{i(\theta _2+\theta _3)} ,\tilde q_1\tilde q_2 e^{i(\pm \theta _1 - \theta _3)} ,\tilde q_1\tilde q_2e^{i(\pm \theta _4-\theta _2)} ;q\right)_{\infty } }{\left( \tilde q_1\tilde q_2e^{i(\theta _3\pm \theta _1)}, \tilde q_1\tilde q_2e^{i(\theta _2\pm \theta _4)} ,\tilde q_1\tilde q_2^2 e^{-i(\theta _2+\theta _3)} ,\tilde q_1e^{-i(\theta _2+\theta _3)} ;q\right)_{\infty } } \cdot \\
& \cdot {}_8W_7\left(\frac{\tilde q_1\tilde q_2^2}{q} e^{-i(\theta _2+\theta _3)} ;\tilde q_2e^{i(\theta _1-\theta _2)} ,\tilde q_2e^{i(-\theta _3\mp\theta _4)} ,\tilde q_2e^{i(-\theta _1-\theta _2)} ,\tilde q_1e^{-i(\theta _2+\theta _3)} ;q,\tilde q_1e^{i(\theta _2+\theta _3)} \right) .
\end{split}
\end{equation}
We apply now \eqref{8W7-MB}, using the variables $y_i$ defined in \eqref{eq:y_variables} (even though we do not take $q \to 1^-$), and express the results via $\Gamma _q$ and $\Gamma $, while using the familiar identity $\Gamma (1-z)\Gamma (z)= \frac{\pi }{\sin(\pi z)} $. We find for the R-matrix
\begin{equation}
\begin{split}
&  R_{\theta _4,\theta _2} \begin{bmatrix}\theta _3 & l_2 \\ \theta _1 & l_1\end{bmatrix} = -\frac{(1-q)^{-2} }{\Gamma (1+l_2-l_1+iy_1-iy_3)\Gamma (l_1-l_2+iy_3-iy_1)} \frac{1}{\left(q,\frac{q\tilde q_2}{\tilde q_1} e^{i(\theta _1-\theta _3)} ,\frac{\tilde q_1}{\tilde q_2} e^{i(\theta _3-\theta _1)} ;q\right)_{\infty } } \cdot \\
& \cdot \sqrt{\frac{\left(\tilde q_1 e^{i(\theta _3 \pm \theta _2)} ,\tilde q_1e^{i(-\theta _1\pm \theta _4)} ,\tilde q_2e^{i(\theta _1\pm \theta _2)} ,\tilde q_2e^{i(-\theta _3\pm \theta _4)} ;q\right)_{\infty } }{\left( \tilde q_1 e^{i(-\theta _3\pm \theta _2)} ,\tilde q_1 e^{i(\theta _1\pm \theta _4)} ,\tilde q_2 e^{i(-\theta _1 \pm \theta _2)} ,\tilde q_2 e^{i(\theta _3\pm \theta _4)} ;q\right)_{\infty } } } \cdot \\
& \cdot  \int_{\mathcal{C}} \frac{ds}{2\pi i} q^s \Gamma (1+s)\Gamma (-s)\Gamma (s+1-l_1+l_2+iy_1-iy_3)\Gamma (-s+l_1-l_2-iy_1+iy_3) \cdot \\
& \cdot \frac{\Gamma _q(s+l_2+iy_1-iy_2)\Gamma _q(s+l_2-iy_3\pm iy_4)\Gamma _q(s+l_2+iy_1+iy_2)}{\Gamma _q(s+1)\Gamma _q(s+l_1+l_2+iy_1-iy_3)\Gamma _q(s+2l_2)\Gamma _q(s+1+l_2-l_1+iy_1-iy_3)} .
\end{split}
\end{equation}
The contour $\mathcal{C}$ of integration is a deformation of the contour going along the imaginary axis, such that the poles that come from Gamma functions with $s$ arguments (with a positive coefficient) are to the left of the contour, and those that come from Gamma functions with $(-s)$ are to the right of the contour (a usual Mellin-Barnes prescription). 
Then we shift the integration variable $ s \to s-l_2+iy_3-iy_4$ (with no other contributions because of the Mellin-Barnes prescription) and express all the $q$-Pochhammers in terms of $\Gamma _q$, to get
\begin{equation} \label{eq:R_matrix_integral_form}
\begin{split}
&  R_{\theta _4,\theta _2} \begin{bmatrix}\theta _3 & l_2 \\ \theta _1 & l_1\end{bmatrix} = - \frac{1 }{(q;q)_{\infty } ^3 (1-q)^3} \cdot \frac{\Gamma _q(1+l_2-l_1+iy_1-iy_3)\Gamma _q(l_1-l_2+iy_3-iy_1)}{\Gamma (1+l_2-l_1+iy_1-iy_3)\Gamma (l_1-l_2+iy_3-iy_1)} \cdot \\
& \cdot \sqrt{\frac{\Gamma _q(l_1-iy_3\pm iy_2) \Gamma _q(l_1+iy_1\pm iy_4) \Gamma _q(l_2-iy_1\pm iy_2)\Gamma _q(l_2+iy_3 \pm iy_4)}{\Gamma _q(l_1+iy_3 \pm iy_2) \Gamma _q(l_1-iy_1\pm iy_4) \Gamma _q(l_2+iy_1 \pm iy_2) \Gamma _q(l_2-iy_3\pm iy_4)} } \cdot \\
& \cdot \int_{\mathcal{C}}  \frac{ds}{2\pi i} q^{s-l_2+iy_3-iy_4} \frac{\Gamma (s+1-l_2+iy_3-iy_4)\Gamma (s+1-l_1+iy_1-iy_4)}{\Gamma _q(s+1-l_2+iy_3-iy_4) \Gamma _q(s+1-l_1+iy_1-iy_4)}  \cdot \\
& \cdot \frac{\Gamma _q(s)\Gamma _q(s-2iy_4)\Gamma _q(s+iy_1+iy_3-iy_4 \pm iy_2)\Gamma (-s+l_2-iy_3+iy_4)\Gamma (-s+l_1+iy_4-iy_1)}{\Gamma _q(s+l_1+iy_1-iy_4)\Gamma _q(s+l_2+iy_3-iy_4)}  .
\end{split}
\end{equation}

With the identification between our $y$ variables and the $k$ variables in the case of the Schwarzian that we expect from comparing \autoref{fig:diagrams_4_point_function_OTO} to \cite{Mertens:2017mtv},
\begin{equation}
y_1 \leftrightarrow k_1,\qquad y_2 \leftrightarrow k_t,\qquad y_3 \leftrightarrow k_4,\qquad y_4 \leftrightarrow k_s,
\end{equation}
one sees immediately that \eqref{eq:R_matrix_integral_form} (without the $q$-dependent constant in front) reduces to (B.28) in \cite{Mertens:2017mtv} as $q \to 1^-$. We can notice that the $q$-deformation of it is rather nontrivial, and in particular cannot be obtained via a simple replacement of the Gamma functions by their $q$-deformations. In \autoref{sec:quantum-def}, we will show that the deformed R-matrix \eqref{eq:R_matrix_full_expr} is related to a $6j$-symbol of the quantum group $\mathcal{U}_{q^{1/2}}(su(1,1))$.

\subsection{Chaos}

The goal in this subsection is to get the Lyapunov exponent as a function of $\lambda $, in the limit $\lambda\rightarrow 0^+$ and low energies. The leading order is just the maximal chaos exponent, but we will go beyond that and show how to compute the subleading corrections in $\lambda$.

Let us start by writing the full crossed 4-point function \eqref{eq:OTO_in_terms_of_R} in terms of $\Gamma $ and $\Gamma _q$ functions.  Using \eqref{eq:Q_Pochhammer_Gamma} we have
\begin{equation} \label{eq:OTO_4_point_function_chaos}
\begin{split}
& \langle \tr e^{-\beta _4 H}
\contraction[2ex]{}{M_2}{e^{-\beta _3 H} M_1 e^{-\beta _2 H} }{M_2}
\contraction{M_2 e^{-\beta _3 H}}{M_1}{e^{-\beta _2 H} M_2 e^{-\beta _1 H} }{M_1}
M_2 e^{-\beta _3 H} M_1 e^{-\beta _2 H} M_2 e^{-\beta _1 H} M_1 \rangle_J = \\
& =\frac{(q;q)_{\infty } ^6 (1-q)^{2+2l_1+2l_2} }{\Gamma _q(2l_1)\Gamma _q(2l_2)}  \int _0^{\pi } \prod _{j=1} ^4 \left[ \frac{d\theta _j}{2\pi } \frac{1}{\Gamma _q(\pm 2i y_j)} \right] e^{-\sum _j \beta _j E_j}  \cdot \\
& \cdot \sqrt{\Gamma _q(l_1 \pm iy_1 \pm iy_4)\Gamma _q(l_1\pm iy_2 \pm iy_3) \Gamma _q(l_2 \pm iy_1 \pm iy_2) \Gamma _q(l_2 \pm iy_3 \pm iy_4)}  \cdot R ,
\end{split}
\end{equation}
where $R$ is given by \eqref{eq:R_matrix_integral_form}. Recall that $y_i=(\pi-\theta_i)/\lambda$.

For the calculation in this section, it will be convenient to change to energies as integration variables recalling that $E={2\cos(\theta)\over \sqrt{1-q}} $. We then write \eqref{eq:OTO_4_point_function_chaos} as
\begin{equation} \label{eq:OTO_for_chaos_energy_variables}
\begin{split}
& \langle \tr e^{-\beta _4 H}
\contraction[2ex]{}{M_2}{e^{-\beta _3 H} M_1 e^{-\beta _2 H} }{M_2}
\contraction{M_2 e^{-\beta _3 H}}{M_1}{e^{-\beta _2 H} M_2 e^{-\beta _1 H} }{M_1}
M_2 e^{-\beta _3 H} M_1 e^{-\beta _2 H} M_2 e^{-\beta _1 H} M_1 \rangle_J = \\
& = \frac{(q;q)_{\infty } ^6 (1-q)^{4+2l_1+2l_2} }{(4\pi )^4 \Gamma _q(2l_1)\Gamma _q(2 l_2)} \int _{-2/\sqrt{1-q} } ^{2/\sqrt{1-q} } \prod _{j=1} ^4 dE_j\,  e^{-\sum _j \beta _j E_j + L} ,
\end{split}
\end{equation}
where
\begin{equation}
\begin{split}
&L= L_1+L_2+L_3+L_4,\\
& L_1 = \log(R), \qquad  L_2 = \log \left( \prod _{j=1} ^4 \frac{1}{\Gamma _q(\pm 2iy_j)} \right) ,\qquad  L_3 = \log\left( \prod _{j=1} ^4\frac{1}{\sin (\lambda y_j)} \right) ,\\
& L_4 = \log \sqrt{\Gamma _q(l_1 \pm iy_1 \pm iy_4)\Gamma _q(l_1\pm iy_2 \pm iy_3) \Gamma _q(l_2 \pm iy_1 \pm iy_2) \Gamma _q(l_2 \pm iy_3 \pm iy_4)} .
\end{split}
\end{equation}

We will work in a regime where the differences in energies are small compared to the average energy. It is then convenient to define $\bar \alpha $, $\bar \beta $, and $\bar \omega $ though
\begin{equation}\label{VarA}
y_2=y_1+{\bar\beta},\ \ y_3=y_1+{\bar\omega},\ \ y_4=y_1+{\bar \alpha} .
\end{equation}
Similarly, we measure the energies above the ground state, defining the corresponding variables $\alpha ,\beta ,\omega $ through
\begin{align}\label{VarB}
E_1 & = M - \frac{2}{\sqrt{1-q} },\ \ E_2=E_1+\beta,\ \  E_3=E_1+\omega,\ \ E_4=E_1+\alpha.
\end{align}
Of course, \eqref{VarA} and \eqref{VarB} contain the same information, but it will be convenient in various formulas to use one or the other.

The interesting regime of temperatures (in which the following calculation can be performed) is
\begin{equation} \label{eq:chaos_range_of_energies}
\lambda ^{3/2} \ll T \ll \sqrt{\lambda }
\end{equation}
and we will restrict to small $\lambda  \ll 1$.
We will see that this range is accessible by restricting to relatively large energies in the integral, but with small energy differences (compared to the average energy), that is
\begin{equation} \label{eq:approx_for_chaos_energies1}
 \alpha ,\beta ,\omega  \ll M ,
\end{equation}
and
\begin{equation} \label{eq:approx_for_chaos_energies2}
T \ll \alpha, \beta,\omega .
\end{equation}
This regime mimics the bulk scattering of shock waves in the background of a black hole \cite{Dray:1984ha,Shenker:2013pqa,Jackson:2014nla}.

We will see later that the relation between the temperature and the averaged energy is
\begin{equation}
T\sim \sqrt{M}\lambda^{3/4},
\end{equation}
so that the small temperature condition $T \ll \sqrt{\lambda } $ in \eqref{eq:chaos_range_of_energies} translates to
\begin{equation} \label{eq:approx_for_chaos_energies3}
M \ll \lambda^{-1/2} .
\end{equation}
In terms of the barred variables, the range of energies specified by the above conditions \eqref{eq:approx_for_chaos_energies1}, \eqref{eq:approx_for_chaos_energies2}, \eqref{eq:approx_for_chaos_energies3}, and giving the dominant contribution, is summarized simply by
\begin{equation} \label{eq:approx_for_chaos}
1 \ll \bar \alpha ,\bar \beta ,\bar \omega  \ll y_1 \ll 1/\lambda  .
\end{equation}

The steps in performing the computation of the Lyapunov exponent are the following:
\begin{itemize}
\item  As in \cite{Mertens:2017mtv,Lam:2018pvp}, we begin by reducing the $s$ integral in \eqref{eq:R_matrix_integral_form} to the contribution of its $s=0$ pole, and specializing to $l_1=l_2=0$.
\item Next we need to estimate $\Gamma$ and $\Gamma_q$ in the relevant ranges \eqref{eq:approx_for_chaos_energies1}, \eqref{eq:approx_for_chaos_energies2}, \eqref{eq:approx_for_chaos_energies3} and get an estimation of the crossed 4-point function.
\item Using a saddle point, we evaluate the relation between the typical energy $M$ and the temperature $T$,  and relations between external time differences and the saddle point energies.
\item Finally, we extract the chaos exponent.
\end{itemize}

\medskip
\textbf{Reduction to the $s=0$ pole.} The $q$-Gamma function has a pole at zero argument similarly to the usual Gamma function
\begin{equation}
\Gamma _q(z) = \frac{1-q}{-z \log(q)} + O(1) .
\end{equation}
By the same arguments as in \cite{Lam:2018pvp}, for small $\lambda $ and large time separation $t$ in $\langle \tr e^{-\beta H} \cdot M_1(0) M_2(t) M_1(0) M_2(t) \rangle_J$, the $s=0$ pole of the first $\Gamma_q(s)$ in the last line of \eqref{eq:R_matrix_integral_form} gives the dominant contribution.\footnote{The argument of \cite{Lam:2018pvp} uses the fact that the Gamma function goes to zero when its argument has large (in magnitude) imaginary part. This is also true here; as $y_1$ is taken to be large, but still $y_1 \le 1/\lambda $, $\Gamma _q(c \pm iy_1) $ goes to zero. One could worry that the corrections to this statement may modify our correction to the maximal Lyapunov exponent, but the corrections here are $O(\epsilon _2\epsilon _3)$ in the notations of \eqref{eq:expansion_parameters}, while the correction we get eventually is $O(\epsilon _1)$.}

Moreover, the chosen operators should not affect the chaotic behavior, and so the Lyapunov exponent should not depend on $l_1$, $l_2$.  Therefore we will set for convenience $l_1=l_2=0$. Taking the $s=0$ residue, \eqref{eq:R_matrix_integral_form} reduces to
\begin{equation} \label{eq:R_matrix_saddle_point}
\begin{split}
&R \approx \frac{1}{(q;q)_{\infty } ^3 (1-q)^2 \log(q)}  \frac{\Gamma _q(1+iy_{1-3} )\Gamma _q(iy_{3-1} )}{\Gamma (1+iy_{1-3} )\Gamma (iy_{3-1} )} 
\sqrt{\frac{\Gamma _q(-iy_{3\pm 2} )\Gamma _q(iy_{1 \pm 4} ) \Gamma _q(-iy_{1 \pm 2} ) \Gamma _q(iy_{3 \pm 4} )}{\Gamma _q(iy_{3 \pm 2} ) \Gamma _q(-iy_{1\pm  4} ) \Gamma _q(iy_{1 \pm 2} ) \Gamma _q(-iy_{3 \pm 4} )} }  \cdot \\
& \cdot q^{iy_{3-4} } \frac{\Gamma (1+iy_{3-4} )\Gamma (1+iy_{1-4} )}{\Gamma _q(1+iy_{3-4} )\Gamma _q(1+iy_{1-4} )}
\frac{\Gamma _q(-2iy_4)\Gamma _q(iy_{1+3-4\pm 2} )\Gamma (iy_{4-3} )\Gamma (iy_{4-1} )}{\Gamma _q(iy_{1-4} )\Gamma _q(iy_{3-4} )} ,
\end{split}
\end{equation}
where the shorthand notation $y_{a \pm b \pm \cdots } =y_a \pm y_b \pm \cdots $ is used.

\medskip

\textbf{Estimation of the crossed 4-point function.}  For starters, from now on we will not keep track of additive $q$-dependent constants in the various $L_j$'s, as these will affect only the overall normalization, whereas we are interested in the time dependence.

While ideally we would like to perform the calculation for every $\lambda $, we restrict ourselves to getting the leading correction for small $\lambda $, manifested in \eqref{eq:approx_for_chaos_energies2} and \eqref{eq:approx_for_chaos}.

Next we need to evaluate the various $\Gamma$ and $\Gamma_q$ in \eqref{eq:OTO_for_chaos_energy_variables}, in the regime \eqref{eq:approx_for_chaos}. The approximation \eqref{eq:approx_for_chaos} means that we need to estimate $\Gamma (z)$, $\Gamma _q(z)$, and $\Gamma _q(z+\epsilon )$ for $1 \ll |z| \ll 1/\lambda $ close to or on the imaginary axis, and $\epsilon \ll z$.\footnote{The condition $\bar \alpha ,\bar \beta ,\bar \omega  \gg 1$ in \eqref{eq:approx_for_chaos} is needed since we need to estimate $\Gamma _q(z)$ with $z \sim \bar \alpha ,\bar \beta ,\bar \omega $.} In \autoref{app:asymptotic_Gamma}, these functions are expanded in $1/z$, $\lambda z$ and $\epsilon /z$.
The resulting expansions that we will use here are \eqref{eq:Stirling_approx}, \eqref{eq:log_Gamma_q_approx}, \eqref{eq:log_Gamma_q_z_epsilon}, \eqref{eq:log_Gamma_q_z_epsilon_diff}, and \eqref{eq:log_Gamma_q_z_epsilon_sum}.
Indeed, in the approximation \eqref{eq:approx_for_chaos} our expansion parameters are
\begin{equation} \label{eq:expansion_parameters}
\begin{split}
 & \epsilon _1 = \lambda y_1 \approx \sqrt{M} \lambda ^{1/4} ,\\
 & \epsilon _2 = \frac{\max(\bar \alpha ,\bar \beta ,\bar \omega )}{y_1} \approx \frac{\max(\alpha ,\beta ,\omega )}{2M} ,\\
 & \epsilon _3=\frac{1}{\min(\bar \alpha ,\bar \beta ,\bar \omega )} \approx \frac{2\sqrt{M} \lambda ^{3/4} }{\min(\alpha ,\beta ,\omega )} ,
 \end{split}
\end{equation}
which translate into the three expansion parameters above differently in the various Gamma functions.

As we use both sets of variables $(y_1,\bar \beta ,\bar \omega ,\bar \alpha )$ and $(M,\beta ,\omega ,\alpha )$, it is useful to write the relation between them:
\begin{equation} \label{eq:relation_barred_unbarred_variables}
\begin{split}
& y_1 = \frac{2}{\lambda } \arcsin \frac{(1-q)^{1/4} \sqrt{M} }{2} = \frac{\sqrt{M} }{\lambda ^{3/4} } \left( 1 + O\left(\epsilon _1^2,\epsilon _1\epsilon _2 \epsilon _3\right) 
\right),\\
& \bar \alpha  = \frac{2}{\lambda } \left(  \arcsin \frac{(1-q)^{1/4} \sqrt{M+\alpha } }{2} -  \arcsin \frac{(1-q)^{1/4} \sqrt{M} }{2} \right) =
\frac{\alpha }{2\sqrt{M} \lambda ^{3/4} } \left( 1 + O\left(\epsilon _1^2,\epsilon _2 \right)
\right) ,
\end{split}
\end{equation}
and similar equations to the second one for $\bar \beta ,\bar \omega $ (obtained by replacing correspondingly $\alpha \mapsto \beta $ and $\alpha  \mapsto \omega $).

Now we can evaluate the asymptotic form of the crossed 4-point function, using the splitting to the $L_j$ pieces before, each one in turn.

The first piece is the R-matrix.
Using \eqref{eq:Stirling_approx}, \eqref{eq:log_Gamma_q_approx}, \eqref{eq:log_Gamma_q_z_epsilon}, and \eqref{eq:log_Gamma_q_z_epsilon_diff} in \eqref{eq:R_matrix_saddle_point}, after some calculation we arrive at
\begin{equation}
\begin{split}
& L_1 = i(\bar \omega -\bar \alpha -\bar \beta )\log(2y_1) + i\bar \alpha  \log |\bar \alpha | + i\bar \beta  \log |\bar \beta | - i(\bar \omega -\bar \alpha ) \log |\bar \omega -\bar \alpha |  - \\
&- i (\bar \omega -\bar \beta ) \log |\bar \omega -\bar \beta | + \left(i(\bar \omega -\bar \alpha -\bar \beta )-\frac{1}{2} \right)\log \left(i(\bar \omega -\bar \alpha -\bar \beta )\right) - \frac{\pi }{2} (\bar \alpha +\bar \beta +\bar \omega )- \\
& - 2\pi y_1 - \log(2y_1) + 2\lambda y_1^2+\lambda y_1(\bar \omega +\bar \beta +\bar \alpha ) + i(\bar \omega -\bar \alpha -\bar \beta )+ \frac{\lambda }{2} (\bar \alpha ^2+\bar \beta ^2+\bar \omega ^2) - \frac{\bar \alpha +\bar \beta +\bar \omega }{4y_1} +\cdots .
\end{split}
\end{equation}
The corrections to this formula (as well as the other $L_j$ below) go as $\frac{1}{\bar \alpha } ,\lambda ^2\bar \alpha ^3,\frac{1}{y_1} ,\lambda ^2y_1^3,\frac{\bar \alpha ^2}{y_1} $ (and $\bar \alpha \to \bar \beta ,\bar \omega $) and smaller terms (as already mentioned, we drop the $q$-dependent constants).

The measure in \eqref{eq:OTO_4_point_function_chaos} is evaluated using \eqref{eq:log_Gamma_q_z_epsilon_sum} (where $y_1 > 0$)
\begin{equation}
\begin{split}
& L_2 = 4 \log(2y_1)+8\pi y_1 - 8\lambda y_1^2+ \\
& \qquad + 2\pi (\bar \alpha +\bar \beta +\bar \omega )-4\lambda y_1(\bar \alpha +\bar \beta +\bar \omega )-2\lambda (\bar \alpha ^2+\bar \beta ^2+\bar \omega ^2) + \frac{\bar \alpha +\bar \beta +\bar \omega }{y_1} +\cdots .
\end{split}
\end{equation}
Clearly it is straightforward to expand $L_3$ and get
\begin{equation}
L_3 = -4\log(\lambda y_1)- \frac{\bar \alpha +\bar \beta +\bar \omega }{y_1} + \cdots .
\end{equation}
(The corrections to $L_3$ are even smaller than the ones for $L_1$.)

The last remaining piece to evaluate is the square root term in \eqref{eq:OTO_4_point_function_chaos}. Using \eqref{eq:log_Gamma_q_z_epsilon_sum} we find
\begin{equation}
\begin{split}
& L_4 = -2 \log(2y_1)-4\pi y_1+4\lambda y_1^2+ \frac{1}{2}  \Big[ 4\lambda y_1 (\bar a+\bar \beta +\bar \omega ) -\pi \bar \alpha -\pi |\bar \alpha |-\pi \bar \beta -\pi |\bar \beta |- \\
& - \pi (\bar \beta +\bar \omega )-\pi |\bar \beta -\bar \omega |-\pi (\bar \omega +\bar \alpha )-\pi |\bar \alpha -\bar \omega | - \frac{1}{2} \log \left(\bar \alpha ^2\bar \beta ^2(\bar \beta -\bar \omega )^2(\bar \alpha -\bar \omega )^2\right) + \\
& + 2\lambda (\bar \alpha ^2+\bar \beta ^2+\bar \omega ^2) - \frac{\bar \alpha +\bar \beta +\bar \omega }{y_1} \Big] + \cdots .
\end{split}
\end{equation}

\textbf{Saddle point equations.} Finally, we would like to write the saddle point equations for \eqref{eq:OTO_for_chaos_energy_variables}, without the $q$-dependent factor in front of the integral, and with $l_1=l_2=0$ (ignoring the constant factors $\Gamma _q(2l_j)$), as mentioned above.

Each of these equations is first written in terms of derivatives with respect to the integration variables $E_j$, which are subsequently translated to $M ,\alpha ,\beta ,\omega $, and finally to derivatives with respect to the variables used to evaluate $L$ above: $y_1,\bar \alpha ,\bar \beta ,\bar \omega $.

The purpose of the first saddle point equation we write, is to find the relation between the typical energy $M$ and the temperature $T$. Denoting the Euclidean times of the operator insertions as in \autoref{fig:diagrams_4_point_function_OTO}, we have the relations
\begin{equation}
\beta _1=\tau _3-\tau _1,\quad \beta _2=\tau _2-\tau _3,\quad \beta _3=\tau _4-\tau _2,\quad \beta _4=\frac{1}{T} +\tau _1-\tau _4 .
\end{equation}
No matter how $1/T$ is split between the different $\beta _j$, we clearly always have $\sum _j \beta _j=1/T$. Therefore, summing the saddle point equations of \eqref{eq:OTO_for_chaos_energy_variables} for all $E_j$ gives the temperature
$\frac{1}{T} = \sum_{j=1}^4 \pder{L}{E_j}  = \pder{L}{M}$.
Using the results for $L_j$ and \eqref{eq:relation_barred_unbarred_variables}, we get
\begin{equation}
\begin{split}
& \frac{1}{T} = \\
& =  \frac{1}{\lambda ^{3/4} } \Bigg[ \frac{1}{2 \sqrt{M} } \left(1+
 O\left( \epsilon _1^2,\epsilon _1\epsilon _2\epsilon _3 \right)
 \right) \left( 2\pi  - 4\lambda y_1 + \frac{i(\bar \omega -\bar \alpha -\bar \beta )}{y_1} +
 O \left( \epsilon _2\epsilon _3,\epsilon _1\epsilon _2\right)
 \right) - \\
& - \frac{\alpha }{4M^{3/2} } \left(1 
+ O \left(\epsilon _1^2,\epsilon _2\right) 
\right) \cdot 
\left( i \log \frac{|\bar \alpha |\cdot |\bar \omega -\bar \alpha |}{2y_1(\bar \omega -\bar \alpha -\bar \beta )} +
O\left(1\right)
\right) - \\
& - \frac{\beta }{4M^{3/2} } i \log \frac{|\bar \beta | \cdot |\bar \omega -\bar \beta |}{2y_1 (\bar \omega -\bar \alpha -\bar \beta )} - \frac{\omega }{4M^{3/2} } i \log \frac{2y_1 (\bar \omega -\bar \alpha -\bar \beta )}{|\bar \omega -\bar \alpha |\cdot |\bar \omega -\bar \beta |}  + \cdots \Bigg].
\end{split}
\end{equation}
In the last line we have written the terms for $\beta ,\omega $ similarly to those of $\alpha $ but without all the higher order corrections (for which the dots at the end stand). In short, we have
\begin{equation}
\begin{split}
& \frac{1}{T} =\frac{1}{2\sqrt{M} \lambda ^{3/4} } \Bigg[ 2\pi -4\lambda y_1 - \frac{\alpha }{2M} i \log \frac{|\alpha | \cdot |\omega -\alpha |}{4M(\omega -\alpha -\beta )} -\\
& \quad - \frac{\beta }{2M} i \log \frac{|\beta | \cdot |\omega -\beta |}{4M(\omega -\alpha -\beta )} -\frac{\omega }{2M} i \log \frac{4M(\omega -\alpha -\beta )}{|\omega -\alpha |\cdot |\omega -\beta |} +O(\epsilon _2) + (\text{smaller terms}) \Bigg] .
\end{split}
\end{equation}
The $\lambda y_1$ term is a correction of order $\epsilon _1$ to the leading contribution, and the other corrections are of order $\epsilon _2$. These latter corrections give an imaginary contribution, and they originate from the R-matrix. Since $T$ is an external real parameter that we specify, we expect the solutions to the saddle point equations to be such that the sum of these latter corrections is negligible. Alternatively, the relation between $M$ and $T$ that we derive can be found from other correlation functions so that it should not be affected by the R-matrix. We will therefore work in the following up to order $O(\epsilon _1)$. This gives
\begin{equation}
T = \frac{\sqrt{M} \lambda ^{3/4} }{\pi - 2 \sqrt{M} \lambda ^{1/4} +\cdots } .
\end{equation}
This relation can also be inverted, so that
\begin{equation}
\sqrt{M} \lambda ^{3/4} = \pi T - 2\pi \lambda ^{-1/2} T^2 + \cdots .
\end{equation}

To get the Lyapunov exponent, there are two ways to proceed. The first one follows \cite{Mertens:2017mtv}.
Let us choose to put the $1/T$ in $\beta _1$ only, so that $\beta _2,\beta _3,\beta _4$ are purely imaginary:
\begin{equation}
\beta _2 = i(t_2-t_3),\quad \beta _3=i(t_4-t_2),\quad \beta _4=i(t_1-t_4) .
\end{equation}
The saddle point equation for $E_4$ is
\begin{equation} \label{eq:chaos_saddle_point_eq2_raw}
\begin{split}
i(t_1-t_4) = \pder{L}{E_4} = \pder{L}{\alpha } = \pder{\bar \alpha }{\alpha } \pder{L}{\bar \alpha }  =
\frac{1}{2\sqrt{M} \lambda ^{3/4} } \left(1 + 
O\left(\epsilon _1^2,\epsilon _2\right)
\right) 
\cdot \Bigg[ i \log \frac{|\bar \alpha |\cdot |\bar \omega -\bar \alpha |}{2y_1(\bar \omega -\bar \alpha -\bar \beta )} +
O(1)
\Bigg] .
\end{split}
\end{equation}
Note 
that at order $\epsilon _1$ there are no corrections in the first round brackets in \eqref{eq:chaos_saddle_point_eq2_raw}. Therefore, at this order, the saddle point equation is simply
\begin{equation} \label{eq:chaos_saddle_point_eq2}
t_1-t_4 \approx \frac{1}{2 \sqrt{M} \lambda ^{3/4} } \log \frac{|\alpha |\cdot |\omega -\alpha |}{4M(\omega -\alpha -\beta )} .
\end{equation}

Lastly, we write the sum of the saddle point equations for $E_2$ and $E_3$. Similarly to before,
\begin{equation}
\begin{split}
& i(t_4-t_3) = \pder{L}{E_2} + \pder{L}{E_3} = \pder{L}{\beta } + \pder{L}{\omega } = \\
& = \frac{1}{2\sqrt{M} \lambda ^{3/4} } \left( 1+
O\left(\epsilon _1^2,\epsilon _2\right)
\right) \cdot 
\Bigg[ i \log \frac{|\bar \beta |\cdot |\bar \omega -\bar \beta |}{2y_1 (\bar \omega -\bar \alpha -\bar \beta )} +O(1)\Bigg] + \\
& + \frac{1}{2\sqrt{M } \lambda ^{3/4} } \left(1+
O\left(\epsilon _1^2,\epsilon _2\right)
\right) \cdot 
\Bigg[ i \log \frac{2y_1 (\bar \omega -\bar \alpha -\bar \beta )}{|\bar \omega -\bar \alpha |\cdot |\bar \omega -\bar \beta |} +O(1) \Bigg] .
\end{split}
\end{equation}
Again working up to order $O(\epsilon _1)$, as in the previous saddle point equation,
we may approximate this by
\begin{equation} \label{eq:chaos_saddle_point_eq3}
t_4-t_3 \approx \frac{1}{2\sqrt{M} \lambda ^{3/4} } \log \frac{|\beta |}{|\omega -\alpha |} .
\end{equation}
If we think about \autoref{fig:diagrams_4_point_function_OTO} as a scattering process, where the interior of the circle corresponds to the bulk, then $t_1-t_4$ is the difference in the times of the incoming states, and $t_3-t_4$ is the time delay of the outgoing state. This is the picture used in \cite{Mertens:2017mtv}, and equations \eqref{eq:chaos_saddle_point_eq2}, \eqref{eq:chaos_saddle_point_eq3} are the same as equations (5.26) and (5.29) of \cite{Mertens:2017mtv}. 
One way to proceed is to express $t_4-t_3$ as a function of $t_1-t_4$ as in \cite{Mertens:2017mtv}, leading to
\begin{equation}
t_4-t_3 \approx -\sign(\omega -\alpha ) \frac{|\alpha |}{2\sqrt{M} \lambda ^{3/4} } e^{2\sqrt{M } \lambda ^{3/4} (t_4-t_1-t_R)} ,\qquad t_R = \frac{\log (4M)}{2\sqrt{M} \lambda ^{3/4} } +\cdots.
\end{equation}
From this, we can read off the Lyapunov exponent $\lambda _L$, finding
\begin{equation} \label{eq:Lyapunov_exponent_through_M}
\begin{split}
\lambda _L &= 2\sqrt{M} \lambda ^{3/4} . 
\end{split}
\end{equation}
Alternatively, following \cite{Lam:2018pvp}, it could be enough for us to use only the first saddle point equation in order to get  the relation between $M$ and $T$. Then we are left with some integral which depends on the external times and the value of $M=M(T)$ (but otherwise the remaining integral does not depend explicitly on the temperature). There are then no more $O(\epsilon _1)$ corrections, as we saw in the other two saddle point equations. As a result, the 4-point function at this order takes the same form as in the Schwarzian case, with a modified relation $M(T)$. In particular, the Lyapunov exponent is given by \eqref{eq:Lyapunov_exponent_through_M}, so that we get
\begin{equation} \label{eq:Lyapunov_exponent}
\lambda _L = 2\pi T - 4\pi \lambda ^{-1/2} T^2+\cdots .
\end{equation}

There are several consistency checks on this equation. First, the leading value for small $\lambda $ is precisely the maximal value $2\pi T$, consistent with the Schwarzian and the more standard SYK. Second, the correction to it is negative, which is necessary by the upper bound on chaos \cite{Maldacena:2015waa}. Also, note that since at leading order $T= \sqrt{M} \lambda ^{3/4} /\pi = \sqrt{\lambda } \epsilon _1/\pi \ll \sqrt{\lambda } $, the correction (second term) in \eqref{eq:Lyapunov_exponent} is indeed small compared to the leading behavior.

\section{Quantum deformation of SL$(2)$}
\label{sec:quantum-def}

In this section we would like to offer some representation-theoretic interpretation of our findings. In the usual SYK scaling, the low energy behaviour is encoded by the Schwarzian theory, which is determined by $SU(1,1)$. Here we would like to suggest a stronger statement that the double-scaled model {\it in its entirety} is given by the Hamiltonian reduction of a quantum particle moving on a non-compact quantum group SU$_{q^{1/2}}^{ext}(1,1)$\footnote{Notice our normalization of $q$. We use $q^{1/2}$ and not $q$ for all the representation-theoretic definitions throughout this section in order to make notations of appendix \ref{sec:quantum_groups}, standard in literature of quantum groups, consistent with the notations in the rest of the paper.}, see \autoref{sec:quantum_groups} for definitions. The first of two pieces of partial evidence that we will provide is that the spectrum of functions on this Hamiltonian reduction has a finite range of energies, similar to the entire double-scaled model. This is to be contrasted with the case of the Schwarzian which encodes the low energy physics and does not have a bound on the energy within it. The second piece of evidence is that the R-matrix, defined in \autoref{sec:diag_rules}, is related to the $6j$-symbol of the dual quantum group $\mathcal{U}_{q^{1/2}}(su(1,1))$, see \eqref{R-6j}. This leads the way to conjecture that the full double-scaled model is dual to some gravitational theory on a non-commutative $AdS_2$. 

To specify the setup, we first need to settle several issues inherent to dealing with non-compact quantum groups, see \autoref{sec:quantum_groups} for more details. On an algebraic level, the quantum group SU$_{q^{1/2}}(1,1)$ can be defined as a Hopf algebra with an involution\footnote{The involution defining the real form SU$_{q^{1/2}}(1,1)$ exists only for $0<q<1$, see \autoref{sec:quantum_groups}. Notice that, being completely equivalent for the undeformed case, the $q$-deformation of SU$(1,1)$ admits a $*$-structure only for $0<q<1$, whereas the deformation of SL$(2, \mathbb{R})$ -- only for $|q|=1$.}, whose relations can be guessed by $q$-deforming those of the algebra of polynomials on SU$(1,1)$. However, it turns out that, as soon as one goes from this formal description to a more detailed one, e.g. tries to (co-)represent the algebra on some Hilbert space, this fails: the algebra is ``too small'' to allow for enough well-defined self-adjoint operators in its tensor products \cite{woronowicz1991}. The solution is to slightly extend the algebra of functions one starts with \cite{Korogodsky1994}: instead of the group SU$(1,1)$ we take a normalizer SU$^{ext}(1,1)$ of SU$(1,1)$ inside SL$(2,\mathbb{C})$ which contains SU$(1,1)$ as an index $2$ subgroup, see the definitions in \autoref{sec:quantum_groups}.

The appropriate notion of a quantum group, taking into account issues of this type, goes via operator algebras and is due to Kustermans-Vaes \cite{Kustermans1999, Kustermans2000}. Starting with a $q$-analogue of the algebra of essentially bounded functions $L^{\infty}(G, d\mu)$ (the von Neumann algebraic approach to quantum groups) or the algebra of continuous functions $\mathcal{C}(G)$, perhaps with some restrictions (the $C^*$-algebraic approach to quantum groups), one can carefully reconstruct all the pieces of needed structure and, in particular, an analogue of the Hilbert space of square-integrable functions $L^2(G, d\mu)$ on the group. Geometrically, the former approach represents the $q$-analogue of the group manifold as a measure space, and the latter -- as a topological space.
Since introducing and carefully treating von Neumann algebraic quantum groups would require a disproportional amount of technical details for our present discussion, we will keep it on a purely Hopf-algebraic level and moreover we will not always distinguish between SU$_{q^{1/2}}(1,1)$ and SU$_{q^{1/2}}^{ext}(1,1)$.

To do a quantum Hamiltonian reduction on SU$_{q^{1/2}}^{ext}(1,1)$, one can proceed in two closely related manners. First, one can write an analogue of a Plancherel decomposition of a biregular corepresentation for SU$^{ext}_{q^{1/2}}(1,1)$ and choose a basis in each irreducible unitary corepresentation of this decomposition, which is compatible with the Hamiltonian constraints we want to impose. The constraints then just become constraints on labels of the appropriate matrix elements of corepresentations of SU$^{ext}_{q^{1/2}}(1,1)$, so picking them out
we would like to identify them as wavefunctions of the reduced model. This approach is more computational, for example one would need to show the completeness of the basis in the reduced Hilbert space by hand.
A slightly more elegant way to look at the reduction is to $q$-deform harmonic analysis on a double coset space, where the subspace of functions $L^2(G, d\mu_G)$ on the group, equivariant with respect to the left- and right action of chosen subgroups, is analyzed, and to look at the reduced space of functions constant (up to characters) on two-sided cosets $H \backslash G \slash K$. The analogues of groups $H$ and $K$ for our case are one-parametric groups generated by so called twisted primitive elements, we will specify them shortly.
In undeformed case, using Gauss decomposition, for almost all elements one can choose a representative in the space of equivariant functions on the group to be a function on a one-dimensional coordinate space. A $q$-deformed version of the reduced space of all such functions would then describe our model and, in particular, a $q$-deformed version of the restriction of a Laplace-Beltrami operator on the group manifold will be its Hamiltonian. Although finding the reduced measure and showing completeness would be much easier in this second picture, its implementation in the deformed case would require some control over the relevant quantum Gelfand pairs.\footnote{The analysis of such problem for any double coset with subgroups $H, K$ generated by $Y_s, Y_t$ (see further) is done in \cite{Koelink:2001}, but only on the level of Hopf $*$-algebras.  The operator-algebraic reduction for the simplest case of $s=t=\infty$, i.e. spherical Fourier transform and the corresponding Plancherel-Godement formula, is described in \cite{Caspers2011}.}
In the following we outline the necessary steps within the first approach.

The $q$-analogues of the $J^+$ and $J^-$ generators of the undeformed case are given (up to overall normalization) by elements $Y_{s=+1}$ and $Y_{s=-1}$ of the dual quantum group $\mathcal{U}_{q^{1/2}}(su(1,1))$ (see \autoref{sec:quantum_groups} for definitions):
\begin{align*}
Y_s \colonequals q^{1/4} E -q^{-1/4} F+\frac{s+s^{-1}}{q^{-1/2}-q^{1/2}}(K-K^{-1}).
\end{align*}
$Y_s$ are called twisted primitive elements
\cite{Koornwinder-zonal, Koelink:2001} and are peculiar, in particular since they behave similarly to Lie algebra elements: the coproduct is
\begin{align}
\Delta(Y_s)=K \otimes Y_s + Y_s \otimes K^{-1},
\end{align}
see also appendix \ref{app:u_qsu(1,1)}.
The elements $Y_sK$ are self-adjoint elements of $\mathcal{U}_{q^{1/2}}su(1,1)$, so there exist corresponding one-parametric subgroups of SU$_{q^{1/2}}(1,1)$.
One can indeed check that in the limit $q \to 1^-$ (and after making a similarity transform to the equivalent real form SL$(2,\mathbb{R})$), the two subgroups reduce to the subgroups of upper- and lower-triangular matrices inside SL$(2,\mathbb{R})$.
The spectrum of $Y_s K$ (in the principal unitary series of $\mathcal{U}_{q^{1/2}}su(1,1)$) is continuous
and, for $s=\pm 1$, they are deformations of the parabolic elements of the undeformed Lie algebra (i.e. the ones corresponding to a conjugacy class of upper/lower triangular matrices in the SL$(2,\mathbb{R})$ group).
In particular, the Al Salam-Chihara polynomials used in our calculations and their non-terminating counterparts  have an interpretation as generalized eigenvectors of 
$Y_s K$ \cite{GroeneveltBilinearSF}.

As we review in \autoref{sec:quantum_groups}, a quantum group can be made into its own corepresentation using the coproduct. These corepresentations are called left and right regular corepresentations
and play a role analogous to that of left and right regular representations of $G$ on the space of functions on the group $L^2(G)$ (up to operator-algebraic issues ignored here).
The $q$-analogue of the decomposition of left regular representation of the group looks as \cite{Groenevelt2010}:
\begin{align}\label{left-reg-corep-main}
W \cong \bigoplus_{p\in q^{\mathbb{Z}/2}} \left(\int_{-1}^{1\, \oplus} W_{p,x} dx \oplus \bigoplus_{x\in\sigma_{discr}}W_{p,x}\right),
\end{align}
where $x\in[-1;1]$, $p \in q^{\mathbb{Z}/2}$.
Speaking precisely, it states the decomposition of a left regular corepresentation of the von Neumann algebraic quantum group (generated by multiplicative unitary of this quantum group) into its irreducible corepresentations. We will treat this decomposition a bit schematically: the left-hand side is Hopf-algebraically just a left regular corepresentation of the Hopf algebra SU$^{ext}_{q^{1/2}}(1,1)$ (see \autoref{sec:quantum_groups}), whereas every corepresentation $W_{p,x}$ on the right-hand side decomposes into a direct sum of representations of the dual group $\mathcal{U}_{q^{1/2}}(su(1,1))$. For the continuous part of the decomposition (corepresentations under the direct integral symbol), which is our interest, this takes two steps. First, 
each of the spaces $W_{p,x}$ above decomposes into a direct sum of carrier spaces of the irreducible corepresentations as
\begin{align}
&W_{p,x}= W_{p,x}^1 \oplus W_{p,x}^2\quad \text{when } x\neq 0 \text{ or } p \in q^{\mathbb{Z}+1/2}\\
&W_{p,0}=W_{p,0}^{1,1} \oplus W_{p,0}^{1,2} \oplus W_{p,0}^{2,1} \oplus W_{p,0}^{2,2}\quad \text{when }  p \in q^{\mathbb{Z}}\nonumber.
\end{align}
These splittings have to do with the fact that one extends SU$(1,1)$ to SU$^{ext}(1,1)$ in order to deform it and thus are not relevant for our present discussion.
Second, the irreducible corepresentations from the continuous part of the regular decomposition further decompose as
\begin{align}\label{W-to-pi}
&W^j_{p,x} \cong \pi_{b(-x), \epsilon(p)} \oplus \pi_{b(x), \epsilon(p)}, \quad j=1,2,\,\,\, x \neq 0\\
&W^{j,k}_{p,0} \cong \pi_{b(0), 0}, \quad j,k=1,2\nonumber
\end{align}
where
\begin{align*}
& b(x)=-\frac{\arccos x}{ \ln q}, \,\,\, x \in [-1;0) \cup (0;1], \quad \quad b(0)=-\frac{\pi}{2\ln q}\\
& \epsilon(p)=\log_q|p| \text{ mod }1.
\end{align*}
Here $x\in[-1;1]$ is the eigenvalue of the Casimir element in the principal unitary representations of $\mathcal{U}_{q^{1/2}}(su(1,1))$, see \autoref{sec:quantum_groups}. In the classical limit, each irreducible corepresentation of the group von Neumann algebra surviving in the right-hand side of \eqref{left-reg-corep-main} bijectively corresponds to a unitary irrep of the group. There is a similar decomposition of the right regular corepresentation.

The decomposition \eqref{left-reg-corep-main} still contains many unitarily equivalent corepresentations.\footnote{We remind that the two corepresentations $U_i \in A \otimes B(H_i)$, $i=1,2$ are called unitarily equivalent if there is a unitary operator $T: H_1 \to H_2$ intertwining them, i.e. such that $(1\otimes T)U_1=U_2 (1\otimes T)$.  }
To resolve the degeneracy, one can combine the left and right regular corepresentations of SU$_{q^{1/2}}^{ext}(1,1)$, i.e. look at a $q$-analogue of decomposition for so called biregular representation of $G\times G$ acting (from the left and from the right) on the space of functions $L^2(G)$ into a direct integral of matrix coefficients $\pi \otimes \bar{\pi}$ of irreducible representations. For a quantum group the biregular corepresentation is given via left- and right- regular corepresentations as $(\chi(V))_{13} (W)_{24}$, where $\chi(V)=\Sigma V \Sigma$ and $\Sigma$ is a flip of factors in tensor product, see the definitions in \autoref{sec:quantum_groups}.\footnote{See also \cite{Caspers2010QuantumGA} for precise definitions of the conjugate (contragradient) and irreducible corepresentations.}

The $q$-analogue of double-sided Plancherel decomposition of $L^2(G)$ is due to Desmedt \cite{Desmedt}. It decomposes the biregular corepresentation $(\chi(V))_{13} (W)_{24}$ (acting on the space of Hilbert-Schmidt operators acting on the GNS space of the Haar weight) of the (von Neumann algebraic) quantum group SU$^{ext}_{q^{1/2}}(1,1) \otimes \text{SU}^{ext}_{q^{1/2}}(1,1)$ into direct sum/integral over irreducible unitary corepresentations of SU$^{ext}_{q^{1/2}}(1,1)$.
A precise formulation \cite{Desmedt, Caspers-thesis} again requires a language of operator-algebraic quantum groups and is a bit too technical for our reviewing purposes. The essential part is the existence (under certain conditions, satisfied by SU$^{ext}_{q^{1/2}}(1,1)$) of an isomorphism which
intertwines between 'the space of functions on the quantum group' and 'span of their matrix elements', i.e. an analogue of Fourier transform. The so called Plancherel measure $d\mu(U)$ (for the continuous part of the spectrum) is given via specialization of Askey-Wilson measure \cite{Caspers-thesis}.

The Plancherel formula resolves the degeneracy in the irreducible corepresentations encountered in the left regular corepresentation \eqref{left-reg-corep-main},
giving a decomposition of the Hilbert space of states of a quantum particle on a quantum group SU$^{ext}_{q^{1/2}}(1,1)$. 
A detailed description of non-equivalent corepresentations is again not important, as soon as we know that the Hopf algebra $\mathcal{U}_{q^{1/2}}(su(1,1))$ naturally acts in the corepresentations of SU$^{ext}_{q^{1/2}}(1,1)$. Its action in the regular corepresentation
is a $q$-deformed equivalent of the action of Laplace-Beltrami operator on the space $L^2(G)$. Under this action, each irreducible corepresentation splits into a sum of $\mathcal{U}_{q^{1/2}}(su(1,1))$ irreducibles and, clearly, the representations with different Casimir eigenvalues cannot be equivalent.
Upon the reduction, one expects that, similarly to the undeformed case, only the principal unitary part of the regular decomposition survives. As the $\mathcal{U}_{q^{1/2}}(su(1,1))$ Casimir (acting in the regular corepresentation) is an appropriate analogue of Laplace-Beltrami operator, upon the reduction it is expected to become the Hamiltonian of our model, so that its eigenvalues (for those representations surviving the reduction) will give the energies of the reduced model. 
Thus looking back at \eqref{W-to-pi}, we see that the spectrum of a $\mathcal{U}_{q^{1/2}}(su(1,1))$ Casimir on the (continuous part of the) biregular corepresentation is bounded to the interval $[-1;1]$ and simple.

Finally, to apply the formulas from above and do the reduction, one needs to specify (generalized) bases in the corepresentations of the quantum group. Like in the undeformed case, one can make various choices of the basis states ('Whittaker vectors') here, depending on which matrix elements one wants to single out.
E.g. the analogue of mixed parabolic basis in the undeformed SL$(2,\mathbb{R})$ case, used in \cite{Blommaert:2018oro}, is the basis which diagonalizes operators corresponding to self-adjoint $\mathcal{U}_{q^{1/2}}(su(1,1))$ elements $Y_{s=+1}K$ and $Y_{t=-1}K$. After fixing the representation labels to the same value, one performs the reduction by computing the corresponding matrix elements in terms of $q$-Whittaker functions. The reduced model is expected to be a variant of 'minisuperspace' version of a $q$-Liouville model, see in particular \cite{Etingof99} for the algebraic aspects of such Hamiltonian reduction in the complexified case.  
We will not go into further details here. The practically essential thing to remark is that the matrix elements and $3j$-symbols relevant for the computation look as direct $q$-deformations of the undeformed expressions, see \cite{Blommaert:2018oro}.

The second piece of evidence to a relation of our model with the discussed quantum group comes via examining the R-matrix from \autoref{sec:diag_rules}. When calculating correlation functions in a particle-on-a-group model, $6j$-symbols of the group play an important role.
Our R-matrix \eqref{eq:R_matrix_full_expr}, similarly to undeformed case,
apparently
has an interpretation as a $6j$-symbol of the dual quantum group $\mathcal{U}_{q^{1/2}}(su(1,1))$. To support this claim, let us specialize $l_1, l_2$ to positive integer values and use a standard convention for the $6j$-symbol
$$\begin{Bmatrix}j_1 & j_2 & j \\ j_4 & j_3 & j' \end{Bmatrix},$$
where the two coupling schemes under consideration are: first, $j_1$ and $j_2$ are coupled into $j$ and then the latter with $j_3$ into $j_4$, and, second, $j_2$ and $j_3$ are coupled into $j'$ and then the latter with $j_1$ (into $j_4$ as well). We then notice that, according to \cite{Groenevelt2006}, a $6j$-symbol of $\mathcal{U}_{q^{1/2}}(su(1,1))$ for the tensor product $D^+_{l_2}\otimes \pi_{\theta_2/\ln q, 1/2} \otimes D_{l_1}^{-}$ reads:\footnote{To arrive at the expression for $6j$-symbol in the first line, we multiply the Racah coefficient in \cite{Groenevelt2006}, Thm 8.5 by the square roots of corresponding measure factors, substitute the listed arguments and moreover send $\theta_k \mapsto -\theta_k$, $k=1, \dots ,4$. Explicitly, our $6j$-symbols are then normalized as:
\begin{align*}
\int_{-\pi}^{\pi} \frac{d\theta_1}{2}\frac{\left(q, e^{\pm 2 i \theta_1};q\right)_{\infty}}{2\pi}&\begin{Bmatrix}l_2^+ & \frac{\theta_2}{\ln q}, \frac{1}{2} & \frac{\theta_1}{\ln q}, \frac{1}{2} \\ \frac{\theta_4}{\ln q}, \frac{1}{2} & l_1^- & \frac{\theta_3'}{\ln q}, \frac{1}{2} \end{Bmatrix}^+ \begin{Bmatrix}l_2^+ & \frac{\theta_2}{\ln q}, \frac{1}{2} & \frac{\theta_1}{\ln q}, \frac{1}{2} \\ \frac{\theta_4}{\ln q}, \frac{1}{2} & l_1^- & \frac{\theta_3}{\ln q}, \frac{1}{2} \end{Bmatrix}\\
&=\frac{\left(q;q\right)_{\infty}}{\left(e^{\pm 2 i \theta_3};q\right)_{\infty}} \times 2\pi \left[\delta(\theta_3-\theta_3')+\delta(\theta_3+\theta_3')\right].
\end{align*}
}
\begin{align}\label{R-6j}
 \begin{Bmatrix}l_2^+ & \frac{\theta_2}{\ln q}, \frac{1}{2} & \frac{\theta_1}{\ln q}, \frac{1}{2} \\ \frac{\theta_4}{\ln q}, \frac{1}{2} & l_1^- & \frac{\theta_3}{\ln q}, \frac{1}{2} \end{Bmatrix} &= \frac{\varphi_{e^{i\theta_3}}\left(e^{i\theta_1}; \tilde q_2 e^{i\theta_2}, \tilde q_1 e^{\pm i \theta_4}, q \tilde q_2^{-1} e^{i\theta_2}; q \right)}{\left[ \left( \tilde q_1 e^{i(\pm \theta _2 \pm \theta _3)} ,\tilde q_1 e^{i(\pm \theta _1 \pm \theta _4)} ,\tilde q_2 e^{i(\pm \theta _1 \pm \theta _2)} ,\tilde q_2 e^{i(\pm \theta _3 \pm \theta _4)} ;q\right)_{\infty } \right]^{1/2}}\nonumber\\
& \equiv  R^{(q)} _{\theta _4\theta _2} \begin{bmatrix}\theta _3 & l_2 \\ \theta _1 & l_1 \end{bmatrix},
\end{align}
where $\varphi_{\gamma}(x; a,b,c,d ; q)$ is the Askey-Wilson function, defined in \autoref{app:special_functions}.
See \cite{Groenevelt2006} for the computation of various other $6j$-symbols, all of which are given either by an Askey-Wilson function or by its polynomial analogues.

\section{Regimes of large $N$ and large $p$ in SYK-like models\label{sec:Largeq}}

Finally we would like to comment on the two distinct large $N$ -- large $p$ settings for which the SYK model has been discussed:
\begin{itemize}
\item A two-stage limit in which first $N\rightarrow\infty,\ p$ fixed, and then $p\rightarrow\infty$. This is the standard discussion in SYK \cite{Jia:2018ccl,Maldacena:2016hyu,Tarnopolsky:2018env}.
\item The double-scaled limit in which $N\rightarrow\infty,\ p\rightarrow\infty$,  $p^2/N$ fixed, used in \citep{Berkooz:2018qkz, Cotler:2016fpe} and the present paper.
\end{itemize}
In this section we will explain why these limits are not the same, and we will show which terms are picked out, or missed, by each of the limits.
For concreteness, we will center our discussion on two quantities:
the partition function (in \autoref{sec:Largeq_Z}) and the full
propagator for a single Majorana fermion (in \autoref{sec:Largeq_G}).

We will discuss these quantities in the high temperature/short times regimes $\sqrt{\lambda}\,\beta<1$ and $\sqrt{\lambda} t<1$. In practice, we will do it by expanding in powers of the Hamiltonian. 
We will  in addition write the expressions as a perturbative expansion in $\lambda$ and $p^{-1}$. This is motivated by the fact that both the fixed $p,\ N\rightarrow\infty$ (fixed $p,\ \lambda\rightarrow 0$) limit, and the fixed $\lambda,\ N\rightarrow\infty$ (fixed $\lambda,\ p\rightarrow\infty$) limit exist.  There is no first principle proof that this is a good expansion, but it was verified to some low orders. 
In this case, we can write the partition function and correlation function as
\begin{equation}
\begin{split}
& \mathcal{Z}=\frac{\left\langle \mbox{tr}e^{-\beta H}\right\rangle _{J}}{\mbox{tr}\mathbb{I}}=1+\Sigma_{m,n=0}^{\infty}\Sigma_{k=1}^{\infty}\frac{\lambda^{m}\beta^{k}a_{k}^{(m,n)}}{p^{n}}\\
& G\left(\tau\right)=\frac{\left\langle \mbox{tr}e^{-\left(\beta-\tau\right) H}\psi_{i}e^{\tau H}\psi_{i}\right\rangle _{J}}{\left\langle \mbox{tr}e^{-\beta H}\right\rangle _{J}}=\mbox{sign}\tau\left(1+\frac{1}{p}\Sigma_{m,n=0}^{\infty}\Sigma_{k=1}^{\infty}\frac{\lambda^{m}P_{k}^{(m,n)}[\beta-\tau,\tau]}{p^{n}}\right),
\end{split}\label{eq:ExpansionShape}
\end{equation}
where $P_{k}^{(m,n)}\left[x,y\right]$ are homogeneous polynomials
of degree $k$, which are symmetric in their two arguments. 

In these
expressions it becomes clear that the two limits capture complementary information: the two-stage limit accesses $p^{-1}$
corrections for the leading order in $\lambda$, while in the double-scaling limit we obtain all-order corrections in $\lambda$, but miss
all the sub-dominant terms in $p^{-1}$.

\subsection{Partition function at high temperature \label{sec:Largeq_Z}}

One of the observations in \cite{Jia:2018ccl} is that, in the normalization
that sets $\mbox{tr} H^{2}=1$, the $2k$-th cumulant $\kappa_{2k}$
of the SYK partition function is suppressed by $N^{1-k}$ at finite
$p$, which become higher powers of $\lambda$ in our conventions.
This implies that at finite $p$, infinitesimal $\lambda$, the dominant
contributions to high moments $M_{2k}=\left\langle \text{tr} H^{2k}\right\rangle _{J}$
can be derived from low order perturbative computations of the very
few first. This is, of course, in agreement with the statement in \cite{erdHos2014phase} that the distribution converges to a Gaussian, in the limit $\lambda\rightarrow 0$. This convergence, however, is only pointwise (in energy), which means that we can apply this in the perturbative-in-$H$ regime only, but this is sufficient to see the differences between the fixed $p$ and double-scaled limits. 

From the explicit computation in equations (3.12) and (3.14) in \cite{Jia:2018ccl}, which are a $1/N$ expansion at finite $p$,
we extract, in our notation,
\begin{equation}
\begin{split}
& \mathcal{Z}=1+\sum_{k=1}^{\infty}\beta^{2k}\frac{M_{2k}}{(2k)!}\\
& \frac{M_{2k}}{(2k)!}=\frac{1}{2^{k}k!}\left(1+\frac{1}{3}\binom{k}{2}\kappa_{4}+\frac{1}{15}\binom{k}{3}\kappa_{6}+\mathcal{O}\left(\lambda^{3}\right)\right)\\
& \kappa_{4}=-\lambda+\frac{\lambda^{2}}{2}-\frac{\lambda^{2}}{p}+\frac{\lambda^{2}}{2p^{2}}+\mathcal{O}\left(\lambda^{3}\right);\;\kappa_{6}=6\lambda^{2}-\frac{2\lambda^{2}}{p}+\mathcal{O}\left(\lambda^{3}\right).
\end{split}\label{eq:Largeq_Gia}
\end{equation}

Table \ref{tab:Z_Compare} compares these results with the chord diagrams. The contribution of a chord diagram with $n$ intersections is $e^{-n\lambda}$, up to $\mathcal{O}\left(p^{-1}\right)$ corrections. All that's left is a counting exercise: the diagrams relevant for $M_0$ and $M_2$, with less than two chords, cannot have intersections; with two chords, relevant for $M_4$, the number of diagrams containing $\left\{ 0,1\right\} $
intersections is, respectively, $\left\{ 2,1\right\} $; and with
three chords, relevant for $M_6$, the number of diagrams containing $\left\{ 0,1,2,3\right\} $
intersections is $\left\{ 5,6,3,1\right\} $. 
\begin{table}
\begin{centering}
\begin{tabular}{|c|c|c|}
\hline 
Momenta & Two-stage limit & Double-scaling limit \tabularnewline
\hline 
\hline 
$M_{0}$ & 1 & 1\tabularnewline
\hline 
$M_{2}$ & 1 & 1\tabularnewline
\hline 
$M_{4}$ & $3-\lambda+\frac{\lambda^{2}}{2}-\frac{\lambda^{2}}{p}+\frac{\lambda^{2}}{2p^{2}}+\mathcal{O}\left(\lambda^{3}\right)$ & $2+e^{-\lambda}+\mathcal{O}\left(\frac{1}{p}\right)$\tabularnewline
\hline 
$M_{6}$ & $15-15\lambda+\frac{27\lambda^{2}}{2}-\frac{17\lambda^{2}}{p}+\frac{15\lambda^{2}}{2p^{2}}+\mathcal{O}\left(\lambda^{3}\right)$ & $5+6e^{-\lambda}+3e^{-2\lambda}+e^{-3\lambda}+\mathcal{O}\left(\frac{1}{p}\right)$\tabularnewline
\hline 
$\ldots$ & $\ldots$ & $\ldots$\tabularnewline
\hline 
\end{tabular}
\par\end{centering}
\caption{\label{tab:Z_Compare}Low order comparison between the iterated limit
results of \cite{Jia:2018ccl} and the predictions of chord expansions
in the double-scaling limit. All expressions are compatible with the
existence of a unified double expansion in $p^{-1}$ and $\lambda$.}
\end{table}

Let us address a more interesting aspect in what follows: the origin of the $1/p$ corrections within the chord diagram picture, as there should be a systematic way of carrying out this expansion. This is an important question because we can think of the exact results in $\lambda$ and an expansion in  $1/p$ as loose analogue of exact results in $\alpha'$ and an expansion in $1/M_p$, respectively, in a quantum gravity background.
The reason for this analogy is that at finite $\alpha'$ the number of states in quantum gravity, say in a BH, is governed by $M_p$, and here it is governed by $N \sim p^2$ at finite $\lambda$. However, already at $1/M_p \rightarrow 0$ we can have a nontrivial background, which is the role played by the fixed $\lambda$ model.

In the exact computation, at finite $N$ and $p$, when the random coupling are Gaussian, it remains true that after
averaging over disorder we are left with paired instances of terms
in the Hamiltonian that can be represented by chords, and that each
chord $i$ picks up a set of $p$ indices $I_{i}$ from $N$. Several new effects kick in, however, at finite $p$ and fixed $\lambda$: a)  The distribution of the size of pairwise intersection is no longer Poisson,  b) it is no longer true in general that
intersections of several such choices $I_{i}\cap I_{j}\cap I_{k}\cap\ldots$
can be neglected, and possibly others.

To see where $1/p$ corrections come from, we can proceed as follows \cite{a:2018kvh}. Suppose that our diagram $\mathcal{D}$ contains $k$
chords. After assigning a set of $p$ indices out of $N$ to each
one, the total $N$ indices have been divided into the following sets:
a set of $s_{\emptyset}$ indices that have not been picked up by any
chord, sets of $s_{i}$ indices, $i\in\left\{ 1,\ldots,k\right\} $
that are present only in chord $i$, sets of $s_{ij}$ indices present
both in chord $i$ and chord $j$ but in no other chord, and so on.
Let us denote generic subindices $ijk\ldots$ by the label $n$. Any choice of positive integers for $\{s_{n}\}$ satisfying the constraints
$\sum_{n}s_{n}=N$ and $\sum_{n\mid i\in n}s_{n}=p$ $\forall i$ corresponds to a valid configuration.
In practice, we rewrite $s_{\emptyset}$ and each $s_{i}$ in
terms of other $s_{n}$ using the constraints, 
\begin{equation}
\begin{split}
& s_\emptyset(s)=N-k\,p+\sum_{i<j} s_{ij}+2\sum_{i<j<k} s_{ijk}+3\sum_{i<j<k<l} s_{ijkl}+\ldots\\
& s_i(s)=p-\sum_{n\ni i,n\neq i} s_n,
\end{split}
\end{equation}
which leaves us with
the free variables $s=\left\{ s_{12},s_{13},\ldots,s_{123},s_{124},\ldots,s_{123\ldots k}\right\} $.
The probability distribution over $s$ is given by
\begin{equation}
\mathcal{P}\left(s\right)=\frac{\left(\begin{array}{c}
N\\
s_{\emptyset}\left(s\right),\left\{ s_{i}\left(s\right)\right\} ,s
\end{array}\right)}{\left(\begin{array}{c}
N\\
p
\end{array}\right)^{k}}=
 \frac{N!}{s_\emptyset(s)!}\prod_{i=1}^{k}\left(\frac{\left(N-p\right)!}{N!}\frac{p!}{s_i(s)!}\right)\prod_{n\in s}\frac{1}{s_{n}!}.\label{eq:Exact_Combinatorial}
\end{equation}
This exact expression includes the two types of corrections mentioned before:
\begin{itemize}
\item In order to see more transparently the corrections to the Poisson distribution, let us assume for the time being that $s_n=0$ if $n$ contains three or more indices. Then, using in each of the factorial quotients in (\ref{eq:Exact_Combinatorial}) the expansion 
\begin{equation}
\frac{N!}{\left(N-a\right)!}=N^{a}\exp {\left(-\sum_{n=1}^{\infty}\sum_{i=0}^{n}\frac{a^{n+1-i}B_{i}}{n\left(n+1\right)N^{n}}{\left(\begin{array}{c}n+1\\i\end{array}\right)}\right)}=N^{a}e^{\left(-\frac{a^{2}}{2N}+\frac{a}{2N}-\frac{a^3}{6N^2}+\ldots\right)},\label{eq:FactorialAsymptote}
\end{equation}
where $B_i$ denotes the $i$-th Bernoulli number, we obtain
\begin{equation}
\mathcal{P}\left(s\right)=\left(\prod_{i<j} \frac{1}{s_{ij}!} \left(\frac{p^2}{N}\right)^{s_{ij}}e^{-\frac{p^2}{N}}\right)\exp \left(-\frac{(k+1)_3}{6}\frac{p^3}{N^2}+\sum_{i<j}\frac{s_{ij}}{p}-\sum_{i\neq j,k} \frac{s_{ij}s_{ik}}{p}+\ldots\right).
\label{eq:P_2Intersections}
\end{equation}
\item Let us now generalize the previous result allowing for non-vanishing three-index entries $s_{ijk}$:
\begin{equation}
\mathcal{P}\left(s\right)=
\left(\prod_{i<j} \frac{1}{s_{ij}!} \left(\frac{p^2}{N}\right)^{s_{ij}}e^{-\frac{p^2}{N}}\right) 
\left(\prod_{i<j<k} \frac{1}{s_{ijk}!}\left(\frac{p^3}{N^2}\right)^{s_{ijk}} e^{-\frac{p^3}{N^2}}\right)
\exp \left(\Phi(s)\right).
\label{eq:P_3Intersections}
\end{equation}
The exponent $\Phi(s)$ summarizes corrections to the Poisson distribution of the form observed in (\ref{eq:P_2Intersections}) \footnote{Some terms of order $\frac{p^3}{N}$ remain in $\Phi(s)$ and correct the Poisson distribution of the $s_{ij}$ elements.
Higher orders of $s_n$ will behave in an analogous manner. }
\end{itemize}
At the end of the day, after recasting $N^{-1}$ as $\lambda p^{2}$,
we are left with a series in $\lambda$ and $p^{-1}$ for each diagram.

The generalization of the expression (\ref{eq:Exact_Combinatorial})
to the case where different chords
can pick different number of indices, of considerable relevance for
discussing amplitudes for observables, is straightforward.

\subsection{Fermion propagator at high temperature \label{sec:Largeq_G}}

In the fixed $p$, large $N$ setting, SYK becomes dominated by the set of
diagrams that were dubbed watermelon diagrams. In leading order in $N$,
one can establish the following Schwinger-Dyson equation between the
self-energy and the full fermion propagator: 
\begin{equation}
\Sigma(\tau)=\frac{\lambda}{2p}G^{p-1}(\tau).\label{eq:Dyson}
\end{equation}
By definition, $\Sigma$ is related to $G$ also by 
\begin{equation}
\frac{1}{G\left(\omega\right)}=-\frac{i\omega}{2}-\Sigma\left(\omega\right).\label{eq:SelfEnergyDef}
\end{equation}
Together, (\ref{eq:Dyson}) and (\ref{eq:SelfEnergyDef}) provide a closed
equation for $G(\tau)$ that can be solved order by order in $p^{-1}$ under the Ansatz
\begin{equation}
G\left(\tau\right)=\text{sign}\tau\left(1+\frac{g_{1}(\tau)}{p}+\frac{g_{2}(\tau)}{p^{2}}+\ldots\right).
\end{equation}
The first order of this expansion was obtained in \cite{Maldacena:2016hyu} (we refer the interested reader to \cite{Tarnopolsky:2018env} for higher order corrections). It is better expressed in terms of the variable $x\equiv \frac{\pi v}{2} \frac{2\tau-\beta}{\beta}$ in the interval $x\in\left[-\frac{\pi v}{2},\frac{\pi v}{2}\right]$:
\begin{equation}
e^{g_{1}(x)}=\frac{\cos^{2}\left(\frac{\pi v}{2}\right)}{\cos^{2}x};\;\mbox{\ensuremath{v} given by }\frac{\pi^{2}v^{2}}{\lambda^{2}\beta^{2}}=\cos^{2}\frac{\pi v}{2.}\label{eq:Largeq_Maldacena}
\end{equation}
Beyond $x\in\left[-\frac{\pi v}{2},\frac{\pi v}{2}\right]$ the {function $g_1$ should be continued periodically.

Let us now turn towards the chord diagram expansion. In this language, the two insertions
of the fermion $\psi_{i}$ can be represented by a chord that carries
a single index $i$. Intersections between regular chords and the
fermionic chord produce the exact factor
\begin{equation}
\tilde{q}=1-\frac{\lambda}{p}.
\end{equation}
Thus, the commutator between the fermion chord's end and a regular
chord's end is of order $p^{-1}$, and therefore that is the order at which corrections
to the free theory will appear. One might worry that our results are not reliable when we compute at this order: we have explicitly stated in previous discussions that the evaluation of chord diagrams as a product of a $q$ factor for each chord intersection neglects terms that are precisely of this same order. However, these type of $p^{-1}$ corrections  appear in the same form in the correlator and the
partition function. Schematically, if we denote by $A$ the terms we have access to in our approximation and by $B$ the terms we deprecate, we can write
\begin{equation}
G=\frac{A_{0}+B_{0}p^{-1}+A_{1}p^{-1}+B_{1}p^{-2}+\ldots}{A_{0}+B_{0}p^{-1}+\ldots}=1+\frac{A_{1}}{A_{0}}p^{-1}+\mathcal{O}\left(p^{-2}\right).
\end{equation}
For this particular observable, therefore, the $\mathcal{O}(p^{-1})$ correction is not sensitive to the sub-dominant terms that we neglect in the chord diagram setup.

The comparison between chord diagram predictions and the predictions of the Schwinger-Dyson method that (\ref{eq:Largeq_Maldacena}) summarizes becomes more transparent when we series expand the two-point function result in the high temperature/small times limit. 
The result of this operation is contrasted order by order in table \ref{tab:G_compare} to the results of the Schwinger-Dyson approach. The leading order in $\lambda$ and $1/p$ agrees. As expected, the Schwinger-Dyson results capture only the leading $\lambda$ factor of the $1-q$ factors sourced by the commutators between chord endpoints, but in exchange they would provide access to $p^{-2}$ corrections that we cannot reliably obtain from the chord diagram perspective, if we compute its higher order corrections \cite{Tarnopolsky:2018env}.

\begin{table}
\begin{centering}
\begin{tabular}{|c|c|c|}
\hline 
Order in $\hat{\beta},\tau$ & Chord diagrams & Schwinger-Dyson
\tabularnewline
\hline 
\hline 
0 & 1 & 1 
\tabularnewline
\hline 
2 & $-\frac{\lambda}{p}P_{2}\left(\tau\right)$ & $-\frac{\lambda}{p}P_{2}\left(\tau\right)$ 
\tabularnewline
\hline 
4 & $\frac{\lambda}{p}\left(1-e^{-\lambda}\right)P_{4}\left(\tau\right)$ & $\frac{\lambda^{2}}{p}P_{4}\left(\tau\right)$
\tabularnewline
\hline 
6 & $-\frac{\lambda}{p}\left(1-e^{-\lambda}\right)^{2}P_{6}\left(\tau\right)$ & $-\frac{\lambda^{3}}{p}P_{6}\left(\tau\right)$\tabularnewline
\hline 
$\ldots$ & $\ldots$ & $\ldots$ 
\tabularnewline
\hline 
\end{tabular}
\par\end{centering}
\caption{\label{tab:G_compare} Dominant contributions at small $\lambda$ and $p^{-1}$ for the high temperature, low time regime of the two-point function of a single fermion $\psi_i$ in SYK. The polynomials $P_i(\tau)$ have been introduced to keep the notation compact. {Their expressions are: $P_{2}\left(\tau\right)\equiv\hat{\beta}\tau$, 
$P_{4}\left(\tau\right)\equiv\hat{\beta}\tau(\hat{\beta}^{2}+3\hat{\beta}\tau+\tau^{2})/6$ , $P_{6}\left(\tau\right)\equiv\hat{\beta}\tau(9\hat{\beta}^{4}+45\hat{\beta}^{3}\tau+80\hat{\beta}^{2}\tau^{2}+45\hat{\beta}\tau^{3}+9\tau^{4})/180$.}}
\end{table}

\section{Summary and conclusion}\label{sec:LastSec}

In this paper we showed how to construct bi-local operators in the full double-scaled theory, and how to use them to construct 4-point functions. More generally, we showed how to write a set of Feynman rules for the entire model. Using this construction, we  computed corrections to the maximal chaos exponent. We also explained the indications that the model is related to the Hamiltonian reduction of a quantum deformation of SL(2). 

There are several future directions that one can take:

1. The main goal would be to solve the entire model, i.e., have a set of simple rules for the computation of all $n$-point functions, as well as more detailed statistics such as the energy level distribution $\rho(E,E')$. The key is likely to be the suggested relation of the model to ${\cal U}_q(su(1,1))$, if the latter controls the model in a rigid enough way. The starting point would be to carry out the Hamiltonian reduction discussed in \autoref{sec:quantum-def}, and to see if this can be extended to more microscopic data such as level statistics. 

2. The role of the quantum deformation of SU$(1,1)$ is likely to be critical also in understanding the bulk, and in particular whether there is an analogous quantum deformation of $AdS_2$,\footnote{For a work on $dS$ space, see \cite{Banks:2006rx}.} which will be dual to the entire model. Typically, quantum geometry has an inherent minimal ``smearing'' length and hence such a deformation can naturally help in understanding non-localities in gravity, which might be related to the black hole information puzzle. 

3. Finally, being able to solve the model at all energy scales, and under a large set of perturbations, might allow us to study both field theory and gravity aspects of non-thermal states and specific microstates.

\acknowledgments
We would like to thank O.~Aharony, D.~Bagrets, A.~Blommaert, E.~Martinec, T.~Mertens, P.~Narayan, M.~Rangamani, S.~Shenker, G.~J.~Turiaci and H.~Verlinde for useful discussions.
This work is supported by an ISF center of excellence grant (1989/14 and 2289/18).
The research of MI on this project has received funding from the European Research Council (ERC)
under the European Union’s Horizon 2020 research and innovation program (QUASIFT
grant agreement 677368).
MB and VN thank KITP (Santa Barbara) for hospitality during final stages of this work.
This research was supported in part by the National Science Foundation under Grant No. NSF PHY-1748958. 
MB holds the Charles and David Wolfson Professorial chair of Theoretical Physics.
\appendices

\appendix

\section{Poisson distribution of intersections and $q$-binomial coefficients}\label{sec:Lemmas}

\subsection{Poisson distribution of intersections of sets in the large $N$ limit}

Let $A \subset \{1,\cdots ,N\}$ be of size $p$. Then the proportion of sets $B \subset \{1,\cdots ,N\}$ of size $p'$ such that $| A \cap B|=k$ approaches a Poisson distribution for $k$ with parameter $p p ' / N$ under the assumptions $ k \ll p,p' \ll N$.

The precise number of such sets $B$ is given by first choosing which $k$ elements are in common to $A$ and $B$, and then choosing the rest of the $p'-k$ elements for $B$. Therefore the proportion of sets $B$ of size $p'$ with intersection of size $k$ with $A$ is given by (where $\sim $ below means that the approximations mentioned above are used)
\begin{equation}
\begin{split}
& \frac{\binom{p}{k} \binom{N-p}{p'-k} }{\binom{N }{p'} } = \frac{p(p-1)\cdots (p-k+1)}{k!} \frac{p' !}{(p'-k)!} \frac{(N-p-p')!}{(N-p-p'+k)!}  \frac{\binom{N-p}{p'} }{\binom{N}{p'} } \sim \\
& \sim  \frac{p^k}{k!} \frac{p'^k}{N^k} \prod_{j=0} ^{p'-1} \left(1-\frac{p}{N-j} \right) \sim \frac{1}{k!} \left( \frac{pp'}{N} \right)^k\left( 1- \frac{p}{N} \right)^{p'} \sim \frac{1}{k!} \left( \frac{pp'}{N} \right)^k e^{-pp'/N} .
\end{split}
\end{equation}
This is indeed a Poisson distribution with parameter $pp'/N$.

\subsection{Weighted separation of chords into two sets}

Let us show the claim depicted in \autoref{fig:q_binomial}, that summing over the possibilities to separate $l$ chords into $i$ and $l-i$ chords, each possibility assigned a weight of $q$ to the power of the number of intersections, gives $\binom{l}{i} _q$.

Denote the desired object by $V(i,l-i)$. There is a simple recursion relation that we can write for $V(i,j)$ in terms of the total number of chords $i+j$. Consider the last chord out of $i+j$ at the bottom (see e.g.\ \autoref{fig:q_binomial}). If it goes to the second set of $j$ chords, then the sum of those weighted possibilities is just the same as the sum of the weighted possibilities without this last chord. If, on the other hand, this last chord goes to the first set (of size $i$) then the sum over these weighted possibilities is $q^j$ (from crossing the $j$ chords) times the sum of weighted possibilities without this last chord. We get
\begin{equation}
V(i,j) = V(i,j-1)+q^j V(i-1,j) .
\end{equation}
This is one of the known analogues of the Pascal identities to $q$-binomial coefficients and so $V(i,j)= \binom{i+j}{i} _q$ (which is the solution with the correct initial values $V(0,j)=V(i,0)=1$ for all $i,j \ge 0$).

\section{Special functions} \label{app:special_functions}

In this section we assume $|q|<1$ and use the following variables:
\begin{align}
& x,y, x_i\in [-1;1] \quad\quad\quad\quad \theta, \phi, \theta_i \in [0;\pi]\nonumber\\
& x=\cos \theta, \quad \quad  y=\cos \phi, \quad\quad x_i=\cos\theta_i.\nonumber
\end{align}

A $q$-Pochhammer symbol is defined as
\begin{align}
(a;q)_n \colonequals \prod_{k=1}^n\left(1-aq^{k-1}\right) ,
\end{align}
and we use a standard shortcut for their products
\begin{equation}
(a_1,a_2,\cdots ;q)_n = (a_1;q)_n (a_2;q)_n \cdots.
\end{equation}
In these formulas $n$ can be also set to infinity when the product converges.
The $q$-analogue of binomial coefficient is then:
\begin{align}
\binom{n}{m}_q \colonequals \frac{\left(q; q\right)_n}{\left(q; q\right)_m\left(q; q\right)_{n-m}}
\end{align}
which extends to a definition of $q$-multinomial coefficient:
\begin{align}
\binom{n}{n_1, \dots , n_k}_q \colonequals \frac{\left(q; q\right)_n}{\prod_{j=1}^{k}\left(q; q\right)_{n_j}}, \quad \sum_{j=1}^k n_j=n.
\end{align}

A $q$-gamma function is defined via $q$-Pochhammer symbol as:
\begin{align}
\Gamma_q(x)\colonequals \frac{(q;q)_{\infty}}{\left(q^x; q\right)_{\infty}}(1-q)^{1-x}, \quad 0<q<1.
\end{align}
The definition extends to $|q|<1$ by using principal values of $q^x$ and $(1-q)^{1-x}$. One can show that $\lim_{q \to 1^-}\Gamma_q(x)=\Gamma(x)$.

A basic hypergeometric series \cite{gasper2004basic} is defined as
\begin{align}\label{basic-hyper}
\pPq{r}{s}{a_1, \dots , a_r}{b_1, \dots , b_s}{q , z} \colonequals \sum_{k=0}^{\infty} \frac{\left(a_1, \dots , a_r ; q \right)_k}{\left( b_1, \dots , b_s, q ; q \right)_k} \left[ (-1)^k q^{\binom{k}{2}} \right ]^{1+s-r} z^k.
\end{align} 
Setting $0<|q|<1$, this absolutely converges for any $z$ if $q\leq s$ and for $|z|<1$ if $r=s+1$. If one of the upper parameters (and none of the lower, for simplicity) belongs to $q^{\mathbb{Z}_{\leq 0}}$, the series reduces to a finite sum (`terminates').

To give some perspective on orthogonal polynomials encountered in this paper, there is a convenient pattern called $q$-Askey scheme.\cite{KoekoekBook}
It can be thought of as a `map' of frequently used hypergeometric orthogonal polynomials, modelled on a certain four-dimensional manifold having boundaries and corners.\footnote{More precisely, a topologically stratified space: such a space locally looks as $\mathbb{R}^p \times \mathbb{R}^{d-p}_{\geq 0}$, instead of $\mathbb{R}^d$ in a usual manifold. See a detailed description of this manifold with corners in \cite{Koornwinder2009} for the Askey scheme, i.e. a $q\to 1^-$ limit of the $q$-Askey scheme.} Over each point of this four-dimensional space, with coordinates depending on parameters $(a,b,c,d)$, there sits an infinite family of orthogonal polynomials indexed by their degree.
Different strata of this `manifold' represent different levels of the $q$-Askey hierarchy, e.g. Al Salam-Chihara polynomials occupy a part of codimension-2 boundary, whereas continuous $q$-Hermite polynomials sit on one of its several codimension-4 corners. 
Clearly, the corresponding orthogonality measures are also interrelated via such a scheme of degenerations, so that it extends to non-polynomial cases as well.

Continuous $q$-Hermite polynomials $\{H_n(x|q)\}$ are a system of polynomials orthogonal on the interval $x\in[-1;1]$ with the following measure:
\begin{equation} \label{eq:Hermite_orthogonality}
\int_{0}^{\pi} \frac{d\theta }{2\pi } (q,e^{\pm 2i\theta } ;q)_{\infty } H_m(x|q) H_n(x|q) = \delta _{m,n} \, (q; q)_n  .
\end{equation}

There is an explicit hypergeometric expression
\begin{align}
H_n(x|q)= e^{i n\theta} \pPq{2}{0}{q^{-n}, 0}{-}{q; q^n e^{-2i\theta}}.
\end{align}
The continuous $q$-Hermite polynomials satisfy the following recursion relation:
\begin{align} \label{eq:q_Hermite_def}
& 2x H_n(x|q)=H_{n+1}(x|q) + (1-q^n) H_{n-1}(x|q),\\
& H_0(x|q)=1, \quad H_1(x|q)=2x.
\end{align}
As mentioned in \autoref{sec:review}, continuous $q$-Hermite polynomials play an important role in the representation theory of $q$-oscillator algebra.

Ismail-Stanton-Viennot \cite{ISMAIL1987379} give a generalization of the orthogonality statement \eqref{eq:Hermite_orthogonality} for product of any number of $q$-Hermite polynomials:
\begin{align}\label{ISV-multiHermite}
\int_{0}^{\pi} \frac{d\theta }{2\pi } (q,e^{\pm 2i\theta } ;q)_{\infty } \prod_{j=1}^n H_{m_j}(x|q) = \sum_{n_{ij}} \prod_{i=1}^{k} \binom{n_i}{n_{i1}, \dots , n_{ik}}_q \,\, \prod_{1 \leq i < j \leq k} \left(q; q\right)_{n_{ij}} q^{B(n_{rs})},
\end{align}
where one sums over all non-negative symmetric matrices $n_{ij}$ with integer entries satisfying $n_{ii}=0$, $\sum_{i=1}^k n_{ij}=n_j$, $j=1, \dots , k$ and $B(n_{rs})\colonequals \sum_{1\leq i < j < m < l \leq k} n_{im} n_{jl} $. This expression has a combinatorial interpretation as specific counting of line intersections in chord diagrams, see the main text and \autoref{sec:non_integral_forms}, or equivalently as enumerating perfect matchings in a complete graph.

The following bilinear generating function \cite{Szablowski2017} will be particularly useful for us:
\begin{align}\label{qHermite-bilinear-shifted}
\sum_{p=0}^{\infty} \frac{t^p}{(q;q)_p} H_{p+m}(x|q) H_{p+n}(y|q)=\frac{\left(t^2 ;q\right)_{\infty}}{\left(t e^{i(\pm\theta\pm\phi)} ;q\right)_{\infty}}\, Q_{m,n}(x,y|t,q),
\end{align}
with
\begin{align}
Q_{m,n}(x,y|t,q)=Q_{n,m}(y,x|t,q)\colonequals \sum_{s=0}^n (-1)^s q^{\binom{s}{2}}\binom{n}{s}_q \frac{t^s}{\left(t^2;q\right)_{m+s}} H_{n-s}(y|q)Q_{m+s}\left(x|te^{\mp i\phi}; q\right),
\end{align}
where $Q_n(x|a,b; q)$ is the Al Salam-Chihara polynomial:
\begin{align}\label{ASC}
Q_n(x|a,b; q)\colonequals \frac{\left(ab;q\right)_n}{a^n}\pPq{3}{2}{q^{-n}, a e^{\pm i\theta } }{ab,0}{q , q }.
\end{align}
The basic hypergeometric function $_3\Phi_2$ is defined in \eqref{basic-hyper}.
Clearly, the above sum $Q_{m,n}$ has just one term if $m=0$ or $n=0$.

When $a,b$ are real or complex conjugates such that $\left|a\right|, \left|b\right|<1$, Al Salam-Chihara polynomials $\{Q_n(x|a,b; q)\}$ form an orthogonal system on the interval $x\in[-1;1]$ with respect to the following measure:
\begin{align}\label{ASC-orthogonality}
\frac{\left(q; q\right)_{\infty}}{2\pi}\int_{0}^{\pi} d\theta \frac{\left(ab, e^{\pm 2i\theta};q\right)_{\infty}}{\left(ae^{\pm i\theta}, be^{\pm i\theta};q\right)_{\infty}}\, Q_m\left(x|a,b; q\right)Q_n\left(x|a,b; q\right)= \delta_{m n}\, \left(q, ab;q\right)_n.
\end{align}
Similarly to continuous $q$-Hermite polynomials for $q$-oscillator algebra, Al Salam-Chihara polynomials are useful for representation theory of $\mathcal{U}_q(su(1,1))$.

Connection formulas are important in the theory of orthogonal polynomials, since they relate different bases in function spaces. The connection formula between continuous $q$-Hermite and Al Salam-Chihara polynomials reads \cite{Szablowski2017}:
\begin{align}\label{Hermite-ASC}
H_n(x|q)=\sum_{j=0}^n \binom{n}{j}_q t^{n-j} H_{n-j}(y|q)\, Q_j(x|te^{\pm i\phi}; q).
\end{align}
The two previous formulas imply the following useful fact:
\begin{align}\label{ASCweight-Hermite}
\frac{\left(q; q\right)_{\infty}}{2\pi}\int_{0}^{\pi} d\theta \frac{\left(t^2, e^{\pm 2i\theta};q\right)_{\infty}}{\left(t e^{\pm i\theta \pm i\phi} ;q\right)_{\infty}}\, H_m\left(x|q\right)=t^m H_m\left(y|q\right).
\end{align}

We will sometimes also apply another connection formula expanding monomial as a sum over continuous $q$-Hermite polynomials \cite{Szablowski2017}:
\begin{align}\label{monomial-Hermite}
x^n=\frac{1}{2^n}\sum_{m=0}^{\lfloor\frac{n}{2}\rfloor} c_{m,n}H_{n-2m}(x|q).
\end{align}
The coefficients $c_{m,n}$ are given by\footnote{Notice that the first binomial coefficient in this expression is not $q$-deformed, unlike the second.}
\begin{align}\label{cmn}
c_{m,n}\colonequals \sum_{j=0}^m (-1)^j q^{j+\binom{j}{2}}\frac{n-2m+2j+1}{n+1} \binom{n+1}{m-j} \binom{n-2m+j}{j}_q.
\end{align}
These coefficients are, of course, closely related to moments of continuous $q$-Hermite polynomials.

We will use the standard shortcut for a basic one-variable well-poised hypergeometric series:
\begin{align}\label{8W7}
_8W_7(a; b,c,d,e,f; q,z)\colonequals \sum_{n=0}^{\infty}\frac{\left(a, \pm q a^{1/2}, b,c,d,e,f; q\right)_n}{\left(\pm a^{1/2}, qa/b, qa/c, qa/d, qa/e, qa/f, q; q\right)_n }\, z^n.
\end{align}
With an additional condition of very-well-poisedness
\begin{equation*}
bcdefz=q^2 a^2,
\end{equation*}
this function possesses a $W(D_5)$ symmetry in its parameters, which is a bit hidden in any of the hypergeometric representations \eqref{8W7} (with just an $W(A_5)\equiv S_5$ part manifest). The additional symmetry generator corresponds to a so-called (limiting case of) Bailey transform \cite{gasper2004basic}:
\begin{align}\label{8W7-Bailey}
_8W_7\left(a; b, c, d, e, f; q, \frac{a^2 q^2}{bcdef}\right )= \frac{\left(aq, \frac{aq}{ef}, \frac{\lambda q}{e}, \frac{\lambda q}{f} ; q\right)_{\infty}}{\left(\frac{aq}{e}, \frac{aq}{f}, \lambda q,  \frac{\lambda q}{ef} ; q\right)_{\infty}} \, {}_8W_7\left(\lambda ; \frac{\lambda b}{a}, \frac{\lambda c}{a}, \frac{\lambda d}{a}, e, f ; q, \frac{aq}{ef}\right),
\end{align}
where $\lambda \colonequals q a^2/bcd$
and we require
\begin{align*}
\left | \frac{a q}{ef} \right |<1, \quad \left | \frac{\lambda q}{ef} \right |<1
\end{align*}
for convergence. Iterating this identity, one gets another relation:
\begin{equation} \label{eq:8W7_second_D5}
\begin{split}
& {}_8W_7\left(a;b,c,d,e,f;q,\frac{a^2 q^2}{b c d e f} \right)= \frac{\left( aq,b,\frac{bc\mu }{a} ,\frac{bd\mu }{a} ,\frac{be\mu }{a} ,\frac{bf\mu }{a} ;q\right)_{\infty } }{\left( \frac{aq}{c} ,\frac{aq}{d} ,\frac{aq}{e} ,\frac{aq}{f} ,\mu q,\frac{b\mu }{a} ;q\right)_{\infty } } {}_8W_7\left(\mu ;\frac{aq}{bc} ,\frac{aq}{bd} ,\frac{aq}{be} ,\frac{aq}{bf} , \frac{b\mu }{a} ;q,b\right),
\end{split}
\end{equation}
where $\mu =\frac{q^2 a^3}{b^2 c d e f} $.

The following Mellin-Barnes-Agarwal integral representation for very-well-poised $_8W_7$ is known \cite{gasper2004basic}:
\begin{align}\label{8W7-MB}
&_8W_7(q^A; q^a, q^b, q^c, q^d, q^e ; q, q^B)= \sin \pi\left(a+b+c-A\right)\nonumber\\
& \times \frac{\left(q^{1+A}, q^{a}, q^{b}, q^{c}, q^{1+A-a-b}, q^{1+A-b-c}, q^{1+A-a-c}, q^{1+A-d-e}; q\right)_{\infty}}{\left(q, q^{a+b+c-A}, q^{1+A-a-b-c}, q^{1+A-a}, q^{1+A-b}, q^{1+A-c}, q^{1+A-d}, q^{1+A-e}; q\right)_{\infty}}\\
& \times \int_{-i\infty}^{+i\infty} \frac{ds}{2\pi i} \frac{\pi q^s}{\sin \pi s \sin \pi \left(a+b+c-A+s\right)} \, \frac{\left(q^{1+s}, q^{1+A-d+s}, q^{1+A-e+s}, q^{a+b+c-A+s}; q\right)_{\infty}}{\left(q^{a+s}, q^{b+s}, q^{c+s}, q^{1+A-d-e+s}; q\right)_{\infty}}\nonumber
\end{align}
for the parameters satisfying $B=2+2A-a-b-c-d-e$, such that
\begin{align*}
\Re B > 0, \qquad \Re \left(s\log q - \log \left(\sin \pi s \sin \pi (a+b+c-A+s)\right)\right).
\end{align*}
If the last condition is not satisfied on the entire imaginary line, the contour should be indented according to a usual Mellin-Barnes prescription (i.e. separating poles going to the right from those going to the left). When phrased in terms of gamma and $q$-gamma functions, the above integral representation is immediately seen to reduce in $q\to 1^{-}$ limit to a Mellin-Barnes representation of the corresponding (undeformed) well-poised hypergeometric $_7F_6(1)$, i.e. a Wilson function \cite{Groenevelt2006} (up to appropriate Pochhammer factors).

For $ab=\alpha\beta$, a (non-symmetric) Poisson kernel of Al Salam-Chihara polynomials \cite{ASKEY199625} is given by the above function as:
\begin{align}\label{ASC-bilinear}
&\sum_{n=0}^n \frac{t^{n}}{\left(ab, q; q\right)_n} \,  Q_n\left(x_1|a,b,q\right)\, Q_n\left(x_2|\alpha,\beta, q\right)\\
&=\frac{\left(\frac{\beta t}{a}, \alpha t e^{\pm i \theta_1}, a t e^{\pm i \theta_2} ; q\right)_{\infty}}{\left( a\alpha t, t e^{i(\pm\theta_1 \pm\theta_2)} ; q\right)_{\infty}} \, _8W_7\left(\frac{a \alpha t}{q} ; \frac{\alpha t}{b},a e^{\pm i\theta_1}, \alpha  e^{\pm i\theta_2}; q, \frac{\beta t}{a}\right).\nonumber
\end{align}
Setting $t=1$ and taking the limit of $\alpha /a \to 1^-$ in the last formula gives \cite{ASKEY199625}
\begin{equation} \label{eq:sum_Q_a_eqs_b}
\sum _{n=0} ^{\infty } \frac{1}{(ab,q;q)_n} Q_n(x_1|a,b,q) Q_n(x_2|a,b,q) = \frac{\left(b e^{\pm i \theta _1} ,ae^{\pm i \theta _2} ;q\right)_{\infty } }{\left( ab,q,e^{\pm 2i\theta _2} ;q\right)_{\infty } } 2\pi \delta (\theta _1-\theta _2) .
\end{equation}

Askey-Wilson function plays many roles in representation theory of SU$_q(1,1)$ and $\mathcal{U}_q(su(1,1))$. In particular, it is closely related to $6j$-symbols of $\mathcal{U}_q(su(1,1))$ and is important in harmonic analysis on SU$_q(1,1)$.\cite{Koelink:2001}  To introduce it, we first take complex parameters $a,b,c,d$ (with positive real parts) and define 
\begin{align}
\tilde a \colonequals \sqrt{\frac{abcd}{q}}, \quad \tilde b\colonequals \frac{ab}{\tilde a}, \quad \tilde c\colonequals \frac{ac}{\tilde a}, \quad \tilde d\colonequals \frac{ad}{\tilde a},
\end{align}
where the branch of a square root is taken to be positive for positive reals and with the cut along $(-\infty, 0)$.
Then the Askey-Wilson function\footnote{This definition follows \cite{Groenevelt2006}. There are also other conventions in the literature.} can be expressed via the well-poised function defined above:
\begin{align} \label{eq:AW_8W7_relation}
\varphi_{\gamma}(x; a,b,c,d ; q)&=\varphi_x(\gamma; \tilde a, \tilde b, \tilde c, \tilde d ; q)\colonequals\frac{\left(\frac{q a x^{\pm 1}\gamma}{\tilde d}, \frac{q}{\tilde d \gamma}, a b, a c, \frac{q a}{d}; q\right)_{\infty}}{\left(\tilde a \tilde b \tilde c \gamma; q\right)_{\infty}} \nonumber\\
&\times _8W_7\left(\frac{\tilde a \tilde b \tilde c \gamma}{q}; a x^{\pm 1}, \tilde a\gamma, \tilde b\gamma, \tilde c\gamma ; q, \frac{q}{\tilde d \gamma}\right),
\end{align}
for $|q/\tilde d \gamma|<1$.

\section{Series forms for observables and $q \to 1^-$ limits} \label{sec:non_integral_forms}

In all the various observables we evaluated (the partition function and correlation functions), one can perform the $\theta $ integrations explicitly to obtain multivariable hypergeometric series expansions, including expansions in terms of modified Bessel functions. The latter are known in the literature on special functions as Neumann series \cite{WatsonBook}.

\subsection{Partition function}

We start with the partition function. Using \eqref{monomial-Hermite} and \eqref{eq:Hermite_orthogonality} in \eqref{eq:partition_function_moment_integral_form}, we get
\begin{align}
\langle \tr H^{k}\rangle_J=\chi\left(k \equiv 0 \text{ mod }2\right)\cdot \frac{c_{\frac{k}{2},k}}{(1-q)^{k/2}},
\end{align}
where the $\mathbb{Z}_2^{\times}$ character $\chi $ in front is zero for odd $k$ and one for even $k$, and $c_{m,n} $ is defined in \eqref{cmn}. Exponentiating the last expression, one gets the thermal partition function in terms of modified Bessel functions of the first kind $I_{\nu}(x)$
\begin{align}\label{partition-function}
\langle \tr e^{-\beta H}\rangle_J=\frac{\sqrt{1-q}}{\beta} \sum_{p=0}^{\infty} (-1)^p \, q^{p+\binom{p}{2}} \left(2p+1\right) I_{2p+1}\left(\frac{2\beta}{\sqrt{1-q}}\right).
\end{align}

\subsection{2-point function} \label{subsection:2pf_evaluation}

We saw that, as expected, the 2-point function moments \eqref{eq:2pf_moments} depend only on $k_1+k_3$; therefore we can take from now on $k_3=0$.

To display the answer for a two-point function, let us introduce
\begin{align*}
k_+=\max(k_1,k_2), \,\, k_-=\min(k_1,k_2), \,\, \Delta k=|k_1-k_2|.
\end{align*}
Using the expansion \eqref{monomial-Hermite} and orthogonality \eqref{eq:Hermite_orthogonality}, the two-point function moments \eqref{eq:2pf_moments} can then be calculated as
\begin{align}
& \langle \tr
\contraction{}{M}{H^{k_2}}{M}
 M H^{k_2} M H^{k_1} \rangle_J= \nonumber \\
& =\chi\left(\Delta k \equiv 0 \text{ mod } 2\right) \cdot \sum_{m=0}^{\lfloor \frac{k_-}{2} \rfloor}  \tilde q^{k_- -2m}\left(q;q\right)_{k_- -2m} \frac{c_{m+\frac{\Delta k}{2},k_+}c_{m,k_-}}{(1-q)^{\frac{k_++k_-}{2}}}\\
& =\chi\left(\Delta k \equiv 0 \text{ mod } 2\right) \cdot \sum_{m=0}^{\lfloor \frac{k_-}{2} \rfloor} \sum_{p_+=0}^{m+\frac{\Delta k}{2}}\sum_{p_-=0}^{m} \tilde q^{k_- -2m}\left(q;q\right)_{k_- -2m} \binom{k_++1}{m-p_++\frac{\Delta k}{2}}\binom{k_-+1}{m-p_-}\nonumber\\
&\times \prod_{\epsilon=\pm} \biggl [\frac{(-1)^{p_{\epsilon}}}{\left(1-q\right)^{k_{\epsilon}}} q^{p_{\epsilon}+\binom{p_{\epsilon}}{2}} \frac{k_-+2p_{\epsilon}-2m+1}{k_{\epsilon}+1} \binom{k_-+p_{\epsilon}-2m}{p_{\epsilon}}_q \biggl ],\nonumber
\end{align}
where as before the character $\chi$ in front is $1$ when $k_1$, $k_2$ are of the same parity and zero otherwise.

The expression for a thermal two-point function is more compact. Introduce
\begin{align}
\beta_1=\beta/2-it, \quad \beta_2=\beta/2+it.
\end{align} 
The thermal two-point function is then given by:
\begin{align} \label{eq:2_point_function_Bessel}
& \langle \tr
\contraction{}{M}{e^{-\beta _2 H}}{M}
 M e^{-\beta _2 H} M e^{-\beta _1 H} \rangle_J= \langle \tr e^{-\beta H/2} M(t) e^{-\beta H/2} M(0) \rangle_J  = \nonumber \\
&
=\int _0^{\pi } \prod_{j=1}^2  \left\{\frac{d\theta_j}{2\pi}\left(q, e^{\pm 2i\theta_j};q\right)_{\infty} e^{-\frac{2\beta_j \cos \theta_j}{\sqrt{1-q}}} \right\} \frac{\left(\tilde q^2 ;q\right)_{\infty}}{\left(\tilde q\, e^{i\left(\pm \theta_1 \pm \theta_2\right)};q\right)_{\infty}}= \\
&= (1-q) \sum_{k, p_1, p_2=0}^{\infty}\frac{\tilde q^k}{\left(q;q\right)_k}\prod_{j=1,2} \biggl[ \frac{(-1)^{p_j}}{\beta_j}\,  \frac{q^{p_j+\binom{p_j}{2}}\left(q;q\right)_{k+p_j}}{\left(q;q\right)_{p_j}} \, (k+2p_j+1)\, I_{k+2p_j+1}\left(\frac{2\beta_j}{\sqrt{1-q}}\right) \biggl ] .\nonumber
\end{align}

\subsection{Uncrossed 4-point function}

Let us now find an alternative expression to \eqref{4-pt-q1q2TO-int}, with no $\theta $ integrals.
For this we should go back to \eqref{4pt-q1q2-ASC} and proceed similarly to \autoref{subsection:2pf_evaluation}, where the 2-point function was computed. In more detail, we use the formula \eqref{monomial-Hermite} for $x_j^{k_j}$, $j=1,4$ first and take the corresponding two integrals by using \eqref{ASC-orthogonality}, \eqref{Hermite-ASC}, \eqref{ASCweight-Hermite}. Then we similarly expand the two remaining monomials, take the integral in $x_3$ using \eqref{ASCweight-Hermite} and arrive at
\begin{align}
\langle  \tr H^{k_4}
\contraction{}{M_2}{H^{k_3}}{M_2}
M_2 H^{k_3} M_2
H^{k_2}
\contraction{}{M_1}{H^{k_1}}{M_1}
M_1 H^{k_1} M_1 \rangle_J &= \prod_{j=1}^4 \sum_{m_j}^{\left\lfloor\frac{k_j}{2}\right\rfloor} \frac{c_{m_j, k_j}}{(1-q)^{\frac{k_j}{2}}} \cdot \tilde q_1^{k_1-2m_1} \tilde q_2^{k_3-2m_3}\\
&\times  \int _0^{\pi } \frac{d\theta _2}{2\pi } (q,e^{\pm 2i \theta _2} ;q)_{\infty }  \prod _{j=1}^4  H_{k_j-2m_j}(x_2|q).\nonumber
\end{align}
The remaining integral can be calculated by using \eqref{ISV-multiHermite}, which gives a closed-form expression for the 4-point function. As the result is a bit heavy and not thus illuminating, let us immediately go to the thermal 4-point function. Summing up and changing the summation indices we get:
\begin{align}
& \langle \tr\, \left( e^{-\beta_4 H}
\contraction{}{M_2}{e^{-\beta_3 H}}{M_2}  M_2 e^{-\beta_3 H} M_2 e^{-\beta_2 H}
\contraction{}{M_1}{e^{-\beta_1 H}}{M_1}  M_1 e^{-\beta_1 H} M_1\right) \rangle_J \nonumber\\
& = (1-q)^2 \sum_{k_{jl}=0: 1 \leq j<l \leq 4}^{\infty} \,\,\,\, \sum_{p_j=0: 1 \leq j \leq 4}^{\infty} \tilde q_1^{k_1(k_{jl})} \tilde q_2^{k_3(k_{jl})} q^{k_{12}k_{34}}\nonumber \\
&\times \prod_{1\leq j< l \leq 4}\frac{1}{\left(q;q\right)_{k_{jl}}} \prod_{j=1}^{4} \biggl[ \frac{(-1)^{k_j(k_{jl})+p_j}}{\beta_j}\,   \frac{q^{p_j+\binom{p_j}{2}}\left(q;q\right)_{k_j(k_{jl})+p_j}}{\left(q;q\right)_{p_j}}\\ 
& \times\left(k_j(k_{jl})+2p_j+1\right)\, I_{k_j(k_{jl})+2p_j+1}\left(\frac{2\beta_j}{\sqrt{1-q}}\right) \biggl ].\nonumber
\end{align}
To shorten the above formulas, we introduced
\begin{align*}
& k_1(k_{jl})\colonequals k_{12}+k_{13}+k_{14}, \quad k_2(k_{jl})\colonequals k_{12}+k_{23}+k_{24},\\
& k_3(k_{jl})\colonequals k_{13}+k_{23}+k_{34}, \quad k_4(k_{jl})\colonequals k_{14}+k_{24}+k_{34}.
\end{align*}
The last sum is a $10$-fold sum over products of four modified Bessel functions of the first kind and factors including powers of $q$ and $q$-Pochhammer symbols. From the point of view of special function theory, the above expansion is an instance of a Neumann-type expansion in modified Bessel functions.\cite{WatsonBook} The above sum can be also written as a limit of a particular bibasic multiple hypergeometric series.\cite{gasper2004basic}

\subsection{$2n$-point functions} \label{app:2n_point_function}

Let us now look at a (thermal, uncrossed) $2n$-point correlator
\begin{equation*}
\langle \tr\, \left( e^{-\beta_{2n} H}
\contraction{}{M_n}{e^{-\beta_{2n-1} H}}{M_n}  M_n e^{-\beta_{2n-1} H} M_n \cdots e^{-\beta_2 H}
\contraction{}{M_1}{e^{-\beta_1 H}}{M_1}  M_1 e^{-\beta_1 H} M_1\right) \rangle_J ,
\end{equation*}
where we brand the intersections of $H$-chords with each of the operator chords $M_1, \dots M_n$ by different weights $\tilde q_1, \dots , \tilde q_n$.

First, we can use the same prescription as before to get 
\begin{align}
& \langle \tr H^{k_{2n}}
\contraction{}{M_n}{H^{k_{2n-1}}}{M_n}
M_n H^{k_{2n-1}} M_n \cdots
H^{k_2}
\contraction{}{M_1}{H^{k_1}}{M_1}
M_1 H^{k_1} M_1\rangle_J=\\
&\sum _{l_j=0, 1 \leq j \leq 2n-1} ^{\infty }\,\,\, \sum _{p_m=0, 2 \leq m \leq n} ^{\infty }\,\,\,  \prod_{m=1}^n \biggl [ \tilde q_m^{l_{2m-2}+l_{2m-1}} P_{p_m,m}^{(l_{2m-2}+p_m)} T^{k_{2m-1}}_{l_{2m-1},l_{2m-2}} T^{k_{2m}}_{l_{2m}+p_{m+1},l_{2m-1}+p_m} \biggl ],\nonumber
\end{align}
where for notational convenience we also use $l_0, l_{2n}, p_1, p_{n+1} \equiv 0$. After substituting the explicit expressions of matrix elements this becomes
\begin{align}
&\sum _{l_j=0, 1 \leq j \leq 2n-1} ^{\infty }\,\,\, \sum _{p_m=0, 2 \leq m \leq n} ^{\infty }\,\,\,\prod_{j=1}^{2n}\biggl[ \frac{1}{\left(q;q\right)_{l_j}} \int _0^{\pi } \frac{d\theta _j}{2\pi } (q,e^{\pm 2i\theta _j} ;q)_{\infty } \left( \frac{2x_j}{\sqrt{1-q} } \right)^{k_j}  \biggl ] \\
&\times \prod_{m=1}^n \biggl [ \frac{\tilde q_m^{l_{2m-2}+l_{2m-1}} \,(\tilde q_m^2;q)_{p_m}   }{(q;q)_{p_m}}\,  H_{l_{2m-2}} (x_{2m-1}|q) H_{l_{2m-1}}(x_{2m-1}|q) H_{l_{2m-1}+p_m}(x_{2m}|q) H_{l_{2m}+p_{m+1}}(x_{2m}|q) \biggl ].  \nonumber 
\end{align}

Calculating the sums and integrals as we did before, in particular using formulas \eqref{qHermite-bilinear-shifted} \eqref{monomial-Hermite}, \eqref{ISV-multiHermite}, one can rewrite this expression as:
\begin{align}\label{2npt-q1q2-sum}
& \langle \tr\, \left( e^{-\beta_{2n} H}
\contraction{}{M_n}{e^{-\beta_{2n-1} H}}{M_n}  M_n e^{-\beta_{2n-1} H} M_n \cdots e^{-\beta_2 H}
\contraction{}{M_1}{e^{-\beta_1 H}}{M_1}  M_1 e^{-\beta_1 H} M_1\right) \rangle_J \nonumber\\
& = (1-q)^n \sum_{k_{jl}=0: 1 \leq j<l \leq 2n}^{\infty} \,\,\,\, \sum_{p_j=0: 1 \leq j \leq 2n}^{\infty} \,\, \prod_{l=1}^{n} \tilde q^{k_{2l-1}(k_{jl})} \cdot q^{B(k_{jl})}\nonumber \\
&\times \prod_{1\leq j< l \leq 2n}\frac{1}{\left(q;q\right)_{k_{jl}}} \cdot \prod_{j=1}^{2n} \biggl[ \frac{(-1)^{k_j(k_{jl})+p_j}}{\beta_j}\,   \frac{q^{p_j+\binom{p_j}{2}}\left(q;q\right)_{k_j(k_{jl})+p_j}}{\left(q;q\right)_{p_j}}\\ 
& \times\left(k_j(k_{jl})+2p_j+1\right)\, I_{k_j(k_{jl})+2p_j+1}\left(\frac{2\beta_j}{\sqrt{1-q}}\right) \biggl ],\nonumber
\end{align}
which gives an explicit form of a thermal $2n$-point function in our model.
The summation goes over $2n$ indices $p_j$, $j=1, \dots, 2n$ and $n(2n-1)$ indices $k_{jl}$, $1 \leq j < l \leq 2n$. The latter can be organized as entries of a symmetric $2n \times 2n$ matrix with zeros on the diagonal, i.e. $k_{jj}=0$, $j=1, \dots 2n$. To facilitate writing down of the above formula, we also defined
\begin{align*}
k_l(k_{jl})\colonequals \sum_{j=1}^{2n} k_{jl}
\end{align*} 
and introduced the notation
\begin{align*}
B(k_{jl}) \colonequals \sum_{1 \leq j^{(1)} < j^{(2)} < j^{(3)} < j^{(4)} \leq 2n} k_{j^{(1)} j^{(3)}} k_{j^{(2)} j^{(4)}}.
\end{align*} 

The expression \eqref{2npt-q1q2-sum} is a $n(2n+1)$-fold sum over products of $2n$ modified Bessel functions of the first kind and factors including powers of $q$ and $q$-Pochhammer symbols.
This expansion is again an instance of a Neumann-type expansion \cite{WatsonBook}, as well as a limiting form of a bibasic multiple hypergeometric series \cite{gasper2004basic}.

\subsection{$q \to 1^-$ limit of the partition function}

One of the advantages of obtaining the explicit expressions above in terms of Bessel functions (and no remaining integrations) is that some of the limits of the various observables are easy to get, as we will demonstrate for the partition function and the 2-point function. Additional limits are found in \cite{Berkooz:2018qkz}.

Consider the partition function \eqref{partition-function} in the limit $q \to 1^-$, or since $q=e^{-\lambda } $, $\lambda  \to 0^+$.
When the argument $w$ of the Bessel function $I_{\alpha  } (w)$ is large, the asymptotic form is independent of the index $\alpha  $, and we are left with evaluating a simpler sum. Let us first specify what is the regime in which this is valid. The asymptotic form we would like to use is
\begin{equation} \label{eq:asymptotic_approx_Bessel}
I_{\alpha   } (w) \sim \frac{e^{w} }{\sqrt{2\pi w} } 
\end{equation}
when $|w| \to \infty $ and $| \arg w| < \frac{\pi }{2} -\delta $ (for some $\delta >0$). This requires $\beta  \gg \sqrt{\lambda } $ in the limit of small $\lambda $. The sum in \eqref{partition-function} contains a decaying factor $q^{p(p+1)/2} $, meaning that the major contribution comes from $| p| \le 1/\sqrt{\lambda } $.
However, it is important that in order to use the above approximation of the Bessel function, the index $\alpha  $ of the Bessel function should not be very large, or more precisely we should have $|\alpha  | \ll \sqrt{|w|} $.\footnote{This can be seen for example by the higher orders in the asymptotic expansion of $I_{\alpha } (w)$.} This means that we need to be in the regime $|p| \ll \beta ^{1/2} \lambda ^{-1/4} $. Using the range of indices $p$ contributing to the sum as was just found, this gives the additional stronger condition $\beta  \gg \lambda ^{-1/2} $. This latter condition is therefore the condition for the validity of the following approximation.

The $\beta $ dependence of the partition function is actually obtained immediately in this range, but let us also evaluate the $\lambda $ dependence (for small $\lambda $). The remaining sum to calculate is
\begin{equation}
\sum _{p=0} ^{\infty } (-1)^p q^{p(p+1)/2} (2p+1) = \sum _{p=-\infty } ^{\infty } (-1)^p q^{p(p+1)/2} p .
\end{equation}
This sum is closely related to theta functions. For the arguments of theta functions we use the notations
\begin{equation}
z = e^{2\pi i\nu } ,\qquad q=e^{2\pi i \tau } .
\end{equation}
We can then write
\begin{equation}
\begin{split}
& \sum _{p=-\infty } ^{\infty } (-1)^p q^{p(p+1)/2} p = \sum _{p =-\infty } ^{\infty } (-1)^p q^{p^2/2} z^p p = \frac{1}{2\pi i} \pder{}{\nu } \sum _{p=-\infty } ^{\infty } (-1)^p q^{p^2/2} z^p = \\
& = \frac{1}{2\pi i} \pder{}{\nu } \theta _{01} (\nu ,\tau ),
\end{split}
\end{equation}
subsequently taking $\tau = \frac{i\lambda }{2\pi } $ and $\nu =\frac{\tau }{2} $. In the $q \to 1^-$ limit, it is useful to use the modular transformation of the theta function to get
\begin{equation}
\begin{split}
& \frac{1}{2\pi i} \pder{}{\nu } \left[ (-i\tau )^{-1/2} e^{-\pi  i \nu ^2 / \tau } \theta _{10} \left( \frac{\nu }{\tau } ,-\frac{1}{\tau } \right) \right] = \\
& = \frac{\partial _{\nu } }{2\pi i} \Bigg [2 (-i\tau )^{-1/2} e^{-\pi i\nu ^2/\tau } e^{-\pi i/(4\tau )} \cos \left(\frac{\pi \nu}{\tau}\right) \cdot  \\
& \qquad \qquad \cdot \prod _{n=1} ^{\infty }  \left( 1-e^{-\frac{2\pi in}{\tau} } \right)\left(1+e^{\frac{2\pi i\nu }{\tau } - \frac{2\pi in}{\tau } } \right) \left( 1+ e^{- \frac{2\pi i\nu }{\tau } - \frac{2\pi in}{\tau } } \right) \Bigg] \at{\nu =\frac{\tau}{2}} =\\
&= \frac{i}{\tau } (-i\tau )^{-1/2} e^{-\pi i\tau /4} e^{-\pi i/(4\tau )} \prod _{n=1} ^{\infty } \left( 1- e^{- \frac{2\pi in}{\tau } } \right)^3 .
\end{split}
\end{equation}
In the limit $\lambda  \to 0$ this is approximated by
\begin{equation}
\sim \left( \frac{2\pi }{\lambda } \right)^{3/2} e^{-\frac{\pi ^2}{2\lambda } } .
\end{equation}
Together with the approximation of the Bessel function, we get that for $\beta  \gg \lambda ^{-1/2} $,
\begin{equation}
\begin{split}
\langle \tr e^{-\beta H}\rangle_J \sim \frac{\sqrt{2} \pi }{\beta ^{3/2} \lambda ^{3/4} } e^{\frac{2\beta }{\sqrt{\lambda } } - \frac{\pi ^2}{2\lambda } } .
\end{split}
\end{equation}

\subsection{$q \to 1^-$ limit of the 2-point function}

The main difference in the 2-point function \eqref{eq:2_point_function_Bessel} compared to the partition function, is that while as before $|p_j| \le 1/\sqrt{\lambda } $ dominate the sum, assuming that $\tilde q$ is of the same order of magnitude as $q$, for the index $k$ we have the larger range $|k| \le 1/\lambda $. (If $\tilde q$ is not of the order of $q$, the modification of these arguments is straightforward; note also that we do not assume here that $\tilde q$ is an integer power of $q$, which is a useful assumption in other approaches \cite{Berkooz:2018qkz}.) Then applying the $|\alpha| \ll \sqrt{|w|}$ condition for the approximation \eqref{eq:asymptotic_approx_Bessel} gives the regime of validity $|\beta_j|  \gg \lambda ^{-3/2} $. Also, the argument $w$ of the Bessel function should not approach the imaginary axis; together we get the conditions $|t| \ll \beta $ and $\beta \gg \lambda ^{-3/2} $ for the following approximation.

Now we will be interested in the time and temperature dependence and will not evaluate the remaining sum. Applying \eqref{eq:asymptotic_approx_Bessel} to \eqref{eq:2_point_function_Bessel} gives the simple result (the function $\mathcal{C} (q,\tilde q)$ can be written via $q$-Bessel functions; importantly, it is independent of $t,\beta $):
\begin{equation}
\langle \tr e^{-\beta H/2} M(t) e^{-\beta H/2} M(0) \rangle_J \sim \mathcal{C} (q,\tilde q) \frac{e^{\frac{2\beta }{\sqrt{1-q} } } }{(\beta ^2 + 4t^2)^{3/2} } ,
\end{equation}
which is compatible with \cite{Berkooz:2018qkz}.

\section{Asymptotic forms of Gamma functions} \label{app:asymptotic_Gamma}

In this appendix we specify the asymptotic forms of $\Gamma (z)$ and $\Gamma _q(z)$ that we use in the main text. The range of arguments $z$ that we are interested in is
\begin{equation}
1 \ll |z| \ll 1/\lambda .
\end{equation}
Moreover, all the arguments $z$ of these functions in our case are either purely imaginary or approach the imaginary axis. The Stirling formula for the Gamma function is still true for such arguments
\begin{equation} \label{eq:Stirling_approx}
\log \Gamma (z) = \left(z-\frac{1}{2} \right)\log(z) - z + \text{const} + O\left(\frac{1}{z} \right) .
\end{equation}

For the $\Gamma _q$ function there is an analogue of Stirling formula, derived by Moak \cite{moak1984}, which gives the asymptotic expansion for large $z$:
\begin{equation} \label{eq:Moak_formula_app}
\begin{split}
& \log \Gamma _q(z) \sim \left(z-\frac{1}{2} \right) \log \frac{1-q^z}{1-q} + \frac{Li_2(1-q^z)}{\log (q)} + \sum _{k=1} ^{\infty } \frac{B_{2k} }{(2k)!} \left( \frac{\log (q)}{q^z-1} \right)^{2k-1} q^z p_{2k-3} (q^z) + \\
& \qquad +  (q\text{-dependent const}) ,
\end{split}
\end{equation}
with $Li_2(x)$ the dilogarithm function, $B_k$ the Bernoulli numbers, and $p_k$ a polynomial of degree $k$, the form of which we will not need.
However, \eqref{eq:Moak_formula_app} is not valid for purely imaginary arguments. Nevertheless, for large imaginary $z$ satisfying $|z| < \frac{\pi }{\lambda } $ (which is the case for a particular $y_j$ in the main text since $\theta  \in [0,\pi ]$), the formula by Moak holds (this range approaches the full imaginary axis as $\lambda  \to 0$, making the Stirling formula valid over the entire imaginary axis, as should be the case). As mentioned above, we will even restrict ourselves to arguments $z$ satisfying $|z| \ll 1/\lambda $.

In all the asymptotic forms of logarithms of Gamma functions below, we will not keep track of additive $q$-dependent constants, consistently with the goal in the main text.

As we are interested in small $\lambda $, we may expand the formula \eqref{eq:Moak_formula_app} in $1/z$ and $\lambda z$ to get
\begin{equation} \label{eq:log_Gamma_q_approx}
\begin{split}
\log \Gamma _q(z) = \left(z- \frac{1}{2} \right)\log(z)-z\left(1-\frac{3\lambda }{4} \right)- \frac{z^2 \lambda }{4} +\cdots .
\end{split}
\end{equation}
(The corrections relative to the $z$ term in the formula are $O\left(\frac{1}{z^2} ,\lambda ^2 z^2\right)$; we did not quote the $1/z$ correction which is a constant, but it will not affect anything we do as we mentioned, so that the corrections relevant to our results start at $1/z^2$).

We will also encounter $\Gamma _q$ functions with small corrections to $z$. In this case we can expand in $1/z$, $\lambda z$ and $\epsilon /z$ and find
\begin{equation} \label{eq:log_Gamma_q_z_epsilon}
\log \Gamma _q(z+\epsilon ) = \log \Gamma _q(z) + \epsilon  \log(z) - \frac{\epsilon }{2z} -\frac{\lambda z\epsilon }{2} -\frac{\lambda \epsilon ^2}{4} +\frac{3\lambda \epsilon }{4} + \cdots .
\end{equation}
(Relative to the $z$ term, the corrections are $O\left(\frac{1}{z^2} ,\lambda ^2z^2, \frac{\epsilon ^2}{z^2} \right)$.)

Since the Gamma functions appear with complex arguments, we need to be careful with the branch cut of the $\log $ function in the asymptotic expansion of the Gamma functions. The principal branch should be used (meaning that the phase of $z$ in $\log (z)$ is chosen to be in $(-\pi ,\pi ]$). In the calculation of the crossed 4-point function, every $\Gamma _q$ function with an argument that is not purely imaginary is divided by a $\Gamma $ function with the same argument, so the logs cancel. We will encounter many occurences of $\log \Gamma _q(z+\epsilon ) \pm \log \Gamma _q(-z-\epsilon )$, and so it will be useful to write the simpler expressions for them. As was just argued, we will need them only for purely imaginary arguments. Using \eqref{eq:log_Gamma_q_z_epsilon}, we get for $z \in i \mathbb{R} $
\begin{equation} \label{eq:log_Gamma_q_z_epsilon_diff}
\begin{split}
\log \Gamma _q(z+\epsilon )-\log \Gamma _q(-z-\epsilon ) \overset{z \in i \mathbb{R} }{=}  z \log(-z^2) - 2z\left(1-\frac{3\lambda }{4} \right) + \epsilon  \log (-z^2) + \frac{3\lambda \epsilon }{2} + \cdots .
\end{split}
\end{equation}
For positive (negative) $z/i$, there is an additional term of $-i\pi /2$ ($+i\pi /2$ respectively), and we have dropped those as they are constants. The role of the branch cut is more important in the following formula,\footnote{For instance, the sign function resulting from this in \eqref{eq:log_Gamma_q_z_epsilon_sum} is important since without it the right-hand side is not invariant under $(z,\epsilon ) \to (-z,-\epsilon )$ while the left-hand side is.} where again $z \in i \mathbb{R} $
\begin{equation} \label{eq:log_Gamma_q_z_epsilon_sum}
\begin{split}
\log \Gamma _q(z+\epsilon )+\log \Gamma _q(-z-\epsilon ) \overset{z \in i \mathbb{R} }{=} - \frac{1}{2} \log(-z^2) + i\pi \epsilon \sign \frac{z}{i} -\pi |z| - \frac{z^2\lambda }{2} -\lambda z\epsilon -\frac{\lambda }{2} \epsilon ^2 - \frac{\epsilon }{z} +\cdots .
\end{split}
\end{equation}

\section{Quantum groups}
\label{sec:quantum_groups}

Special functions of hypergeometric type described in \autoref{app:special_functions} are known to play a major role in the representation theory of quantum groups. Like their $q=1$ cousins, they appear in variety of contexts, in particular the same special functions can play very different representation-theoretic roles.
In this appendix we outline a Hopf-algebraic definition of the quantum groups used in the main text.
Namely, we will define a quantum deformation of the (extension of the) group SU$(1,1)$ and its dual\footnote{In order to keep track of various structures,
let us adopt a convention:
for a generic abstract object $X$ (Lie group, algebra, Hopf algebra, Hopf $*$-algebra), we denote its concrete realization related to SU$(1,1)$ by $X_1$, the one related to $G=\text{SU}^{ext}(1,1)$ -- by $X_e$ and the one relevant for $\text{SU}^{ext}_q(1,1)$ -- by $X_q$.}.
Useful mathematical references are \cite{KasselBook, Timmermann}.

\subsection{Hopf algebra of functions on a Lie group}
\label{app:HopfLie}

1. {\it Definitions} \\
Let us start with a simple non-deformed example. The elements of the real form of the group SL$(2,\mathbb{C})$, called SU$(1,1)$, can be written as
\begin{align} \label{G_1}
   G_1=\text{SU}(1,1)=
  \left\{\begin{pmatrix} a&c\\ \bar{c}&\bar{a} \end{pmatrix} | a,c \in \mathbb{C}, |a|^2-|c|^2=1 \right\}.
\end{align}
In view of the coming quantum deformation it is also necessary \cite{Korogodsky1994} to consider a slightly bigger group\footnote{The $q$-deformation of SU$(1,1)$ cannot be consistently defined on the operator-algebraic level \cite{woronowicz1991}, see also section \ref{sec:quantum-def}.} instead of SU$(1,1)$, which we will denote $G_e=\text{SU}^{ext}(1,1)$:
\begin{align}\label{G_e-embedded}
G_e =\text{SU}^{ext}(1,1)=
  \left\{\begin{pmatrix} a&c\\ \epsilon \bar{c}& \epsilon\bar{a} \end{pmatrix} | a,c \in \mathbb{C}, \epsilon\in\{\pm 1\}, |a|^2-|c|^2=\epsilon \right\} \subset \text{SL}\,(2, \mathbb{C}).
\end{align}
The group $\text{SU}^{ext}(1,1)$ is a normalizer of SU$(1,1)$ inside SL$(2,\mathbb{C})$ and contains SU$(1,1)$ as an index $2$ subgroup. In particular, $\text{SU}^{ext}(1,1)$ as a set can be written as disjoint union of two connected components contained in SL$(2,\mathbb{C})$: one of them is SU$(1,1)$, the other is SU$(1,1)\times \begin{pmatrix} 0 & 1\\ -1 & 0 \end{pmatrix}$.

Consider an algebra of functions $\mathcal{A}_e= \text{Fun}\left(G_e\right)$ defined as follows: for each $g\in \text{SU}^{ext}(1,1)$, a value of an element of $\mathcal{A}_e$  can be found from that of its generators: $\alpha(g)=a, \gamma(g)=c, e(g)=\epsilon$; $\alpha, \gamma, e \in \mathcal{A}_e$ and we are allowed to multiply/add these letters in any order.
$\mathcal{A}_e$ is a commutative associative algebra with identity 1 called algebra of regular functions on the group (under pointwise multiplication).
Multiplication is given by a map $m:\mathcal{A}_e \times \mathcal{A}_e \to \mathcal{A}_e$, $(m(f_1\otimes f_2))(g)=(f_1\otimes f_2)(g,g)=f_1(g)f_2(g)$.
As the algebra $\mathcal{A}_e$ came from the group, we would expect there is more structure to it than just said, at least we would like to see how the group laws of $\text{SU}^{ext}(1,1)$ are reflected in $\mathcal{A}_e$. To spell this out, first one needs a map $\Delta: \mathcal{A}_e \rightarrow \mathcal{A}_e\otimes \mathcal{A}_e$, which is called coproduct.  The coproduct definition for $\mathcal{A}\left(\text{SU}^{ext}(1,1)\right)$ (and any other Lie group algebra) is very simple: for any group algebra element $f\in \mathcal{A}_e$, we define
$\Delta(f)(g_1, g_2)\colonequals f(g_1 g_2)$, $g_i\in \text{SU}^{ext}(1,1)$, which has a property of coassociativity:
\begin{align}\label{coassoc}
(\Delta \otimes \iota) \circ \Delta = (\iota \otimes \Delta) \circ \Delta,
\end{align}
where $\iota$ is an identity map. Indeed, one can check that, for the simple coproduct we just defined, this equation is equivalent to:\footnote{To shorten notations, we will mostly omit the symbol $\circ$ for composition of maps in the following.}
\begin{align*}
\left(((\Delta \otimes \iota)\Delta) f\right)(g_1, g_2, g_3) = f\left((g_1 g_2) g_3\right)=f\left(g_1 (g_2 g_3)\right)= \left( ((\iota \otimes \Delta) \circ \Delta) f \right)(g_1, g_2, g_3).
\end{align*}
Specifically, multiplying two matrices\footnote{In order to be consistent with our other conventions, in particular with the definition of corepresentation \eqref{corep-Udef}, the formulas we write are derived by multiplying these matrices from right to left.} \eqref{G_e-embedded} belonging to $G_e$, the group multiplication translates to the following coproducts on generators of $\mathcal{A}_e$:
\begin{align}\label{G_e-coproduct}
\Delta(\alpha)=\alpha \otimes \alpha + \left(e \gamma^{*}\right)  \otimes \gamma, \quad \Delta(\gamma)= \gamma  \otimes \alpha  +  \left(e \alpha^{*}\right) \otimes \gamma , \quad \Delta(e)=e \otimes e.
\end{align}

To translate the remaining group structure, we also need two other maps, defined for the group algebra of our Lie group as:
\begin{align}\label{counit-antipode}
&\text{counit } \quad \quad \varepsilon:\,\, \mathcal{A}_e \rightarrow \mathbb{C},\quad \varepsilon(f)=f(id),\\
&\text{antipode } \quad S:\,\, \mathcal{A}_e \rightarrow \mathcal{A}_e,\quad S(f)(g)=f(g^{-1}),\nonumber
\end{align}
where $f\in \mathcal{A}_e$, $g\in G_e$ and $id$ is the unit element of our group.
From the group laws these can be seen to satisfy relations
\begin{align}\label{counit-antipode-relations}
&(\epsilon \otimes \iota)  \Delta = \iota = (\iota \otimes \epsilon) \circ \Delta\\
&(m \circ (S \otimes id)\circ \Delta)=\epsilon\,\text{1}=(m \circ (id \otimes S)\circ \Delta),\nonumber
\end{align} 
in our simple case boiling down to
\begin{align*}
&\left(((\epsilon \otimes \iota)  \Delta )f\right)(g) = f(id \cdot g)= f(g)=f(g \cdot id) = \left(((\iota \otimes \epsilon) \Delta)f\right)(g),\\
&\left((m (S \otimes id)\Delta)f\right)(g)=f(g^{-1}g)= f(id)= f(g g^{-1})=\left((m (id \otimes S) \Delta)f\right)(g).
\end{align*}
Again specifying to $\mathcal{A}_e$, we see that in terms of generators this yields:
\begin{align}\label{G_e-counit-antipode}
&\epsilon(\alpha)=\epsilon(e)=1, \quad \epsilon(\gamma)=0\\
&S(\alpha)= e \alpha^{*}, \quad S(\alpha^{*})= e \alpha, \quad S(\gamma)= -\gamma, \quad S(\gamma^{*})= -\gamma^{*}, \quad S(e)= e.\nonumber
\end{align}

Last but not least, one needs to specify how to conjugate elements, which for our example in the present section amounts to choosing matrix entries of group elements to be real or complex. Of course, this was already implicitly done in our definitions \eqref{G_1}, \eqref{G_e-embedded}. Generally, in order to choose a real form for a Hopf algebra one introduces a $*$-structure which is an (antilinear antimultiplicative) involution of this algebra. All the other pieces of algebraic structure listed above should respect this operation.
Unlike our non-deformed case, where the two possible non-compact real forms of SL$(2,\mathbb{C})$, SU$(1,1)$ and SL$(2, \mathbb{R})$, are equivalent under Cayley transformation
(i.e. conformal transformation sending a unit disk to an upper half-plane) -- after the quantum deformation is turned on, this equivalence will stop to hold. A $*$-structure for the deformation of $G_1$ or $G_e$ exists only when $0<q<1$ whereas the one for SL$(2, \mathbb{R})$ --  only for $|q|=1$. \cite{Vaksman1991, Masuda1990a, Masuda1990b} This means that, in our story, we will need the deformation of SU$(1,1)$ (or rather its extension SU$^{ext}(1,1)$) and not the other real form.
We will present more details on this important fact in a discussion of
the real forms for
the dual Hopf algebra $\mathcal{U}_q(sl(2,\mathbb{C}))$, see subsection \ref{app:u_qsu(1,1)}.

All in all, we just outlined a widely encountered mathematical structure called a Hopf $*$-algebra $A_e=\left(\mathcal{A}_e, \Delta, \varepsilon, S\right)$, associated to our locally compact group $G_e$. The listed structures straightforwardly generalize to any Lie group. Looking, perhaps, as unnecessary complication from the viewpoint of classical group theory, this is a step which will now allow us to talk about quantum groups -- it is hard to guess what a non-commutative and non-local `quantum' deformation of a Lie group should look like, but the algebra of functions on the Lie group is rather straightforward to deform. However, as the quantum deformation of $A$ is being turned on, the simple expressions for $\Delta, \varepsilon, S$ should also be deformed to preserve self-consistency. We will specify the precise relations in a moment.
\medskip \\
2. {\it Corepresentations and multiplicative unitaries} \\
Before that, let us remind what corresponds to representations of a group, when we speak the language of its algebra.\footnote{In this small discussion, we keep the Hopf algebra rather general most of the time and only specify to our example of $G_e$ and $A_e$ to illustrate the concepts.}
Namely, it is easy to see that a (left) representation, i.e. a vector space $V$ together with a map $\pi: G\otimes V \rightarrow V $, is traded for a dual map $u: V \rightarrow A \otimes V$, which is called a (left) corepresentation of a Hopf algebra $A$. Similarly to a definition of representation, it is required that
\begin{align}\label{corep-1}
(\iota\otimes u)\circ u=(\Delta\otimes \iota)\circ u, \quad (\epsilon\otimes \iota) \circ u=\iota.
\end{align}

There are two related useful ways to think of this map. 
One of them is to fix a basis $\{v_i\}_i$ in the (finite-dimensional) vector space $V$, which gives us
\begin{align}\label{corep-2}
u(v_j)=\sum_i a_{ij} \otimes v_i, \quad a_{ij}\in A,
\end{align}
i.e. a corepresentation on $V$ is defined by an
$n\times n$-matrix of algebra elements $a_{ij}$ (with the corresponding analogues of conditions \eqref{corep-1}). For example, taking $a_{ij}$ as functions sending a group element $g\in G_e$ to its matrix elements in the group representation $\pi$ on $V$, we see that there is a one-to-one correspondence between representations of a group $G_e$ on $V$ and corepresentations of its group algebra $A_e$ on the same space. Similarly, a corepresentation on an infinite-dimensional separable Hilbert space can be to some extent thought of as an infinite-dimensional matrix of algebra elements. 

Second convenient way to think about a (unitary) corepresentation is as a single (unitary) element $U\in A \otimes \text{Hom}\,(V)$. The operator $U$ is then required to obey
\footnote{We use a standard leg notations, where subscripts on an operator denote the components of the tensor product it acts on.}
\begin{align}\label{corep-Udef}
(\Delta \otimes \iota)(U)=(U)_{13}(U)_{23},
\end{align}
which again amounts to a slightly reformulated condition \eqref{corep-1}.

Two special corepresentations of $A$ defined by coproduct
are called regular corepresentations. They are co-analogues of the left (right) regular representations of a Lie group $G$. As in the classical case, both these regular corepresentations are highly reducible. We will now construct the operators $U$ defining them.

A unitary operator $W$ on $A\otimes A$
called (left) multiplicative unitary is defined via
\begin{align}\label{left-mult-unit}
W^*\left(a \otimes b\right)=\Delta(b)(a\otimes 1),
\end{align}
for any $a,b \in A$.
E.g. in our classical example of $G_e$, this multiplicative unitary acts via\footnote{In our discussion of multiplicative unitaries we freely use an equivalence $\text{Fun}(G)\times \text{Fun}(G) \sim \text{Fun}(G\times G)$ for the space of regular functions on the group, and similarly for the functions of more variables.}
\begin{align}
(W^*f)(g_1,g_2)=f(g_1,g_1g_2), \quad (Wf)(g_1,g_2)=f(g_1,g_1^{-1}g_2).
\end{align}
From here we can see that this map is indeed closely related to the definition of the left action of the group.
Left regular corepresentation of a Hopf algebra\footnote{In a proper locally compact quantum group setting, this corepresentation acts on the GNS space of the (left) Haar weight.} $A$  is then defined by $U_{left \,\, reg}=W$. One can check that this reduces to the usual definition in the classical case of Lie groups:
\begin{align*}
\left((\Delta \otimes \iota)(W)\right)&\left(f(g_1, g_2, g_3)\right)=f\left((g_1, g_2, (g_1 g_2)^{-1} g_3\right)=f\left((g_1, g_2, g_2^{-1} g_1^{-1} g_3\right)\\
&=\left(W\right)_{13}\left(f(g_1, g_2, g_2^{-1}g_3)\right)=\left(W\right)_{13}\left(W\right)_{23}\left(f(g_1, g_2, g_3)\right).
\end{align*}
The coproduct can be recovered from $W$ via
\begin{align}
\Delta(a)=W^*(1\otimes a)W\quad \text{ for all } a\in A,
\end{align} 
as an operator acting on $A \otimes A$, so that left-hand side acts by multiplication. Indeed, acting on a function of two variables $f\in \text{Fun}(G_e\times G_e)$, right-hand side yields:
\begin{align*}
\Delta(a)f(g_1, g_2) = W^*(1\otimes a) W ) f(g_1, g_2) = a(g_1 g_2) f(g_1, g_2).
\end{align*}

Coassociativity of comultipication implies that multiplicative unitary satisfies a pentagon equation:
\begin{align}
W_{12} W_{13} W_{23} = W_{23} W_{12}.
\end{align}
Multiplicative unitary can be even axiomatically defined as a unitary operator satisfying this equation (so that the coproduct is obtained from it later). In fact, this is what happens for the (operator-algebraic) quantum group SU$^{ext}_q(1,1)$, see \cite{Koelink2003}. It's straightforward to check that the pentagon equation is satisfied for the case of the Hopf algebra of functions on a Lie group:
\begin{align*}
\left((W_{12} W_{13} W_{23})f\right)(g_1, g_2, g_3) = f(g_1, g_1^{-1}g_2, g_2^{-1}g_3)=\left(( W_{23} W_{12})f\right)(g_1, g_2, g_3).
\end{align*}

In a full analogy, the right regular corepresentation (of a unimodular quantum group) is defined using the operator $V$, called right multiplicative unitary, which acts as
\begin{align}\label{right-mult-unit}
V\left(a \otimes b\right)=\Delta(a)(1\otimes b)
\end{align}
for any $a,b \in A$, and $U_{right \,\, reg}=\chi\left(V\right)=\Sigma V \Sigma$, where $\Sigma$ is a flip of factors in tensor product. E.g. in our classical example of $G_e$, this second multiplicative unitary acts via
\begin{align*}
\left(\chi(V)f\right)(g_1, g_2)=f(g_1, g_2 g_1), \quad \left(\chi(V^*)f\right)(g_1, g_2)=f(g_1, g_2 g_1^{-1}).
\end{align*}
We leave it as an exercise to check that this operator $\chi(V)$ defines a corepresentation and satisfies pentagon equation.

The analogue of a biregular representation\footnote{In a locally compact group setting, this corepresentation acts on the space of Hilbert-Schmidt operators on the GNS space of the Haar weight, provided the quantum group is unimodular.} of undeformed Lie groups is given by combining left and and right regular corepresentations as
\begin{align}\label{biregular}
\tilde{U}\colonequals(\chi(V))_{13} (W)_{24}.
\end{align}
Again it's easy to check that this operator satisfies pentagon equation (where both sides act on a tensor product of six copies of $A$) and thus is a well-defined multiplicative unitary operator.

\subsection{Quantum group SU$^{ext}_q(1,1)$}
\label{su_q(1,1)}

Now we define a quantum group Hopf algebra $\mathcal{A}_q=(\mathcal{A}_q, \Delta, \epsilon, S)$ as a (non-commutative and non-cocommutative) deformation of the algebra $A_e$ considered in the previous subsection \ref{app:HopfLie} \footnote{All the four elements of the quadruple get deformed in the process, but we put a subscript $q$ only in $\mathcal{A}$ to avoid unnecessary cluttering. It should be always clear from the context which coproduct, counit or antipode we use in each case.}.  As before, in addition to a usual (unital associative) algebra structure $A_q$ (over $\mathbb{C}$), there are maps $\Delta: A_q \to A_q \otimes A_q $, $\epsilon: A_q \to \mathbb{C}$, $S: A_q \to A_q$, called coproduct, counit and antipode, which satisfy the same symbolic conditions (\eqref{coassoc}, \eqref{counit-antipode-relations}) as in our example above. Analogously, to get a Hopf $*$-algebra defining a real form, this should be supplemented by a real structure, so that all the five components are mutually compatible.
See e.g. \cite{KasselBook, Timmermann} for a detailed introduction to Hopf-algebraic quantum groups (and some aspects of the operator-algebraic setup).

We start with the first item of a quadruple $A_q$.
A coordinate ring of $\text{SU}^{ext}_q(1,1)$, denoted $\mathcal{A}\left(\text{SU}^{ext}_q(1,1)\right)=\mathcal{A}_q$, is generated by three elements $\alpha, \gamma, e$ and their conjugates
obeying
\begin{align}
& \alpha^{*}\alpha-\gamma^{*}\gamma=e, \alpha \alpha^{*}- q^2 \gamma^{*}\gamma=e,  \quad \gamma^{*}\gamma=\gamma \gamma^{*},\nonumber\\
& \alpha \gamma=q\gamma\alpha, \quad \alpha \gamma^{*}=q\gamma^{*}\alpha\\
& e^{*}=e, \quad e^2=1, \quad \alpha e =e \alpha, \quad \gamma e =e \gamma. \nonumber
\end{align}
If we send $q\to 1^-$ in these formulas, we recover $A_e$. Like there, the real structure $*$ is already implicit in this definition.

The coproduct then reads (on generators) as:
\begin{align}
\Delta(\alpha)=\alpha \otimes \alpha + q \left(e \gamma^{*}\right) \otimes \gamma  , \quad \Delta(\gamma)= \gamma \otimes \alpha + \left(e \alpha^{*}\right)\otimes \gamma , \quad \Delta(e)=e \otimes e,
\end{align}
compare to formulas \eqref{G_e-coproduct},
whereas counit and antipode are:
\begin{align*}
&\epsilon(\alpha)=\epsilon(e)=1, \quad \epsilon(\gamma)=0\\
&S(\alpha)= e \alpha^{*}, \quad S(\alpha^{*})= e \alpha, \quad S(\gamma)= -q \gamma, \quad S(\gamma^{*})= -q^{-1} \gamma^{*}, \quad S(e)= e,
\end{align*}
which also can be easily detected as a $q$-deformation of \eqref{G_e-counit-antipode}.
One can, of course, check all the previously listed properties \eqref{coassoc}, \eqref{counit-antipode-relations} of these items. A (formal) quantum group SU$_q(1,1)$ can be obtained from here by projecting the element $e$ down to 1.

Finer details of this quantum group, such as Haar measure and its (co-)actions on various Hilbert spaces, require an operator-algebraic formulation in an essential way \cite{Koelink2003, Groenevelt2010}, see also a short subsection \ref{app:operator-alg}. In particular, one can show that the analogous deformation of SU$(1,1)$, a would-be-quantum group SU$_q(1,1)$, cannot exist on an operator-algebraic level.\cite{woronowicz1991}

Formulas \eqref{left-mult-unit}, \eqref{right-mult-unit}, \eqref{biregular} define left, right and bi-regular corepresentations of this quantum group, in analogy to the undeformed case.

\subsection{Quantum group $\mathcal{U}_q(su(1,1))$}
\label{app:u_qsu(1,1)}

1. {\it Definition of the complex form} \\
The second quantum group we shall need is a real form $\mathcal{U}_q(su(1,1))$ of the group $\mathcal{U}_q(sl(2, \mathbb{C}))$, which is a deformation of the universal enveloping algebra $\mathcal{U}(su(1,1))$. The Hopf algebra $\mathcal{U}_q(sl(2, \mathbb{C}))$ is generated by
the elements $K^{\pm 1}, E, F$, obeying the relations:
\begin{align}\label{Uq-def}
K K^{-1}=K^{-1} K=1, \quad K E K^{-1}= q E, \quad K F K^{-1}= q^{-1} F, \quad EF - FE =\frac{K^2-K^{-2}}{q-q^{-1}}.
\end{align}
The coproduct is defined on the generators
as:
\begin{align}
&\Delta(E)=K\otimes E+E\otimes K^{-1}, \quad \Delta(F)=K\otimes F+F\otimes K^{-1},\nonumber\\
&\Delta(K)=K\otimes K, \quad  \Delta(K^{-1})=K^{-1}\otimes K^{-1},
\end{align}
and the rest of the items as:
\begin{align*}
& \epsilon(K)=\epsilon(K^{-1})=1, \quad \epsilon(E)=\epsilon(F)=0,\\
& S(K)=K^{-1}, \quad S(E)=-q^{-1} E, \quad S(F)=-qF, \quad S(K^{-1})=K.
\end{align*}
The homomorphisms $\Delta, \epsilon$ and antihomomorphism $S$ are then extended to the whole algebra.

A Casimir element of $\mathcal{U}_q(sl(2, \mathbb{C}))$ is defined as
\begin{align}\label{uqsl2-Casimir}
\Omega \colonequals \frac{\left(q^{-1}-q\right)^2 FE -qK^2-q^{-1} K^{-2}}{2}
\end{align}
It is a self-adjoint element of $\mathcal{U}_q(sl(2, \mathbb{C}))$ and generates its center.
\medskip \\
2. {\it Real structure} \\
As in the case of the Lie algebra $sl(2, \mathbb{C})$, to define our Hopf $*$-algebra $\mathcal{U}_q(su(1,1))$ we now need to choose a real structure on the algebra $\mathcal{U}_q(sl(2, \mathbb{C}))$. Let us pause for a second and see what are possible choices here. For this we allow ourselves to look at complex $q$, such that $q^{\pm 1}$ are not roots of unity. A $*$-structure is by definition an (antilinear anti)automorphism $*$, $*^2=\iota$ of the algebra, such that the counit and antipode are equivariant with respect to this operation.

The supply of such automorphisms (up to equivalence\footnote{In a usual vein, two such different $*$-operations, $*_1$ and $*_2$ are said to be equivalent if there is a Hopf algebra automorphism $\rho$ of $\mathcal{U}_q(sl(2, \mathbb{C}))$, such that $\rho \circ *_1=*_2 \circ \rho$.}) for our Hopf algebra is rather limited, which follows from a structural theorem known as Poincare-Birkhoff-Witt decomposition. It states that $\mathcal{U}_q(sl(2,\mathbb{C}))$ is an infinite-dimensional vector space, with a basis given by the ordered monomials built out of the generators:
\begin{align}
\left\{E^l K^m F^n| \, m\in \mathbb{Z}, \, l,n \in \mathbb{Z}_{\geq 0}\right\}.
\end{align}
In short, from here one can see that any involution should preserve certain subspaces of elements which does not leave much room for options.\cite{Masuda1990a}

There are two classes of non-equivalent possibilities for $*$ yielding a non-compact real form in our $\mathcal{U}_q(sl(2, \mathbb{C}))$ case. This is similar to the dual Hopf algebra $A_q$ and unlike e.g. $sl(2, \mathbb{C})$ (where the two non-compact real forms are equivalent). The first non-compact real form
$\mathcal{U}_q(su(1,1))$, $0 < q < 1$ corresponds to
\begin{align}\label{Uq-real}
K^*=K, \quad E^*=-F \quad F^*=-E,
\end{align} 
and the other choice
\begin{align}\label{Uq-real2}
K^*=K, \quad E^*=-E, \quad F^*=-F
\end{align} 
corresponds to a real form known as $\mathcal{U}_q(sl(2, \mathbb{R}))$. One can see that compatibility of \eqref{Uq-def} and this second prescription requires $|q|=1$.
One can also easily check that coproduct and counit respect the above involutions.
\medskip \\
3. {\it $q\to 1^-$ limit} \\
Setting $K=e^{(q-1)H/2}$ and sending $q\to 1^-$ in the definition of $\mathcal{U}_q(su(1,1))$ recovers $\mathcal{U} \left(su(1,1)\right).$ Namely, we get the following commutation relations:
\begin{align}
[H, E]=2E, \quad [H,F]=-2F, \quad [E,F]=H,
\end{align}
whereas the coproduct reduces to a usual coproduct $\Delta$ of $su(1,1)$. It describes how the Lie algebra $su(1,1)$ acts in the tensor products of its representations I.e. for $X\in\{E, F, H\}$, $\Delta(X)=X\otimes 1+1 \otimes X$.\footnote{Elements of a Hopf algebra with such coproduct are sometimes called primitive.}
\medskip \\
4. {\it Pairing between }$\text{SU}_q(1,1)${\it and $\mathcal{U}_q(su(1,1))$} \\
Pairing, a non-degenerate duality between Hopf algebras $\mathcal{A}$ and $\mathcal{U}$, is a non-degenerate bilinear form on $\mathcal{U} \times \mathcal{A}$, such that:
\begin{align*}
&\langle \Delta(u), a\otimes b \rangle = \langle u, ab \rangle, \quad
\langle u\otimes v, \Delta(a) \rangle = \langle uv, a \rangle\\
&\langle 1, a \rangle = \epsilon(a), \quad \langle u,1 \rangle = \epsilon(u),\\
&\langle S(u), a \rangle = \langle a, S(u) \rangle, \quad \langle u^*, a \rangle = \overline{\langle u, S(a)^* \rangle}.
\end{align*}
In the undeformed case, say for $A_e$, this pairing reduces to the action of the algebra $\mathcal{U}(su(1,1))$ on its Lie group SU$(1,1)$ via left/right invariant vector fields.
The Hopf $*$-algebras SU$_q(1,1)$ and $\mathcal{U}_q=\mathcal{U}_q(su(1,1))$ can be paired in the above way.\cite{Koelink:2001}

For two such general Hopf $*$-algebras $A$ and $\mathcal{U}$ paired in nondegenerate duality one can define left and right actions of $\mathcal{U}$ on $A$ via 
\begin{align*}
u.a \colonequals (\iota \otimes u)(\Delta(a)), \quad a.u \colonequals (u \otimes \iota)(\Delta(a)), \quad u\in \mathcal{U}_q, a\in \mathcal{A}_q.
\end{align*}
For $X \in \text{Span}\{E, F, K-K^{-1}\}$, $a\in A_q$ is called left- (right-) $X$-invariant if $X.a=0$ ($a.X=0$).
\medskip \\
5. {\it Twisted primitive elements of $\mathcal{U}_q(su(1,1))$} \\
An element $g$ of a Hopf algebra $\mathcal{U}$ is called group-like if $\Delta(g)=g \otimes g$. E.g. elements $K^m\in \mathcal{U}_q(su(1,1))$, $m\in \mathbb{Z}$ are group-like.
An element $u$ of a Hopf algebra $\mathcal{U}$ is called twisted primitive (with respect to a group-like element $g\in\mathcal{U}$, $\Delta(g)=g \otimes g$) if its coproduct looks as
\begin{align}\label{tw-primitive-def}
\Delta(u)=g\otimes u + u \otimes S(g).
\end{align}
The twisted primitive elements behave in many respects similarly to the undeformed Lie algebra elements (which satisfy the relation \eqref{tw-primitive-def} with $g=1$ and are thus called primitive). \cite{Koornwinder-zonal, Koelink:2001}

In our $\mathcal{U}_q$ case, it is easy to see that \eqref{tw-primitive-def} is true for
\begin{align}
Y_s \colonequals q^{1/2} E -q^{-1/2} F+\frac{s+s^{-1}}{q^{-1}-q}(K-K^{-1}).
\end{align}
Using \eqref{Uq-def}, \eqref{Uq-real}, one can check that $Y_s K$ is self-adjoint if $s\in \mathbb{R}$ or $s\in \mathbb{C}, |s|=1$.
\medskip \\
6. {\it Irreducible $*$-representations of $\mathcal{U}_q(su(1,1))$} \\
Some of the admissible unitary $*$-irreps of $\mathcal{U}_q(su(1,1))$ are\footnote{Among these, only discrete, principal unitary and strange series appear in the decomposition of the left regular corepresentation, see also section \ref{sec:quantum-def}.}:
\begin{itemize}
\item Positive/negative discrete series $D^{\pm}_k$, $k\in \mathbb{N}/2$; the Casimir eigenvalue is
\begin{align*}
D^{\pm}_k\left(\Omega\right)=-\cos((2k-1)\ln q).
\end{align*}

Concretely, e.g. positive discrete series can be realized on $\ell^2(\mathbb{Z}_{\geq 0})$ spanned by $\{e_n| n\in \mathbb{Z}_{\geq 0}\}$, via
\begin{align*}
&K \, e_n=q^{k+n}e_n, \quad K^{-1} \, e_n = q^{-k-n}e_n\\
&E \, e_n = \frac{q^{-\frac{1}{2}-k-n}}{q^{-1}-q} \sqrt{\left(1-q^{2n+2}\right)\left(1-q^{4k+2n}\right)}e_{n+1}\\
&F \, e_n = \frac{q^{\frac{1}{2}-k-n}}{q^{-1}-q} \sqrt{\left(1-q^{2n}\right)\left(1-q^{4k+2n-2}\right)}e_{n-1}
\end{align*}
and $e_{-1}\equiv 0$.

\item Principal series $\pi_{b,\epsilon}$, $b\in [0; -\frac{\pi}{2 \ln q}]$ and $\epsilon=0,1/2$ but such that $(b, \epsilon) \neq (0,1/2)$\footnote{Otherwise it splits as $D_{1/2}^+\oplus D_{1/2}^-$ in analogy to its undeformed cousin.}; the Casimir eigenvalue is
\begin{align*}
\pi_{b,\epsilon}\left(\Omega\right)=\cos\left(-2b \ln q\right).
\end{align*}

It can be realized on $\ell^2(\mathbb{Z})$ spanned by $\{e_n| n\in \mathbb{Z}\}$, via
\begin{align*}
&K \, e_n=q^{n+\epsilon}e_n, \quad K^{-1} \, e_n = q^{-n-\epsilon}e_n\\
&E \, e_n = \frac{q^{-\frac{1}{2}-n-\epsilon}}{q^{-1}-q} \sqrt{\left(1+q^{2n+1+2\epsilon+2a}\right)\left(1+q^{2n+1+2\epsilon-2a}\right)}e_{n+1}\\
&F \, e_n = \frac{q^{\frac{1}{2}-n-\epsilon}}{q^{-1}-q} \sqrt{\left(1+q^{2n-1+2\epsilon+2a}\right)\left(1+q^{2n-1+2\epsilon-2a}\right)}e_{n-1}
\end{align*}

\item Finally, there are strange and complementary series of representations.

\end{itemize}

All representations which are not of the strange series have their counterparts in $q\to 1^-$ limit. The strange series disappears in the undeformed case.

\subsection{Locally compact quantum groups}
\label{app:operator-alg}

In conclusion of this section, let us mention that there are further analytical subtleties in treating non-compact quantum groups: in particular, the tensor products of representations of SU$_q(1,1)$ are ill-defined.\cite{woronowicz1991} A more detailed and careful approach goes via so called von Neumann algebraic quantum groups \cite{Kustermans1999, Kustermans2000}, see in particular \cite{Koelink2003, Groenevelt2010} where a properly defined non-compact quantum group SU$_q^{ext}(1,1)$ \cite{Korogodsky1994} is constructed. For classical (i.e. non-quantum) groups, this approach is analogous to considering von Neumann algebra of essentially bounded functions $L^{\infty}(G, d\mu)$ (modulo equivalence of two functions differing on a measure zero set) on the group as a main object of study. In this (von Neumann algebraic) setting of locally compact quantum groups, the integration against Haar measure is a functional with particularly nice properties (n.s.f. weight) on the von Neumann algebra and Hilbert space of square-integrable functions is recovered as a so called GNS space for that Haar weight. This gives an appropriate description of Haar measure for non-compact quantum groups, necessary topology to various function spaces on the quantum group and provides a finer description of quantum algebra (co)representations on Hilbert spaces, making the latter well-defined.
Much of the theory is developed in parallel to the lore of both the undeformed locally compact groups and quantum compact groups, albeit with more technical difficulties. Since in the present paper we do not use this machinery for explicit calculations, we restrain from giving any details.

\bibliographystyle{JHEP}
\bibliography{Double_scaled_SYK}

\end{document}